\documentclass[12pt]{article}
\RequirePackage[OT1]{fontenc}
\RequirePackage{amsthm,amsmath,amssymb,subcaption,graphicx,epstopdf,enumerate,lmodern}
\RequirePackage[round,colon,authoryear]{natbib}
\RequirePackage{bigints}
\RequirePackage[colorlinks,citecolor=blue,urlcolor=blue]{hyperref}
\usepackage{bm,color,fancyvrb,xcolor,mathtools}
\usepackage{amscd}
\usepackage{mathrsfs}
\usepackage{bbm}

\font\tencmmib=cmmib10 \skewchar\tencmmib '60
\newfam\cmmibfam
\textfont\cmmibfam=\tencmmib

\def\lessim{\ \lower4pt\hbox{$
		\buildrel{\displaystyle <}\over\sim$}\ }
\def\gessim{\ \lower4pt\hbox{$\buildrel{\displaystyle >}
		\over\sim$}\ }

\def\blist#1#2#3
{    \begin{list}{}{
			\setlength{\parsep}{0pt}
			\setlength{\leftmargin}{#1 pt}
			\setlength{\listparindent}{0pt}
			\setlength{\itemindent}{\listparindent}
			\setlength{\labelsep}{#3 pt}
			\setlength{\labelwidth}{\leftmargin}
			\addtolength{\labelwidth}{-\labelsep}
			\addtolength{\labelwidth}{\itemindent}
			\setlength{\rightmargin}{#2 pt}   }   }
	
	\def\elist{\end{list}}

\usepackage{tensor}

\newtheorem{theorem}{Theorem}[section]

\newtheorem{lemma}{Lemma}

\renewcommand{\hat}{\widehat}

\renewcommand{\hat}{\widehat}

\newcommand{\bfm}[1]{\ensuremath{\mathbf{#1}}}

\def\ba{\bfm a}

\def\be{\bfm e}   \def\bE{\bfm E}  \def\EE{\mathbb{E}}

     \def\RR{\mathbb{R}}

\def\bu{\bfm u}   \def\bU{\bfm U}  
\def\bv{\bfm v}     
\def\bw{\bfm w}     
\def\bx{\bfm x}   \def\bX{\bfm X}  
\def\by{\bfm y}   \def\bY{\bfm Y}  
\def\bz{\bfm z}

\def\calC{{\cal  C}} 
 
\def\calE{{\cal  E}}

\def\calL{{\cal  L}}

\def\calP{{\cal  P}} 
 
\def\calR{{\cal  R}} 
\def\calS{{\cal  S}}

\DeclareMathOperator{\argmin}{argmin}
\DeclareMathOperator{\argmax}{argmax}

\def\newpage{\vfill\eject}

\newcommand{\vertiii}[1]{{\left\vert\kern-0.25ex\left\vert\kern-0.25ex\left\vert #1 
		\right\vert\kern-0.25ex\right\vert\kern-0.25ex\right\vert}}

\usepackage{tikz}
\usetikzlibrary{decorations.pathreplacing,calc, patterns}

\def\scrE{\mathscr{E}}
\def\scrX{\mathscr{X}}
\def\scrU{\mathscr{U}}

\def\scrW{\mathscr{W}}
\def\scrT{\mathscr{T}}
\def\scrI{\mathscr{I}}

\usepackage{comment}

\usepackage{algorithm2e}

\begin{document}

	\title{On the Multiway Principal Component Analysis$^\ast$}
	\author{Jialin Ouyang and Ming Yuan\\
		Department of Statistics\\
		Columbia University}
	\date{(\today)}
	
	\maketitle

\begin{abstract}
Multiway data are becoming more and more common. While there are many approaches to extending principal component analysis (PCA) from usual data matrices to multiway arrays, their conceptual differences from the usual PCA, and the methodological implications of such differences remain largely unknown. This work aims to specifically address these questions. In particular, we clarify the subtle difference between PCA and singular value decomposition (SVD) for multiway data, and show that multiway principal components (PCs) can be estimated reliably in absence of the eigengaps required by the usual PCA, and in general much more efficiently than the usual PCs. Furthermore, the sample multiway PCs are asymptotically independent and hence allow for separate and more accurate inferences about the population PCs. The practical merits of multiway PCA are further demonstrated through numerical, both simulated and real data, examples.
\end{abstract}

	\footnotetext[1]{
		This research was supported in part by NSF Grants DMS-2015285 and DMS-2052955.}

\section{Introduction}
More and more often in practice, we need to deal with data of rich and complex structures that are more appropriately organized as multiway arrays rather than the usual data matrices. Examples of such multiway data are ubiquitous in many fields such as chemometrics, economics, psychometrics, and signal processing among others \citep[see, e.g.,][]{kroonenberg2008applied, anandkumar2014tensor, zhang2018tensor, chen2020statistical, chen2020semiparametric, han2020tensor, xia2020inference, bi2021tensors, chen2021factor, han2022optimal}. In this paper, we investigate the methodological implications and statistical properties of principal component analysis (PCA) for this type of data and pinpoint the benefits and challenges of doing so.

PCA is among the most popular statistical methods for multivariate data analysis when data are organized as matrices. See, e.g., \cite{tAND84a, pcabook}. With each column vector of a data matrix as an observation, PCA seeks orthogonal linear transformations of these vectors into a new coordinate system so that the variance of each coordinate is maximized successively. It allows us to represent most of the variation in the data by a small number of coordinates and therefore can guide us in reducing the dimensionality. As such, PCA often serves as a critical first step to capture the essential features in a dataset for many downstream analyses and is widely used in many scientific and engineering fields. Moving beyond matrices, for multiway data, each observation itself forms a matrix or more generally a multiway array. For example, when repeated measurements are made across different combinations of location and time, each observation can be more naturally organized as a matrix with each row corresponding to a certain location and each column a time point. To apply PCA to this type of data, it is tempting to neglect the multiway nature of the observations and treat each observation as a vector nonetheless, a practice often referred to as \emph{stringing}. However, as observed in numerous practical applications, appropriately accounting for the additional structure when applying PCA can greatly enhance interpretability and improve efficiency. See, e.g., \cite{kroonenberg2008applied}.

There is a long and illustrious history of developing suitable methods for such a purpose and it can be traced back at least to the pioneering work of Tucker, Harshman, and Carroll in the 1960s. Since then, numerous approaches have also been developed. Examples include \cite{kroonenberg1980principal,de2000multilinear,vasilescu2002multilinear, yang2004two, kong2005generalized, zhang20052d, lu2006multilinear, lu2008mpca, li2010l1, liu2017characterizing,taguchi2018tensor} among many others. See, e.g., \cite{lu2011survey, cichocki2015tensor} for recent surveys of existing techniques. Most of these developments are outside the mainstream statistics literature and often with a strong algorithmic flavor and exploratory data analysis focus. These approaches are intuitive and often yield more interpretable insights than naively applying PCA after stringing. However, their statistical underpinnings are largely unknown. The main goal of this article is to fill in this void. Indeed, as we shall demonstrate, a careful and rigorous statistical treatment allows for a better understanding of the operating characteristics of multiway PCA, leads to improved methodology, and reveals new opportunities and challenges in analyzing multiway data.

More specifically, we focus on a simple and natural approach to multiway PCA: when seeking linear transformations that maximize the variance, we impose the additional constraint that they conform to the multiway structure of the data. Doing so not only allows for enhanced interpretability but also inherits many nice properties of the usual PCA. Just as the usual principal components (PCs) are the eigenvectors of the covariance matrix, the multiway PCs can be identified with certain eigenvectors of the covariance operator. To better understand the impact of multiway structure on our ability to recover and make inferences about the multiway PCs, we also investigate the properties of multiway PCA under a spiked covariance model. 

Statistical properties of the usual PCA are well understood in the classical setting where the sample size is large whereas the number of variables is small \citep[see, e.g.,][]{tAND84a}. More and more often in today's applications, however, the dimensionality can also be large. There are abundant theoretical results concerning the usual PCA in such a high-dimensional setting as well, especially in the context of the spiked covariance model. For example, \cite{johnstone2009consistency} first demonstrated the critical role of dimensionality in PCA by showing that, with fixed signal strength, the sample PCA is consistent if and only if the number of variables is of a smaller order than the sample size. In another influential paper, \cite{paul2007asymptotics} established the asymptotic distribution of sample PCs. Other related treatments include \cite{baik2006eigenvalues, nadler2008finite, johnstone2009consistency, jung2009pca, bai2010spectral, lee2010convergence, benaych2011eigenvalues, bai2012sample, shen2013surprising, kolt14_2, kolt14, wang2017asymptotics, koltchinskii2020efficient} among numerous others. In a sense, our results naturally extend these earlier works to multiway PCA. However, the need to work with higher-order covariance operators rather than covariance matrices creates new and fundamental challenges and requires us to develop a different proof strategy and several new technical tools. More importantly, our analysis also reveals fundamental differences in behavior between the usual PCA and multiway PCA and inspires new methodological development for the latter.

Firstly, we establish the rates of convergence for the sample multiway PCs under mild regularity conditions. These rates explain why it is essential that we account for the inherent data structure when applying PCA to multiway data, and why naively applying PCA after stringing could be problematic. Intuitively, multiway PCA uses fewer parameters than the usual PCA and therefore is easier for estimation. This is described precisely by our result in that the estimation error of multiway PCs is determined by the dimension of each mode of the data array rather than the total number of entries, and therefore multiway PCs can be estimated accurately even if the latter far exceeds the sample size. But a more important observation is that how well a multiway PC can be estimated is determined by the corresponding eigenvalue of the covariance operator, and not the gap between its eigenvalues like the usual PCA. This somewhat surprising finding has far-reaching implications. In particular, it means that for multiway data the PCs can be estimated well even if their corresponding eigenvalues are not simple.

Moreover, to facilitate making statistical inferences about multiway PCs, we derive asymptotic distributions of the sample multiway PCs. Our results again reveal unexpected but important distinctions between multiway PCA and usual PCA. For example, the estimated multiway PCs are asymptotically independent of each other, and their asymptotic distribution is determined by their corresponding eigenvalues instead of eigengaps. Furthermore, we show that bias correction is important for the sample multiway PCs. Similar to the usual PCA, sample multiway PC can exhibit significant bias when the dimension (of each mode) is high. But there is also another source of bias that may arise due to the inherent ambiguity in ordering the PCs in absence of eigengaps. Nonetheless, we show that both types of bias can be eliminated, enabling us to make inferences about and construct confidence intervals for the multiway PCs.

The rest of the paper is organized as follows. In Section \ref{sec:multiway_pca}, we introduce the notion of multiway PCA both at a population level and how it works on a finite sample. Section \ref{sec:stat} investigates the rates of convergence for the sample multiway PCs. Turning our attention to the asymptotic distribution of multiway PCA in section \ref{sec:asy}, we show how to make valid inferences about the multiway PCs. The merits of the multiway PCA and our proposed approaches are further demonstrated through numerical experiments, both simulated and real, in Section \ref{sec:numerical}. We conclude with a summary in Section \ref{sec:summary}. Due to the space limit, all proofs are relegated to supplementary material.

\section{Multiway PCA} \label{sec:multiway_pca}
Multiway PCA can be viewed through the lens of usual PCA with the additional multiway structure imposed on the PCs. Let $\scrX\in \RR^{d_1\times\cdots\times d_p}$ be an order-$p$ random array. To simplify, we shall assume in what follows that $\scrX$ is centered, i.e., $\EE\scrX=0$, unless otherwise indicated. The idea behind PCA is to look for a linear transformation of $\scrX$ that maximizes the variance:
\begin{equation}
\label{eq:usualpc}
\max_{\scrW\in \RR^{d_1\times\cdots\times d_p}: \|\scrW\|_{\rm F}=1} {\rm var}(\langle \scrX, \scrW\rangle).
\end{equation}
Here $\|\scrW\|_{\rm F}=\langle \scrW,\scrW\rangle^{1/2}$ and
$$
\langle \scrX, \scrW\rangle=\sum_{j_1=1}^{d_1}\cdots\sum_{j_p=1}^{d_p} x_{j_1,\ldots,j_p}w_{j_1,\ldots,j_p}.
$$
Denote by $\scrU_1$ the solution to \eqref{eq:usualpc}. The basic premise of multiway PCA is that $\scrU_1$ conforms to the multiway structure underlying $\scrX$ in that it is a rank-one tensor and can be expressed as
\begin{equation}
	\label{eq:rank1pc}
\scrU_1=\bu_1^{(1)}\otimes\bu_1^{(2)}\otimes\cdots\otimes \bu_1^{(p)},
\end{equation}
where $\bu_1^{(q)}$ is a unit length vector $\RR^{d_q}$ and $\otimes$ stands for the outer product, i.e., the $(i_1,\ldots, i_p)$ entry of $\scrU_1$ is given by
$$\left[\scrU_1\right]_{i_1,\ldots, i_p}=u_{1i_1}^{(1)}u_{1i_2}^{(2)}\cdots u_{1i_p}^{(p)}.$$
In other words, $\scrU_1$ is also the solution to 
\begin{equation}
\label{eq:defpc}
\max_{\scrW\in \Theta} {\rm var}(\langle \scrX, \scrW\rangle),
\end{equation}
where $\Theta$ is the collection of all unit length rank-one tensors of conformable dimensions, i.e.,
$$
\Theta=\{\scrW=\bw^{(1)}\otimes\bw^{(2)}\otimes\cdots\otimes \bw^{(p)}: \bw^{(q)}\in \RR^{d_q}, \|\bw^{(q)}\|=1, \forall q=1,\ldots, p\}.
$$
Even if the solution to \eqref{eq:usualpc} is not strictly rank-one as described by \eqref{eq:rank1pc}, imposing such a constraint when seeking variance-maximizing transformation can nonetheless be desirable because of the enhanced interpretability: the additional rank-one constraint allows us to separate the effect along each mode, and help address questions such as ``who does what to whom and when'' which are often central to multiway data analysis. See, e.g., \cite{kroonenberg2008applied} for further discussion and numerous motivating examples.

Subsequent PCs can be defined successively:
\begin{equation}
\label{eq:usualpck}
\max_{\substack{\scrW\in \RR^{d_1\times\cdots\times d_p}: \|\scrW\|_{\rm F}=1\\ \scrW\perp \scrU_l, l=1,\ldots,k-1}} {\rm var}(\langle \scrX, \scrW\rangle).
\end{equation}
As before, we shall consider the case when the solution has rank one. A key requirement in defining PCs is that the $k$th PC is orthogonal to all other PCs, i.e., $\scrW\perp \scrU_l$. In vector case, i.e., $p=1$, this simply means that $\langle \scrW, \scrU_l\rangle=0$. In multiway case, however, there are many different notions of orthogonality. See, e.g., \cite{kolda2001orthogonal} for a detailed discussion on this subject. Each notion has its own subtleties and caveats that may have different statistical implications. In this work we shall focus on the notion of \emph{complete orthogonality}: two rank-one tensors $\scrW_1=\bw_1^{(1)}\otimes\bw_1^{(2)}\otimes\cdots\otimes \bw_1^{(p)}$ and $\scrW_2=\bw_2^{(1)}\otimes\bw_2^{(2)}\otimes\cdots\otimes \bw_2^{(p)}$ are complete orthogonal if and only if $\langle\bw_1^{(q)}, \bw_2^{(q)}\rangle= (\bw_1^{(q)})^\top \bw_2^{(q)}=0$ for all $q=1,\ldots, p$. More specifically, the $k$th multiway PC, denoted by $\scrU_k$, solves
\begin{equation}
\label{eq:defpck}
\max_{\scrW\in \Theta: \scrW\perp_c \scrU_l, \forall l<k} {\rm var}(\langle \scrX, \scrW\rangle),
\end{equation}
where $\perp_c$ stands for complete orthogonality.

As in the case of the usual PCA, multiway PCs can also be equivalently defined using the covariance matrix of ${\rm vec}(\scrX)$. In fact, it is more convenient to think of a covariance operator when it comes to multiway data. More specifically, we shall view
$$
\Sigma:={\rm cov}(\scrX)=\EE (\scrX\otimes \scrX)
$$
as a $d_1\times d_2\times\cdots \times d_p\times d_1\times\cdots \times d_p$ array. Then for any $\scrW\in\Theta$,
$$
{\rm var}(\langle \scrX, \scrW\rangle)=\langle \Sigma, \scrW\otimes\scrW\rangle=\langle \Sigma, \bw^{(1)}\otimes\cdots\otimes \bw^{(p)}\otimes \bw^{(1)}\otimes\cdots\otimes \bw^{(p)}\rangle.
$$
Write
$$
\lambda_k={\rm var}(\langle \scrX, \scrU_k\rangle).
$$
Because of the symmetry of $\Sigma$, 
$$
\scrU_1\otimes\scrU_1=\argmax_{\bw^{(1)}\otimes\cdots\otimes \bw^{(2p)}: \|\bw^{(q)}\|=1}\langle \Sigma, \bw^{(1)}\otimes\cdots\otimes\bw^{(2p)}\rangle
$$ 
so that $\lambda_1\scrU_1\otimes\scrU_1$ is also the best rank-one approximation to $\Sigma$ \citep[see, e.g.,][]{friedland2013best}. Similarly,
$$
\scrU_k\otimes\scrU_k=\argmax_{\substack{\bw^{(1)}\otimes\cdots\otimes \bw^{(2p)}: \|\bw^{(q)}\|=1\\ \bw^{(1)}\otimes\cdots\otimes \bw^{(2p)}\perp_c \scrU_l\otimes \scrU_l, \quad l<k}}\langle \Sigma, \bw^{(1)}\otimes\cdots\otimes\bw^{(2p)}\rangle
$$ 
%Equivalently,
%$$
%\scrU_k\otimes\scrU_k=\argmax_{\bw^{(1)}\otimes\cdots\otimes \bw^{(2p)}: \|\bw^{(q)}\|=1}\langle \check{\Sigma}_k, \bw^{(1)}\otimes\cdots\otimes\bw^{(2p)}\rangle.
%$$ 
%where
%$$
%\check{\Sigma}_k=\Sigma\times_1\calP_k^{(1)}\times_2\cdots\times_p\calP_k^{(p)}\times_{p+1}\calP_k^{(1)}\times_2\cdots\times_{2p}\calP_k^{(p)}
%$$
%and $\calP_k^{(1)}$ is the projection matrix of the linear subspace spanned by $\{\bu_l^{(q)}:1\le l<k\}$.

In vector case, e.g. $p=1$, $\{(\lambda_k,\scrU_k): k\ge 1\}$ are the eigenpairs of the covariance matrix $\Sigma$ and
$$
\Sigma_r:=\sum_{k=1}^r \lambda_k \scrU_k\otimes\scrU_k
$$
is the best rank-$r$ approximation to $\Sigma$, i.e.,
$$
\Sigma_r=\argmin_{A\in \RR^{d_1\times d_1}: {\rm rank}(A)\le r}\|A-\Sigma\|.
$$
When $p>1$, this characterization becomes tenuous because the notion of best low-rank approximation becomes precarious. For matrices, best low-rank approximations can be identified with singular value decomposition thanks to the Eckart-Young theorem. Low-rank approximation to tensors is much more subtle and the best low-rank approximation may not exist in general. See, e.g., \cite{hackbusch2012tensor}. Nonetheless, by construction, $\Sigma_r$ is the so-called best rank-$r$ \emph{greedy orthogonal approximation} to $\Sigma$. See, e.g., \cite{kolda2001orthogonal}. In particular, when the multiway structure does manifest itself in a way such that the usual PCs are rank-one tensors, for example, the solution to \eqref{eq:usualpc} and \eqref{eq:usualpck} has rank one, then $\Sigma_r$ is the best low-rank approximation to $\Sigma$. 

Sample multiway PCs can also be defined in a similar fashion. Specifically, given a sample $\scrX_1,\ldots, \scrX_n$ of independent copies of $\scrX$, $\scrU_k$s can be estimated by maximizing the sample variances:
\begin{equation}
	\label{eq:defsamplepc}
	\hat{\scrU}_k:=\argmax_{\scrW\in \Theta: \scrW\perp_c \scrU_l, \forall l<k} \frac{1}{n}\sum_{i=1}^n \langle \scrX_i, \scrW\rangle^2.
\end{equation}
Let
$$
\hat{\Sigma}=\frac{1}{n}\sum_{i=1}^n \scrX_i\otimes\scrX_i
$$
be the sample covariance operator.  Then $\hat{\scrU}_1$ can be defined via the best rank-one approximation to $\hat{\Sigma}$
$$
\hat{\scrU}_1=\argmax_{\scrW\in \Theta} \langle \hat{\Sigma}, \scrW\otimes\scrW\rangle.
$$
And other PCs can also be equivalently defined as
$$
\hat{\scrU}_k=\argmax_{\scrW\in \Theta: \scrW\perp_c \hat{\scrU}_l, \forall l<k} \langle \hat{\Sigma}, \scrW\otimes\scrW\rangle.
$$
Note that $\hat{\scrU}_k$ can also be identified with the best rank-one approximation to a deflated covariance operator:
$$
\hat{\scrU}_k=\argmax_{\scrW\in \Theta} \langle \check{\Sigma}_k, \scrW\otimes\scrW\rangle,
$$ 
where
$$
\check{\Sigma}_k=\hat{\Sigma}\times_1\hat{\calP}_k^{(1)}\times_2\cdots\times_p\hat{\calP}_k^{(p)}\times_{p+1}\hat{\calP}_k^{(1)}\times_{p+2}\cdots\times_{2p}\hat{\calP}_k^{(p)}
$$
and $\hat{\calP}_k^{(1)}$ is the projection matrix of the linear subspace spanned by $\{\hat{\bu}_l^{(q)}:1\le l<k\}$. Hereafter $\times_q$ represents the mode $q$ product between a tensor $\scrT \in \mathbb{R}^{d_1 \times d_2 \times \dots \times d_k}$ and a matrix $A \in \RR^{m \times d_q}$ so that $\scrT \times_q A\in \RR^{d_1 \times \dots d_{q-1} \times m \times d_{q+1} \dots \times d_k}$ with elements
$$
[\scrT \times_q A]_{i_1 \dots i_{q-1} j i_{q+1} \dots i_k} = \sum_{i_q=1}^{d_q} \scrT_{i_1 \dots i_q \dots i_k} A_{j i_q}.
$$

Computing the best rank-one approximation to a tensor is a classical problem in numerical linear algebra, and casting the sample multiway PCA as such allows us to take advantage of the many existing algorithms for doing so. In this work, we focus on the statistical properties of multiway PCA. Readers interested in further discussions about the computational aspect are referred to, e.g., \cite{zhang2001rank,hackbusch2012tensor, janzamin2019spectral} and references therein.

Similar to the usual PCs, multiway PCs can be used to construct low-rank approximations of the original data. However, there are also fundamental, albeit sometimes subtle, differences between the two types of PCA. The usual sample PCs coincide with the leading singular vectors of the data matrix after appropriate centering and therefore can be computed via singular value decomposition (SVD). In contrast, multiway PCA is, while closely related to, not equivalent to the best low-rank approximations of the original data array in general.  More specifically, consider stacking the observations into a higher-order tensor $\bX\in \RR^{n\times d_1\times\cdots\times d_p}$ whose $i$th frontal slice is $\scrX_i$. In the case when $\scrX$ is a vector, i.e., $p=1$, $\bX$ is a matrix and the sample PC, $\hat{\scrU}_k$ as defined above, is its $k$th right singular vector. It is therefore tempting to do the same and estimate $\scrU_k$s by seeking the best orthogonal low-rank approximation to $\bX$ directly:
\begin{equation}
	\label{eq:odeco}
	\min_{\substack{\scrW_1,\ldots,\scrW_r\in \Theta, \ba_1,\ldots,\ba_r\in \RR^n\\ \scrW_l\perp_c\scrW_k, \ba_l\perp \ba_k, \forall l\neq k}}\left\|\bX-\left(\ba_1\otimes\scrW_1+\cdots+\ba_r\otimes\scrW_r\right)\right\|_{\rm F}
\end{equation}
See, e.g., \cite{harshman1984parafac}. This problem, often known as the tensor SVD problem, has attracted a lot of attention in recent years. See, e.g., \cite{richard2014statistical, hopkins2015tensor, liu2017characterizing, zhang2018tensor, auddy2020perturbation}. 
However, the sample multiway PCs are generally not the solution to \eqref{eq:odeco}. First of all, the difference between the best orthogonal rank-$r$ and rank-$(r-1)$ approximations to $\bX$ is generally not a rank-one tensor and therefore cannot be associated with a multiway PC. See, e.g., \cite{hackbusch2012tensor}. To overcome this challenge, one may consider solving \eqref{eq:odeco} in a greedy fashion, i.e, optimizing \eqref{eq:odeco} over $\scrW_k$ and $\ba_k$ only while fixing the other ones. In general, however, this still results in a different set of PCs because of the extra orthogonality constraint on $\ba_k$s imposed by \eqref{eq:odeco}. As we shall see, this subtle distinction between multiway PCA and low-rank approximations to a data tensor not only means that a treatment different from that for the tensor SVD is needed for multiway PCA but also leads to different statistical behavior between the two.

\section{Rates of Convergence} \label{sec:stat}

A natural question one first asks is how well $\scrU_k$ and its components $\bu_k^{(q)}$s can be estimated by their sample counterparts. We shall now turn our attention to this question and study the rate of convergence for the sample multiway PCs. On the one hand, we provide further justification for the superiority of multiway PCA to the usual PCA with stringing, in addition to enhanced interpretability. On the other hand, our investigation also leads to new insights into the operating characteristics of sample multiway PCA and its intriguing distinction from the usual PCA. To fix ideas, we shall consider the so-called spiked covariance model as a working model for our theoretical development.

Suppose that a random array $\scrX\in \RR^{d_1\times \cdots\times d_p}$ follows a linear factor model:
\begin{equation}
\label{eq:spike}
\scrX=\sum_{k=1}^r \sigma_k\theta_k \scrU_k+\sigma_0\scrE,
\end{equation}
where $(\theta_1,\ldots,\theta_r)^\top\sim N(0,I_r)$ are the random factor loadings, $\scrU_k$s ($\in \Theta$) are unit length rank-one principal components such that $\scrU_k\perp_c \scrU_l$ for any $k\neq l$,  and $\scrE$ is a noise tensor with independent $N(0,1)$ entries. It is worth pointing out that our results and arguments can be extended beyond normality and applied to general subgaussian distributions. We opt for the normality assumption for ease of presentation. Without loss of generality, we shall also assume that eigenvalues of the signal are nontrivial and sorted in non-increasing order, i.e., $\sigma_1\ge \sigma_2\ge\cdots\ge\sigma_r>0$. Note that we do not require $\sigma_k$s to be distinct. It is not hard to see that the covariance operator of the aforementioned $\scrX$ is given by
$$
\Sigma=\sum_{k=1}^r\sigma_k^2\scrU_k\otimes \scrU_k +\sigma_0^2\scrI,
$$
where $\scrI$ is the identity tensor, i.e., $\scrI_{j_1\ldots j_pj_1'\ldots j_p'}=1$ if $j_q=j_q'$ for all $q=1,\ldots, p$ and $0$ otherwise. The spiked covariance model such as \eqref{eq:spike} is widely used as a working model to study PCA in the case of vector observations, i.e., $p=1$. See, e.g., \cite{johnstone2001distribution} and \cite{paul2007asymptotics}.

In this section, we shall establish the rates of convergence of the sample multiway PCs. To this end, denote by $\angle (\bw_1,\bw_2)$ the angle between two vectors $\bw_1$ and $\bw_2$ taking value in $[0,\pi/2]$, and similarly for two arrays $\scrW_1$ and $\scrW_2$, $\angle (\scrW_1,\scrW_2)$ denotes the angle between their vectorizations ${\rm vec}(\scrW_1)$ and ${\rm vec}(\scrW_2)$. 

It is instructive to begin with the classical setting where the dimensionality $d_1,\ldots,d_p$ as well as all other parameters, e.g. $\sigma_0$, $\sigma_k$s and $r$, are held fixed as the sample size $n$ diverges. Our first result shows that the sample PC $\hat{\scrU}_k$ and its components $\hat{\bu}_k^{(q)}$s are root-$n$ consistent in this regime.

\begin{theorem}\label{pr_rate}
Let $\scrX_1,\ldots,\scrX_n$ be independent observations following the spiked covariance model \eqref{eq:spike} with $p>1$ such that $\scrU_k=\bu_k^{(1)}\otimes\cdots\otimes\bu_k^{(p)}$ and $\sigma_k>0$. Assume that all parameters are fixed as the sample size $n$ increases. Let $\hat{\scrU}_k=\hat{\bu}_k^{(1)}\otimes\cdots\otimes \hat{\bu}_k^{(p)}$ be the sample multiway PC as defined by \eqref{eq:defsamplepc}. Then there exists a permutation $\pi$ over $[r]:=\{1,\ldots, r\}$ such that
\begin{equation}
\label{eq:rate}
\max_{1\le q\le p}\sin \angle ( \hat{\bu}_k^{(q)}, \bu_{\pi(k)}^{(q)} )=O_p(n^{-1/2}),
\end{equation}
for all $k\in [r]$, and hence
$$
\sin \angle ( \hat{\scrU}_k, \scrU_{\pi(k)}) =O_p(n^{-1/2}),\qquad k=1,\ldots, r
$$
as $n\to\infty$.
\end{theorem}

The most notable difference between the above result and those for the usual PCA \citep[e.g.,][]{tAND84a} is that fact that the root-$n$ consistency of the sample multiway PCs does not require that the eigenvalues ($(\sigma_k^2+\sigma_0^2)$s or equivalently $\sigma_k$s) of the covariance matrix be simple, i.e., $\sigma_k\neq \sigma_{k+1}$. Note that, without the multiway structural constraint, the usual PCs are only uniquely defined and hence can possibly be estimated if their corresponding eigenvalues are simple. As Theorem \ref{pr_rate} indicates, such a restriction is not necessary for multiway PCA. For multiway PCA, each sample PC is root-$n$ consistent regardless of the other eigenvalues. It is also worth noting that, since we do not require the $\sigma_k$s to be distinct, there is no guarantee that $\hat{\scrU}_k$ estimates $\scrU_k$. This is not a deficiency of multiway PCA, but rather a necessity due to the possible indeterminacy of the $k$th largest eigenvalue. In fact, if $\sigma_{k+1}<\sigma_k<\sigma_{k-1}$, then we can choose $\pi(k)=k$ in Theorem \ref{pr_rate}. In general, Theorem \ref{pr_rate} shows that each of the sample PCs is necessarily a root-$n$ consistent estimate of one of the multiway PCs.

To further understand the operating characteristics and merits of multiway PCA, we now consider the more general case and further highlight the role of dimensionality and signal-to-noise ratio. For brevity, in what follows, we shall assume that $\scrX$ is ``nearly cubic'' in that there exist constants $0<c_1,c_2<\infty$ such that $c_1d\le d_1,\ldots, d_p\le c_2 d$ for some natural number $d$ which may diverge with $n$. General cases can be treated similarly but incur considerably more cumbersome notation and tedious derivation.

\begin{theorem} \label{thm_rate}
Let $\scrX_1,\ldots,\scrX_n$ be independent observations following the spiked covariance model \eqref{eq:spike} with $p>1$ such that $\scrU_k=\bu_k^{(1)}\otimes\cdots\otimes\bu_k^{(p)}$. Let $\hat{\scrU}_k=\hat{\bu}_k^{(1)}\otimes\cdots\otimes \hat{\bu}_k^{(p)}$ be the sample multiway PC as defined by \eqref{eq:defsamplepc}. Suppose that 
\begin{equation}\label{eq:assumption}
r\log r\le c_0\min\{n,d\}\quad {\rm and}\quad  \left(\frac{\sigma_0}{\sigma_r} + \frac{\sigma_0^2}{\sigma_r^2}\right)\cdot \max\left\{\sqrt{\frac{d}{n}}, {\frac{d}{n}}\right\} \le \frac{c_0}{\sqrt{r}},
\end{equation}
for a sufficiently small constant $c_0>0$. Then there exist a constant $C>0$ and a permutation $\pi$ over $[r]$ such that
\begin{equation}
\label{eq:thm_rate}
\max_{1\le q\le p}\sin \angle ( \hat{\bu}_k^{(q)}, \bu_{\pi(k)}^{(q)} ) \le C\left(\frac{\sigma_0}{\sigma_{\pi(k)}} + \frac{\sigma_0^2}{\sigma_{\pi(k)}^2} \right)\cdot \max\left\{\sqrt{\frac{d}{n}}, {\frac{d}{n}}\right\},
\end{equation}
for all $k\in[r]$, and hence
$$
\sin \angle ( \hat{\scrU}_k, \scrU_{\pi(k)}) \le C\left(\frac{\sigma_0}{\sigma_{\pi(k)}} + \frac{\sigma_0^2}{\sigma_{\pi(k)}^2} \right)\cdot \max\left\{\sqrt{\frac{d}{n}}, {\frac{d}{n}}\right\},\qquad k=1,\ldots, r,
$$
with probability tending to one as $n$ diverges.
\end{theorem}

Theorem \ref{thm_rate} can be viewed as a generalization of Theorem \ref{pr_rate}. Its proof is rather involved and we shall brief discuss some of the challenges and the main ideas for resolving them. The proof proceeds by induction over $k$. Special attention is needed to deal with the case when an eigenvalue is not simple or the eigengap is small. This creates difficulty in identifying which multiway PC a sample multiway PC estimates, or equivalently the permutation $\pi$. To this end, we shall define
$$
\pi(1)=\argmax_{1\le l\le r} \left\{ \sigma_l^2 \left| \prod_{q=1}^p \langle \bu_l^{(q)}, \hat{\bu}_1^{(q)} \rangle \right|\right\},
$$
and for $k>1$,
$$
\pi(k):=\argmax_{l \notin \pi([k-1])} \left\{ \sigma_l^2 \left| \prod_{q=1}^p \langle \bu_l^{(q)}, \hat{\bu}_k^{(q)} \rangle \right|\right\}.
$$
To remove the influence of eigengaps altogether, we need to carefully quantify the impact of estimation error of $\hat{\scrU}_1, \ldots, \hat{\scrU}_{k-1}$ on the $k$th sample multiway PC. To this end, we shall derive bounds for both
$$
\max_{1\le q\le p}\sin\angle(\bu_{\pi(k)}^{(q)}, \hat{\bu}_k^{(q)}),
$$
and
$$
\max_{1\le q\le p}\max_{l\notin \pi([k])}\langle\bu_{l}^{(q)}, \hat{\bu}_k^{(q)}\rangle,
$$
and leverage the fact that the latter can be much smaller than the former.

When $d=O(n)$, the convergence rate given in Theorem \ref{thm_rate} is
$$
\sin \angle ( \hat{\scrU}_k, \scrU_{\pi(k)}) \le C\left(\frac{\sigma_0}{\sigma_{\pi(k)}} + \frac{\sigma_0^2}{\sigma_{\pi(k)}^2} \right)\cdot \sqrt{\frac{d}{n}};
$$
and when $d\gg n$, we have
$$
\sin \angle ( \hat{\scrU}_k, \scrU_{\pi(k)}) \le C\left(\frac{\sigma_0}{\sigma_{\pi(k)}} + \frac{\sigma_0^2}{\sigma_{\pi(k)}^2} \right)\cdot \frac{d}{n}.
$$
In particular, $\hat{\scrU}_k$ is consistent, e.g.,
$$
\sin \angle ( \hat{\scrU}_k, \scrU_{\pi(k)})=o_p(1),
$$
whenever $\sigma_{\pi(k)}/\sigma_0\gg \max\{d/n,(d/n)^{1/4}\}$.

Of particular interest here is the role of dimensionality. The rates of convergence given by Theorem \ref{thm_rate} depend on the dimensionality through $d$ rather than the ambient dimension $D:=d_1d_2\cdots d_p$. This is because multiway PCA restricts PCs to be rank-one tensors and therefore has fewer parameters. Such dimensionality reduction is especially important for multiway data. Consider, for example, the case when $\sigma_0$ and $\sigma_k$s are fixed, then by virtue of the results from \cite{johnstone2009consistency}, direct application of the usual PCA after stringing necessarily leads to an inconsistent estimate of $\scrU_k$ whenever $D\gg n$. Yet, our result indicates that multiway PCA is consistent whenever $d\ll n$.

To draw further comparisons with the usual PCA, we now focus on the case when $d\ll n$ and $r$, $\sigma_0,  \sigma_1,\ldots, \sigma_r$ are fixed. As shown by \cite{birnbaum2013minimax}, in this regime, the usual PCA (i.e., $p=1$) satisfies
$$
\sin \angle ( \hat{\scrU}_k, \scrU_{\pi(k)})\asymp \left(\frac{\sigma_0}{\sigma_{\pi(k)}} + \frac{\sigma_0^2}{\sigma_{\pi(k)}^2} \right)\cdot \sqrt{\frac{d}{n}}+{1\over \sqrt{n}}\left(\sum_{k'\neq k}{(\sigma_0+\sigma_{\pi(k)})(\sigma_0+\sigma_{\pi(k')})\over \sigma_{\pi(k)}^2-\sigma_{\pi(k')}^2}\right)
$$
with probability tending to one. Comparing the above rate with that from Theorem \ref{thm_rate}, it is clear that the difference between the two lie at the second term on the right hand side. Its presence for the usual PCA dictates that there should be no ties among $\sigma_k$s. Even if the $\sigma_k$s are all distinct, how well we can estimate a PC crucially depends on the gap between its corresponding eigenvalue and the other eigenvalues when $p=1$. In contrast, the bounds given by Theorem \ref{thm_rate} are determined by $\sigma_{\pi(k)}$ alone and not the eigengap $\min\{\sigma_{\pi(k)-1}^2-\sigma_{\pi(k)}^2, \sigma_{\pi(k)}^2-\sigma_{\pi(k)+1}^2\}$ as in the usual PCA case. 

It is also instructive to compare the convergence rate for the multiway PCA from Theorem \ref{thm_rate} with those for tensor SVD. Recall that
$$
\bX=\sum_{k=1}^r \sigma_k\Theta_k\otimes \scrU_k + \sigma_0 \bE,
$$
where $\Theta_k=(\theta_{1k},\ldots, \theta_{nk})^\top$ is a vector containing the $n$ realizations of $\theta_k$ and $\bE$ is a $n\times d_1\times\cdots\times d_p$ tensor whose $i$th frontal slice is $\scrE_i$. In contrast, $\Theta_k$s are deterministic in a tensor SVD model. If $\Theta_k$s are orthogonal to each other, then $\scrU_k$ can be estimated at the rate of $\sigma_0\|\bE\|/(\sigma_k\|\Theta_k\|)$ which is of the order $(\sigma_0/\sigma_k)\max\{\sqrt{d/n},1\}$. This is a direct consequence of the perturbations bounds from \cite{auddy2020perturbation} and a similar bound was also derived by \cite{richard2014statistical} in the rank-one case, i.e., $r=1$. In our case, however, $\Theta_k$ and $\Theta_l$ are random and in general not orthogonal to each other. As a result, the rates we obtained are different in their dependence on the signal-to-noise ratio $\sigma_k/\sigma_0$. Similar phenomenon has also been observed for the usual PCA \citep[see, e.g.,][]{birnbaum2013minimax}.

\section{Asymptotic Normality and Bias Correction} \label{sec:asy}

We now turn to the distributional properties of multiway PCA. This requires us to further delineate the role of bias in the sample PCs. It is known that the usual PCA is biased when the dimension ($D$) is large when compared with the sample size. See, e.g., \cite{kolt14_2, koltchinskii2020efficient} and the references therein. The same phenomenon is observed for the sample multiway PCs and a non-negligible bias arise when the dimension of each mode ($d$) is large when compared with the sample size. In addition, there is a more subtle source of bias for the sample multiway PCs due to the ambiguity in ordering the multiway PCs in the absence of eigengaps. As noted before, the lack of an eigengap means that the $k$th PC may not necessarily be estimated by the $k$th sample multiway PC. As a more concrete example, consider the case when $r=2$ and $\lambda_1=\lambda_2$. Then  $\scrU_1$ can be estimated by either $\hat{\scrU}_1$ or $\hat{\scrU}_2$, and as Theorem \ref{thm_rate} shows, the rate of convergence remains the same in both cases. But the asymptotic distribution may differ between the two scenarios: $\hat{\scrU}_2$ is required to be orthogonal to $\hat{\scrU}_1$ and estimating $\scrU_1$ by $\hat{\scrU}_2$ may incur extra bias. 

In this section, we shall introduce ways to correct for both types of bias and establish the asymptotic normality of the bias-corrected sample PCs. As is customary in the literature, we shall assume that $r$ and $\sigma_1,\ldots, \sigma_r$ are fixed for brevity. In light of the results from the previous section, the sample PCs are consistent if $d\ll n$ in this setting. We shall therefore focus on this regime in the current section.
%In other words, in absence of an eigengap or more generally if the eigengap is too small, the estimate of $\scrU_1$ may still depend upon the eigengap albeit in a subtle way. 

\subsection{When $d=o(\sqrt{n})$}

When $d$ is not too large, the bias is solely due to the possibility of repeated eigenvalues and thus ambiguity of the ordering of PCs. Indeed if $\sigma_k$s are distinct, then there is no need for bias correction when $d=o(\sqrt{n})$ and all of our results in this subsection will hold for the sample multiway PCs. But in practice, we may not know or want to assume that the eigenvalues are simple. Fortunately, we can remove any possible bias fairly easily by a simple one-step update of the sample PCs. More specifically, we shall consider estimating $\bu_{\pi(k)}^{(q)}$ by $\tilde{\bu}_k^{(q)}$, the leading eigenvector of 
$$
\hat{\Sigma}(\hat{\bu}_k^{(1)},\ldots,\hat{\bu}_k^{(q-1)}, \cdot ,\hat{\bu}_k^{(q+1)},\ldots, \hat{\bu}_k^{(p)}, \hat{\bu}_k^{(1)},\ldots,\hat{\bu}_k^{(q-1)}, \cdot ,\hat{\bu}_k^{(q+1)},\ldots, \hat{\bu}_k^{(p)}).
$$
The additional step frees up the orthogonality constraints imposed on the $k$th sample multiway PC and therefore allows us to suppress any adverse influence of $\hat{\scrU}_1,\ldots, \hat{\scrU}_{k-1}$.

We now consider the asymptotic distribution of the bias-corrected sample PCs. We again start with the classical regime when all parameters are fixed as $n$ increases.

\begin{theorem} \label{thm:low_dim}
Let $\scrX_1,\ldots,\scrX_n$ be independent observations following the spiked covariance model \eqref{eq:spike} with $p>1$ such that $\scrU_k=\bu_k^{(1)}\otimes\cdots\otimes\bu_k^{(p)}$ and $\sigma_k>0$. Assume that all parameters are fixed as the sample size $n$ increases. 
Let $\tilde{\bu}_1^{(q)}, \ldots, \tilde{\bu}_r^{(q)}$ be defined as above. Then there exists a permutation $\pi: [r]\to[r]$ such that
\begin{eqnarray*}
&\sqrt{n} \left[{\rm vec}(\tilde{\bU}^{(q)})-{\rm vec}(\bU^{(q)}_\pi)\right]\\
&\overset{d}{\to} N \left(0,   {\rm diag}\left(\left(\frac{\sigma_0^2}{\sigma_{\pi(1)}^2} + 
\frac{\sigma_0^4}{\sigma_{\pi(1)}^4}\right)\calP_{\bu_{\pi(1)}^{(q)}}^\perp,\ldots, \left(\frac{\sigma_0^2}{\sigma_{\pi(r)}^2} + 
\frac{\sigma_0^4}{\sigma_{\pi(r)}^4}\right)\calP_{\bu_{\pi(r)}^{(q)}}^\perp\right)\right),
\end{eqnarray*}
as $n\to\infty$, where $\tilde{\bU}^{(q)}=[\tilde{\bu}_1^{(q)},\ldots,\tilde{\bu}_r^{(q)}]$, $\bU_\pi^{(q)}=[\bu_{\pi(1)}^{(q)},\ldots,\bu_{\pi(r)}^{(q)}]$ and $\calP_{\bu_k^{(q)}}^\perp=I_{d_q}-\bu_k^{(q)} \otimes \bu_k^{(q)}$.
\end{theorem}

Theorem \ref{thm:low_dim} indicates that
$$
n\cdot{\rm var}\left(\tilde{\bu}_k^{(q)}\right)\to {\sigma_0^2(\sigma_{\pi(k)}^2+\sigma_0^2)\over \sigma_{\pi(k)}^4} \left(I-\bu_{\pi(k)}^{(q)}\otimes \bu_{\pi(k)}^{(q)}\right),
$$
and
$$
n\cdot{\rm cov}\left(\tilde{\bu}_k^{(q)},\tilde{\bu}_l^{(q)}\right)\to 0.
$$
Namely, all estimates of the multiway PCs are asymptotically normal and independent of each other. Note also that the asymptotic distribution of $\tilde{\bu}_k^{(q)}$ does not depend on other eigenvalues or PCs. In other words, it can be estimated to the same precision as if all other components $\scrU_l$, $l\neq k$ are known! This is to be contrasted with the usual PCA where the asymptotic distribution of $\bu_k^{(q)}$ depends on all other eigenvectors and eigenvalues.%This fact has a huge practical implication -- we can directly use it to construct confidence intervals for $\bu_k^{(q)}$ or its functional. Its asymptotic distribution depends only on $\sigma_0$, $\sigma_{\pi(k)}$, and $\bu_k^{(q)}$, all of which can be consistently estimated by their sample counterpart. %This is to be contrasted with the usual PCA where the asymptotic distribution of $\bu_k^{(q)}$ depends on all other eigenvectors and eigenvalues, and therefore can be much harder to approximate in practice.

More specifically, it is well known that in vector case, i.e., when $p=1$, under the additional assumption that $\sigma_1^2,\ldots,\sigma_r^2$ are distinct, the sample PCs satisfy
\begin{eqnarray*}
&n\cdot{\rm var}\left(\hat{\bu}_k^{(1)}\right) \\
&\to \sum_{1\le l\le r, l\neq k}{(\sigma_k^2+\sigma_0^2)(\sigma_l^2+\sigma_0^2)\over (\sigma_k^2-\sigma_l^2)^2} \bu_l^{(1)}\otimes \bu_l^{(1)}+{(\sigma_k^2+\sigma_0^2)\sigma_0^2\over \sigma_k^4} \left(I-\sum_{1\le l\le r}\bu_l^{(1)}\otimes \bu_l^{(1)}\right)
\end{eqnarray*}
and for any $1\le l\le r$ and $l\neq k$,
$$
n\cdot{\rm cov}\left(\hat{\bu}_k^{(1)},\hat{\bu}_l^{(1)}\right)\to -{(\sigma_k^2+\sigma_0^2)(\sigma_l^2+\sigma_0^2)\over (\sigma_k^2-\sigma_l^2)^2}\cdot \bu_k^{(1)}\otimes \bu_l^{(1)}.
$$
See, e.g., \cite{tAND84a}. It is clear that the sample PCs are always correlated with each other. Moreover, note that
$$
{\sigma_0^2(\sigma_k^2+\sigma_0^2)\over \sigma_k^4}\le {(\sigma_k^2+\sigma_0^2)(\sigma_l^2+\sigma_0^2)\over (\sigma_k^2-\sigma_l^2)^2}.
$$
and the strict inequality holds for any $k\neq l\le r$. This suggests that the estimated multiway PCs have smaller variations than the usual PCs with the same set of eigenvalues.

%\subsection{When $d=o(\sqrt{n})$}
We now turn our attention to the more general case when the dimensionality and other parameters are allowed to diverge with $n$. Because the PCs now may have different dimensions for different sample sizes, it is more natural to consider their linear forms, e.g. $\langle \bu^{(q)}_k, \bv\rangle$, for some fixed vector $\bv\in \RR^{d_q}$. If the dimensions are fixed, Theorem \ref{thm:low_dim} immediately suggests that $\langle \tilde{\bu}^{(q)}_k, \bv\rangle$ estimates $\langle \bu^{(q)}_{\pi(k)}, \bv\rangle$, and
$$
\sqrt{n}\left(\langle \tilde{\bu}^{(q)}_k, \bv\rangle-\langle \bu^{(q)}_{\pi(k)}, \bv\rangle\right)\to_d N\left(0,\left(\frac{\sigma_0^2}{\sigma_{\pi(k)}^2} + 
\frac{\sigma_0^4}{\sigma_{\pi(k)}^4}\right)\|\calP_{\bu_{\pi(k)}^{(q)}}^\perp\bv\|^2\right)
$$
The following result shows that this continues to hold as long as $d=o(\sqrt{n})$.

\begin{theorem} \label{thm:linear_form}
Let $\scrX_1,\ldots,\scrX_n$ be independent observations following the spiked covariance model \eqref{eq:spike} with $p>1$ such that $\scrU_k=\bu_k^{(1)}\otimes\cdots\otimes\bu_k^{(p)}$ and $\sigma_k>0$. %Let $\hat{\scrU}_k=\hat{\bu}_k^{(1)}\otimes\cdots\otimes \hat{\bu}_k^{(p)}$ be the sample PC as defined by \eqref{eq:defsamplepc}. 
Assume that \eqref{eq:assumption} holds, $d=o(\sqrt{n})$. Then there exists a permutation $\pi:[r]\to[r]$ such that
$$
\sqrt{n}\left(\langle \tilde{\bu}^{(q)}_k, \bv\rangle-\langle \bu^{(q)}_{\pi(k)}, \bv\rangle\right)\to_d N\left(0,\left(\frac{\sigma_0^2}{\sigma_{\pi(k)}^2} + 
\frac{\sigma_0^4}{\sigma_{\pi(k)}^4}\right)\|\calP_{\bu_{\pi(k)}^{(q)}}^\perp\bv\|^2\right)
$$
as $n\to\infty$.
\end{theorem}

Theorem \ref{thm:linear_form} shows that the same asymptotic behavior of $\tilde{\bu}_j^{(q)}$ as in the fixed dimension case can be expected whenever $d=o(\sqrt{n})$.

\subsection{When $d=o(n)$}

For higher dimension, the simple bias-correction described above is no longer sufficient and a close inspection reveals that $\tilde{\bu}_j^{(q)}$ still incurs a non-negligible bias when $d\gg \sqrt{n}$. Thankfully, both types of bias can be corrected with a sample-splitting approach similar in spirit to the scheme developed by \cite{kolt14_2} for the usual PCA.

Without loss of generality, assume that $n$ is an even number and we randomly split the $n$ observations into two halves: $(\scrX_1,\ldots, \scrX_{n/2})$ and $(\scrX_{n/2+1},\ldots,\scrX_{n})$. Denote by $\hat{\Sigma}^{[1]}$ and $\hat{\Sigma}^{[2]}$ the sample covariance operator based on the two halves of data respectively. Similarly, we shall write $\hat{\scrU}_k^{[h]}=\hat{\bu}_k^{(1),[h]}\otimes\cdots\otimes \hat{\bu}_k^{(p),[h]}$ the $k$th sample PC based on the $h$ ($=1$ or $2$) halves of the data. However, as noted before, $\hat{\scrU}_k^{[1]}$ and $\hat{\scrU}_k^{[2]}$ may not estimate the same PC. To this end, we shall reorder $\hat{\scrU}_k^{[1]}$s and $\hat{\scrU}_k^{[2]}$s with $\hat{\scrU}_k$s (i.e., the estimators derived from the entire dataset) as reference points. Specifically, without loss of generality, we assume that
$$
\hat{\scrU}_k^{[1]}=\argmin_{k\le l\le r} \sin\angle\left(\hat{\scrU}_l^{[1]}, \hat{\scrU}_k\right).
$$
The same procedure is applied to relabel $\hat{\scrU}_k^{[2]}$s. Note also that the sign of a PC is irrelevant in that $\scrU_k$ and $-\scrU_k$ represent the same transformation. We shall therefore also assume hereafter, without loss of generality, that $\langle \hat{\bu}_k^{(q),[1]},\hat{\bu}_k^{(q),[2]}\rangle\ge 0$.

Recall that  
\begin{eqnarray*}
	\sigma_k^2 \bu_k^{(q)} {\bu_k^{(q)}}^{\top} + \sigma_0^2 I_{d_q} =\Sigma(\bu_k^{(1)},\ldots,\bu_k^{(q-1)},\cdot,\bu_k^{(q+1)},\ldots,\bu_k^{(p)},\\
	\bu_k^{(1)},\ldots,\bu_k^{(q-1)},\cdot,\bu_k^{(q+1)},\ldots,\bu_k^{(p)}).
\end{eqnarray*}
We shall then update the sample PC using the above identity with $\Sigma$ and $\bu_k^{(q)}$s estimated from separate halves. Denote by $\check{\bu}_k^{(q),[1]}$ the leading eigenvector of  
\begin{align*}
    \hat{\Sigma}^{[1]}& (\hat{\bu}_k^{(1),[2]},\dots,\hat{\bu}_k^{(q-1),[2]},\cdot,\hat{\bu}_k^{(q+1),[2]},\dots,\hat{\bu}_k^{(p),[2]}, \\
	&\hat{\bu}_k^{(1),[2]},\dots,\hat{\bu}_k^{(q-1),[2]},\cdot,\hat{\bu}_k^{(q+1),[2]},\dots,\hat{\bu}_k^{(p),[2]}),
\end{align*}
and similarly $\check{\bu}_k^{(q),[2]}$ the leading eigenvector of  
\begin{align*}
	\hat{\Sigma}^{[2]}& (\hat{\bu}_k^{(1),[1]},\dots,\hat{\bu}_k^{(q-1),[1]},\cdot,\hat{\bu}_k^{(q+1),[1]},\dots,\hat{\bu}_k^{(p),[1]}, \\
	& \hat{\bu}_k^{(1),[1]},\dots,\hat{\bu}_k^{(q-1),[1]},\cdot,\hat{\bu}_k^{(q+1),[1]},\dots,\hat{\bu}_k^{(p),[1]}).
\end{align*}
To avoid losing efficiency due to sample splitting, we consider a new estimate $\check{\scrU}_k=\check{\bu}_k^{(1)}\otimes\cdots\otimes \check{\bu}_k^{(p)}$ where
$$
\check{\bu}_k^{(q)}={\check{\bu}_k^{(q),[1]}+\check{\bu}_k^{(q),[2]}\over \left\|\check{\bu}_k^{(q),[1]}+\check{\bu}_k^{(q),[2]}\right\|}.
$$
The following theorem shows that we can construct an unbiased estimate of $\langle \bu^{(q)}_{\pi(k)}, \bv\rangle$ by appropriately rescaling $\langle \check{\bu}^{(q)}_k, \bv\rangle$, as long as $d=o(n^{2/3})$.

\begin{theorem} \label{thm:bias_simple}
Let $\scrX_1,\ldots,\scrX_n$ be independent observations following the spiked covariance model \eqref{eq:spike} with $p>1$ such that $\scrU_k=\bu_k^{(1)}\otimes\cdots\otimes\bu_k^{(p)}$ and $\sigma_k>0$. Let $\check{\scrU}_k=\check{\bu}_k^{(1)}\otimes\cdots\otimes \check{\bu}_k^{(p)}$ be the estimated PC as defined above. Assume $r$ and $\sigma_1,\dots,\sigma_r$ are fixed, and $d=o(n^{2/3})$. Then there exists a permutation $\pi:[r]\to[r]$ such that
$$
\sqrt{n}\left((1+b_k^{(q)})\langle \check{\bu}^{(q)}_k, \bv\rangle-\langle \bu^{(q)}_{\pi(k)}, \bv\rangle\right)\to_d N\left(0,\left(\frac{\sigma_0^2}{\sigma_{\pi(k)}^2} + 
\frac{\sigma_0^4}{\sigma_{\pi(k)}^4}\right)\|\calP_{\bu_{\pi(k)}^{(q)}}^\perp\bv\|^2\right)
$$
as $n\to\infty$ where
\begin{equation}
	\label{eq:bias_simple}
	b_k^{(q)}=\sqrt{ 1+ \frac{d_q}{n} \left(\frac{\sigma_0^2}{\sigma_{\pi(k)}^2} + 
		\frac{\sigma_0^4}{\sigma_{\pi(k)}^4}\right) }-1.
\end{equation}
\end{theorem}

It is worth pointing out that when $d=o(n^{1/2})$, the bias correction factor described by \eqref{eq:bias_simple} obeys $b_k^{(q)}=o(n^{-1/2})$ and therefore can be neglected. This agrees with our earlier observation and of course also suggests that sample-splitting is unnecessary if $d\ll n^{1/2}$. When $d\gg n^{1/2}$, bias correction becomes essential. In particular, Theorem \ref{thm:bias_simple} suggests that, as long as $d=o(n^{2/3})$, an explicit bias correction factor can be applied. For higher dimensions, it is unclear if a similar explicit expression exists for the debiasing factor. Nonetheless, we can derive a suitable bias correction factor for all $d\ll n$ via additional sample splitting.

More specifically, we first randomly split the observations into two halves. The first half of the data is then further split into two equal-sized groups to compute the sample covariance operators $\hat{\Sigma}^{[1][1]}$ and $\hat{\Sigma}^{[1][2]}$, then we compute $\hat{\bu}_k^{(q),[1][1]}$ and $\hat{\bu}_k^{(q),[1][2]}$ as the leading eigenvectors of
\begin{align*}
    \hat{\Sigma}^{[1][1]}& (\hat{\bu}_k^{(1),[2]},\dots,\hat{\bu}_k^{(q-1),[2]},\cdot,\hat{\bu}_k^{(q+1),[2]},\dots,\hat{\bu}_k^{(p),[2]}, \\
	&\hat{\bu}_k^{(1),[2]},\dots,\hat{\bu}_k^{(q-1),[2]},\cdot,\hat{\bu}_k^{(q+1),[2]},\dots,\hat{\bu}_k^{(p),[2]}), \\
	\hat{\Sigma}^{[1][2]}& (\hat{\bu}_k^{(1),[2]},\dots,\hat{\bu}_k^{(q-1),[2]},\cdot,\hat{\bu}_k^{(q+1),[2]},\dots,\hat{\bu}_k^{(p),[2]}, \\
	&\hat{\bu}_k^{(1),[2]},\dots,\hat{\bu}_k^{(q-1),[2]},\cdot,\hat{\bu}_k^{(q+1),[2]},\dots,\hat{\bu}_k^{(p),[2]}),
\end{align*}
Similarly, we used the second half of the data to compute $\hat{\bu}_k^{(q),[2][1]}$s, and $\hat{\bu}_k^{(q),[2][2]}$s. As before, we shall sort these estimates in compatible order and sign. Let
\begin{equation}
	\label{eq:bias_general}
	\hat{b}_k^{(q)} = \frac{ \left\|\check{\bu}_k^{(q),[1]}+\check{\bu}_k^{(q),[2]}\right\| }{ \sqrt{\left\langle \hat{\bu}_k^{(q),[1][1]}, \hat{\bu}_k^{(q),[1][2]} \right \rangle}+\sqrt{\left\langle \hat{\bu}_k^{(q),[2][1]}, \hat{\bu}_k^{(q),[2][2]} \right \rangle} } -1.
\end{equation}

\begin{theorem} \label{thm_b}
Let $\scrX_1,\ldots,\scrX_n$ be independent observations following the spiked covariance model \eqref{eq:spike} with $p>1$ such that $\scrU_k=\bu_k^{(1)}\otimes\cdots\otimes\bu_k^{(p)}$ and $\sigma_k>0$. Let $\check{\scrU}_k=\check{\bu}_k^{(1)}\otimes\cdots\otimes \check{\bu}_k^{(p)}$ be the estimated PC as defined above. Assume $r$ and $\sigma_1,\dots,\sigma_r$ are fixed, and $d=o(n)$.
Then there exists a permutation $\pi:[r]\to[r]$ such that
%    \begin{align*}
%    b_k^{(q)}=\sqrt{ 1+ \frac{d_q}{n} \left(\frac{\sigma_0^2}{\sigma_{\pi(k)}^2} + \frac{\sigma_0^4}{\sigma_{\pi(k)}^4}\right) }-1 + O_p\left( \frac{d^{3/2}}{n^{3/2}} \right) + o_p\left( \frac{1}{\sqrt{n}} \right),
%    \end{align*}
%    and
    $$
	\sqrt{n}\left((1+\hat{b}_k^{(q)})\langle \check{\bu}^{(q)}_k, \bv\rangle-\langle \bu^{(q)}_{\pi(k)}, \bv\rangle\right)\to_d N\left(0,\left(\frac{\sigma_0^2}{\sigma_{\pi(k)}^2} + 
	\frac{\sigma_0^4}{\sigma_{\pi(k)}^4}\right)\|\calP_{\bu_{\pi(k)}^{(q)}}^\perp\bv\|^2\right),
	$$
    as $n\to\infty$ where $\hat{b}_k^{(q)}$ is given by \eqref{eq:bias_general}. Moreover,
    \begin{align*}
	    \hat{b}_k^{(q)}=\sqrt{ 1+ \frac{d_q}{n} \left(\frac{\sigma_0^2}{\sigma_{\pi(k)}^2} + \frac{\sigma_0^4}{\sigma_{\pi(k)}^4}\right) }-1 + O_p\left( \frac{d^{3/2}}{n^{3/2}} \right) + o_p\left( \frac{1}{\sqrt{n}} \right).
	\end{align*}
\end{theorem}

%Note that Theorem \ref{thm:bias_simple} is the special case of Theorem \ref{thm_b}, since $O_p\left( \frac{d^{3/2}}{n^{3/2}} \right) = o_p\left( \frac{1}{\sqrt{n}} \right)$ when $d = o(n)$.)}

In light of Theorem \ref{thm_b}, the double sample splitting approach can be employed to derive confidence intervals for linear forms of the multiway PCs as long as $d=o(n)$. This robustness, however, comes at the expense of increased computational cost and could incur a loss of efficiency in finite samples. In practice, one may still prefer the explicit bias correction as described by Theorem \ref{thm:bias_simple} if $d$ is not very large, or the one-step update if $d$ is small.

\subsection{Inference about multiway PCs}
The asymptotic normality we showed earlier in the section forms the basis for making inferences about linear forms $\langle \bu^{(q)}_{\pi(k)}, \bv\rangle$. In particular, one of the most interesting and also simplest examples of linear forms of PCs is their coordinates, i.e., $\bv$ is a column vector of the identity matrix. To derive confidence intervals of or testing hypotheses about $\langle \bu^{(q)}_{\pi(k)}, \bv\rangle$, however, we need to also estimate its variance. Specifically, its asymptotic distribution depends only on $\sigma_0$, $\sigma_{\pi(k)}$, and $\bu_k^{(q)}$, all of which can be consistently estimated by their sample counterpart. Let
$$
\hat{\sigma}_0^2={1\over \prod_{q=1}^p (d_q-r)}\sum_{1\le i_q\le d_q, 1\le q\le p}[\check{\Sigma}_{r+1}]_{i_1\cdots i_pi_1\cdots i_p}
$$
and
$$
\hat{\sigma}_{\pi(k)}^2=\langle\hat{\Sigma}, \hat{\scrU}_k\otimes \hat{\scrU}_k\rangle-\hat{\sigma}_0^2.
$$
The following theorem suggest that the asymptotic normality remains valid if we replace the variance of linear forms $\langle \bu^{(q)}_{\pi(k)}, \bv\rangle$ with these estimates:

\begin{theorem}
	\label{thm:inference}
Let $\scrX_1,\ldots,\scrX_n$ be independent observations following the spiked covariance model \eqref{eq:spike} with $p>1$ such that $\scrU_k=\bu_k^{(1)}\otimes\cdots\otimes\bu_k^{(p)}$ and $\sigma_k>0$. Assume $r$ and $\sigma_1,\dots,\sigma_r$ are fixed.
There exists a permutation $\pi:[r]\to[r]$ such that
\begin{itemize}
\item[(a)] If $d=o(\sqrt{n})$, then
$$
{\sqrt{n}\left(\langle \tilde{\bu}^{(q)}_k, \bv\rangle-\langle \bu^{(q)}_{\pi(k)}, \bv\rangle\right)\over \sqrt{\frac{\hat{\sigma}_0^2}{\hat{\sigma}_{\pi(k)}^2} + \frac{\hat{\sigma}_0^4}{\hat{\sigma}_{\pi(k)}^4}} \ \big\|\calP_{\tilde{\bu}_{\pi(k)}^{(q)}}^\perp\bv\big\| }\to_d N\left(0,1\right),
$$
\item[(b)] If $d=o(n^{2/3})$, then
$$
{\sqrt{n}\left((1+b_k^{(q)})\langle \check{\bu}^{(q)}_k, \bv\rangle-\langle \bu^{(q)}_{\pi(k)}, \bv\rangle\right)\over \sqrt{\frac{\hat{\sigma}_0^2}{\hat{\sigma}_{\pi(k)}^2} + \frac{\hat{\sigma}_0^4}{\hat{\sigma}_{\pi(k)}^4}} \ \big\|\calP_{\check{\bu}_{\pi(k)}^{(q)}}^\perp\bv\big\| }\to_d N\left(0,1\right),
$$
where $b_k^{(q)}$ is given by \eqref{eq:bias_simple}.
\item[(c)] If $d=o(n)$, then
$$
{\sqrt{n}\left((1+\hat{b}_k^{(q)})\langle \check{\bu}^{(q)}_k, \bv\rangle-\langle \bu^{(q)}_{\pi(k)}, \bv\rangle\right)\over \sqrt{\frac{\hat{\sigma}_0^2}{\hat{\sigma}_{\pi(k)}^2} + \frac{\hat{\sigma}_0^4}{\hat{\sigma}_{\pi(k)}^4}} \ \big\|\calP_{\check{\bu}_{\pi(k)}^{(q)}}^\perp\bv\big\| }\to_d N\left(0,1\right),
$$
where $\hat{b}_k^{(q)}$ is given by \eqref{eq:bias_general}.
\end{itemize}
\end{theorem}

Theorem \ref{thm:inference} is an immediate consequence of Slutsky's Theorem and Theorems \ref{thm:linear_form}-\ref{thm_b}. It allows us to make inference or construct confidence intervals for $\langle \bu^{(q)}_{\pi(k)}, \bv\rangle$. Consider, for example, testing hypothesis that
$$
H_0: \langle \bu^{(q)}_{\pi(k)}, \bv\rangle=0\qquad vs \qquad H_a:\langle \bu^{(q)}_{\pi(k)}, \bv\rangle\neq 0,
$$
when $d=o(\sqrt{n})$. We can proceed to reject $H_0$ if and only if
$$
\left|\sqrt{n}\left(\langle \tilde{\bu}^{(q)}_k, \bv\rangle-\langle \bu^{(q)}_{\pi(k)}, \bv\rangle\right)\right|\ge z_{\alpha/2}\sqrt{\frac{\hat{\sigma}_0^2}{\hat{\sigma}_{\pi(k)}^2} + 
\frac{\hat{\sigma}_0^4}{\hat{\sigma}_{\pi(k)}^4}} \big\|\calP_{\tilde{\bu}_{\pi(k)}^{(q)}}^\perp\bv\big\|,
$$
where $z_{\alpha/2}$ is the upper $\alpha/2$ quantile of the standard normal distribution. Theorem \ref{thm:inference} guarantees this is a level-$\alpha$ test asymptotically. Similarly, we can also construct $(1-\alpha)$ confidence interval for $\langle \bu^{(q)}_{\pi(k)}, \bv\rangle$:
$$
\left(\langle \tilde{\bu}^{(q)}_k, \bv\rangle\pm \frac{z_{\alpha/2}}{\sqrt{n}}\sqrt{\frac{\hat{\sigma}_0^2}{\hat{\sigma}_{\pi(k)}^2} + 
\frac{\hat{\sigma}_0^4}{\hat{\sigma}_{\pi(k)}^4}} \big\|\calP_{\tilde{\bu}_{\pi(k)}^{(q)}}^\perp\bv\big\|\right).
$$
In particular, by taking $\bv\in\{\be_1,\ldots, \be_{d_q}\}$, we can use the above formula to derive confidence intervals for the coordinates of $\bu_{\pi(k)}^{(q)}$. Situations with larger $d$ can also be treated accordingly.

\section{Numerical Experiments} \label{sec:numerical}

To complement our theoretical analyses and further demonstrate the merits of multiway PCA, we conducted several sets of numerical experiments.

\subsection{Simulation Studies} \label{sec:simu}

We first present a set of simulation studies to illustrate the finite-sample behavior of the sample PCs. These experiments are specifically designed to assess the role of bias correction, and robustness to deviation from the normal distribution. Throughout this subsection, unless otherwise noted, samples were generated according to the spike covariance model \eqref{eq:spike} with $p=2$, e.g., each $\scrX_i$ is a matrix. Since the two modes are exchangeable, we only focus on the first mode $q=1$ for brevity. We also fixed the number of spikes at $r=2$. In each case, we shall set the singular values $\sigma_1=\sigma_2$. In other words, for each of our examples, the usual PCA (with stringing) will not be able to identify the PCs because of the multiplicity. As mentioned before, without loss of generality and for the sake of brevity, we reordered $\check{\bu}_1^{(1)}$ and $\check{\bu}_2^{(1)}$ such that $\sin \angle (\check{\bu}_1^{(1)},\bu_1)\le \sin \angle (\check{\bu}_2^{(1)},\bu_1)$. In addition, we replaced $\check{\bu}_k^{(1)}$ with $-\check{\bu}_k^{(1)}$ whenever $\langle \check{\bu}_k^{(1)}, \bu_k^{(1)}\rangle<0$. For low-dimensional setup, $\tilde{\bu}_1^{(1)}$ and $\tilde{\bu}_2^{(1)}$ are treated similarly.

\begin{figure}[htpb]
    \centering
  \begin{subfigure}[b]{0.45\textwidth}   
  	\centering
  	 \includegraphics[scale=0.2]{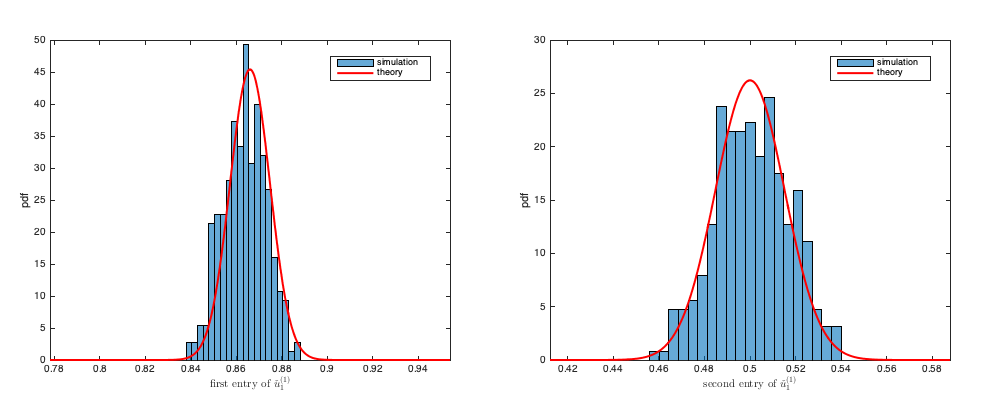}
    \caption{$d_1=d_2=10$}
    \label{f:normal_10_10_400_3}
  \end{subfigure}
 \\
 \begin{subfigure}[b]{0.45\textwidth}   
  	\centering
    \includegraphics[scale=0.2]{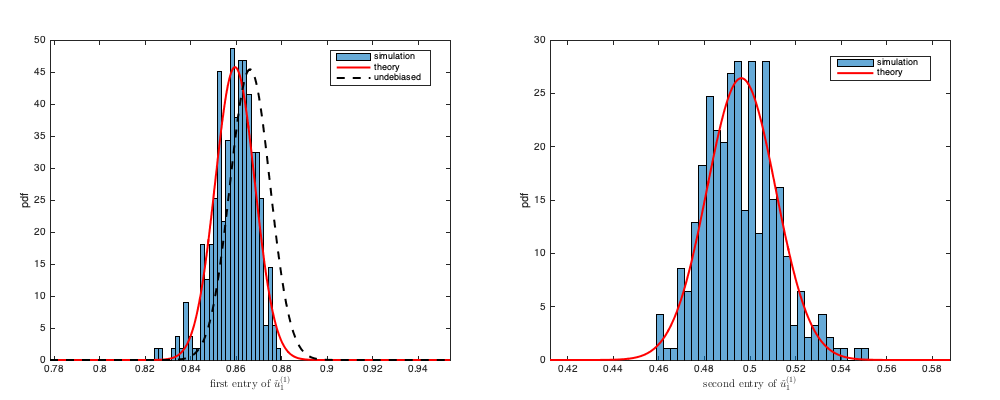}
\caption{$d_1=d_2=50$} 
% The first entry is biased to the left because its true value is positive.
\label{f:normal_50_50_400_3}
\end{subfigure}
    \caption{Multiway PCA for data generated from normal distribution.}
\end{figure}

In the first set of experiments, we considered a low-dimensional setup with $d_1=d_2=10$, $n=200$, $\sigma_1=\sigma_2=2$, and the true PCs were given by
\begin{align}
&\bu_1^{(1)}=(\sqrt{3}/2,1/2,0,\dots,0)^\top,\quad \bu_1^{(2)}=(1,0,\dots,0)^\top, \nonumber \\
&\bu_2^{(1)}=(-1/2,\sqrt{3}/2,0,\dots,0)^\top, \quad \bu_2^{(2)}=(0,1,0,\dots,0)^\top. \label{eq:simu_u}
\end{align}
Figure \ref{f:normal_10_10_400_3} reports the histograms of the first two (nonzero) entries of $\tilde{\bu}_1^{(1)}$ based on 300 simulation runs. The histograms are overlaid with the asymptotic distributions derived in Theorem \ref{thm:low_dim}. The agreement between the two confirms the accuracy of the asymptotic distribution when the dimensionality is low.

To demonstrate the need and effectiveness of bias correction, we increased the dimension to $d_1=d_2=50$. Correspondingly we set $n=400$ and $\sigma_1=\sigma_2=3$. We repeated the experiment another 300 times and as before, Figure \ref{f:normal_50_50_400_3} reports the histograms of the first two entries of $\check{\bu}_1^{(1)}$ along with the asymptotic distribution derived in Theorem \ref{thm:bias_simple}, plotted in red lines. The dashed black line overlaid with the histogram of the first entries corresponds to the asymptotic distribution without bias correction as given by Theorem \ref{thm:low_dim}. It is clear that in this setting, debiasing is necessary and the bias correction of Theorem \ref{thm:bias_simple} indeed leads to a more precise approximation of the finite sample distribution.

\begin{figure}[htpb]
    \centering
  \begin{subfigure}[b]{0.45\textwidth}   
	\centering
    \includegraphics[scale=0.2]{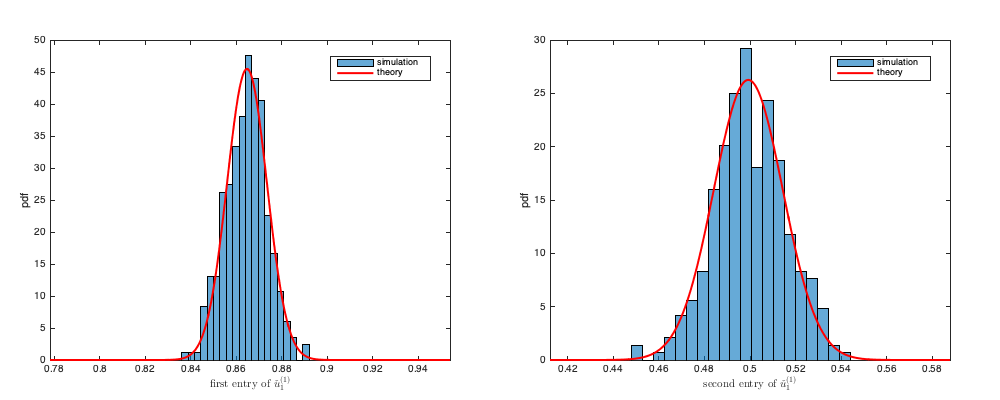}
    \caption{$d_1=d_2=10$}
    \label{f:poi_10_10_400_3}
   \end{subfigure}
   \\
  \begin{subfigure}[b]{0.45\textwidth}   
    \centering
    \includegraphics[scale=0.2]{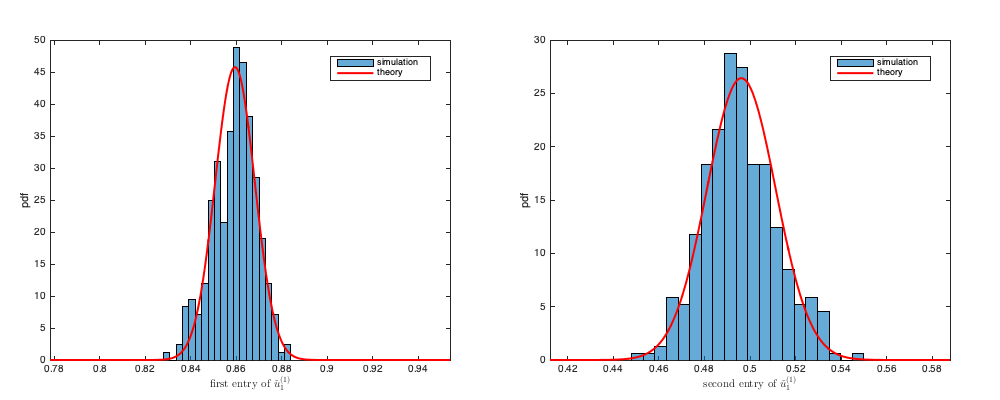}
    \caption{$d_1=d_2=50$}
    \label{f:poi_50_50_400_3}
\end{subfigure}
    \caption{Multiway PCA for data generated from Poisson distribution.}
\end{figure}

Our next set of simulations aims to explore the robustness of our approach to deviation from normality. To this end, $\{\theta_k, k \in [r]\}$ and the entries of $\scrE$ were simulated independently from ${\rm Poisson}(1) -1$ (so that they still have mean $0$ and variance $1$). Again we set $n=400$ and $\sigma_1=\sigma_2=3$. Figures \ref{f:poi_10_10_400_3} and \ref{f:poi_50_50_400_3} summarize results based on 300 runs, for dimensions $d_1=d_2=10$ and $d_1=d_2=50$, respectively. We overlay them with the theoretical asymptotic distributions given by Theorems \ref{thm:low_dim} and \ref{thm:bias_simple}. The results are qualitatively similar to those from the previous setting.%The results suggest that although our asymptotic theories are developed under the Gaussian assumption, they remain valid beyond Gaussian.

\subsection{World Bank Data}

We now consider a real-world data example -- the open source global development data from the World Bank\footnote{https://data.worldbank.org/}. The world Bank offers access to annual country-level data of a number of development indicators. In particular, we shall focus on the following nine most common and important economic and demographic indicators: 
\blist{25}{0}{5}
\item[] \verb+GDP+: gross domestic product (GDP) based on purchasing power parity;
\item[] \verb+Import+: import volume index (year 2000=100);
\item[] \verb+Export+: export volume index (year 2000=100);
\item[] \verb+CO2+: total CO2 emissions, in kilo-ton;
\item[] \verb+CPI+: Consumer price index (year 2010 = 100);
\item[] \verb+Life Span+: Life expectancy at birth;
\item[] \verb+Urban Population+: Urban population, percentage of total population;
\item[] \verb+Tourism+: number of international inbound tourists;
\item[] \verb+Birth Rate+: Birth rate, crude (per 1,000 people).
\elist
Yearly data for these indicators have been recorded and we focus on data from Year 2000 through 2018, as considerable data are missing outside this range. We also discarded countries that have more than 5\% of missing data in our analysis, resulting in a total of 160 countries under consideration. 

These indicators are all positive but of vastly different magnitudes. To this end, a log transformation was first applied. Each log-transformed indicator was then standardized so that the log-transformed indicator has a mean $0$ and a mean absolute deviation $1$ for all countries. The use of mean absolute deviation, instead of variance, for standardization allows more robust analysis in the presence of outlying observations. Denote by $\bX_{k,t,i}$ the resulting indicator $i$ for country $k$ at time $t$. There remain a handful of missing values and for convenience, they are replaced with $0$ in our analysis. The data tensor $\bX$ of dimensions $160 \times 19 \times 9$. Each frontal slice 
$$
\scrX_k = \bX_{k,\cdot,\cdot}
$$
corresponds to a country and is a $19 \times 9$ matrix. Note that its ambient dimension is $19\times 9=171$ and greater than the number of countries so it is problematic to apply the usual PCA with stringing. Accounting for the multiway structure, we can consider the multiway PCs of the form
$$
\scrU_k = \bu_k^{(1)} \otimes \bu_k^{(2)} \in \RR^{19 \times 9}.
$$
These PC carry a clear meaning: each $\scrU_k$ represents a shared \emph{development pattern}, where $\bu_k^{(1)}$ is the corresponding shared \emph{temporal trend}, and $\bu_k^{(2)}$ is the corresponding \emph{comovement pattern}.

\begin{figure}[htpb]
    \centering
    \includegraphics[scale=0.4]{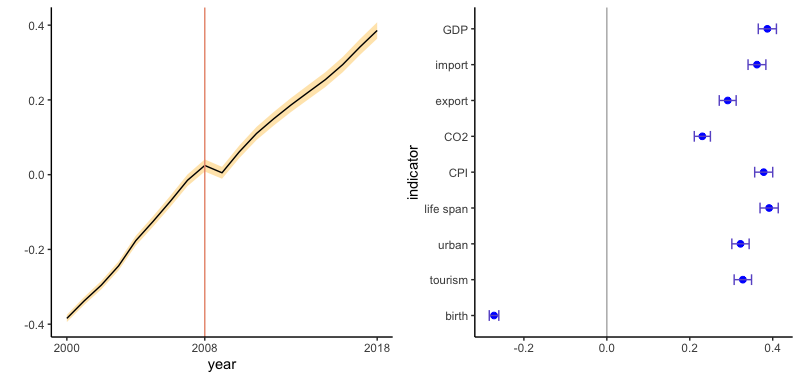}
    \caption{The first PC, general economic development: $\bu_1^{(1)}$ and $\bu_1^{(2)}$ plotted with 95\% confidence intervals.}
    \label{WB1}
\end{figure}

Figure \ref{WB1} plots the estimated leading PC along both modes, namely $\check{\bu}_1^{(1)}$ and $\check{\bu}_k^{(2)}$, together with the $95\%$ confidence intervals for each of their coordinates. It is by far the most significant component, explaining 57.6\% of the total variation. It is also evident from the temporal component that the first PC describes a roughly constant growth trend. The only year with a decrease is 2008 when the Global Financial Crisis took place. Correspondingly, except for the entry corresponding to birth rate, all other entries of $\check{\bu}_k^{(2)}$ are positive. This suggests a general economic development during this period, with the birth rate in decline.

\begin{figure}[htpb]
    \centering
    \includegraphics[scale=0.4]{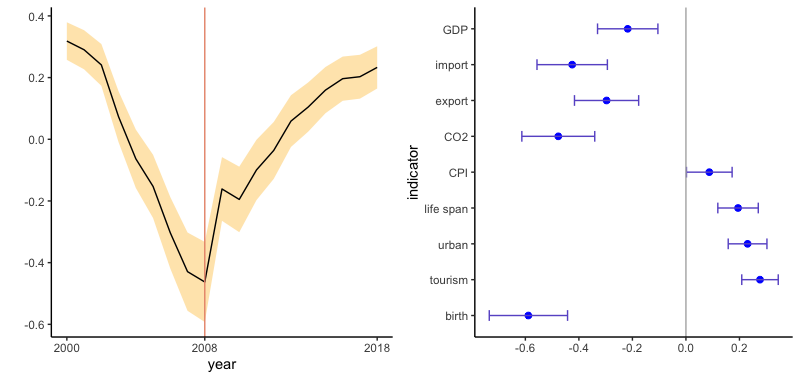}
    \caption{The second PC, life quality: $\bu_2^{(1)}$ and $\bu_2^{(2)}$ plotted with 95\% confidence intervals.}
    \label{WB2}
\end{figure}

Similarly, Figure \ref{WB2} shows the second multiway PC in the two modes along with their $95\%$ confidence bands. This PC captures a change of developmental direction at the year of 2008. In particular, CPI, life span, urban population, and tourism steadily decreased prior to 2008 but reversed course after the financial crisis. In contrast, GDP, import, export, CO2 emission and birth rate followed an opposite pattern. There are many plausible explanations for this pattern. It is possible that the quantitative easing policies applied by most major economies since 2008 led to growth in the domestic market, thus enhancing the life-quality indicators. It is also possible that the growing inequality after 2008, also caused by quantitative easing among other factors \citep[see, e.g.,][]{montecino2015did}, has in turn caused the increase in life quality among the upper and the upper middle class. The tourism indicator is the number of international inbound tourists, which most likely is driven by the upper middle class and beyond. The continuous increase in life expectancy in the USA is also reported to be driven primarily by the well-off \citep[see, e.g.,][]{chetty2016association}.

%\footnote{https://www.vox.com/science-and-health/2018/1/9/16860994/life-expectancy-us-income-inequality}

\begin{figure}[htpb]
    \centering
    \includegraphics[scale=0.4]{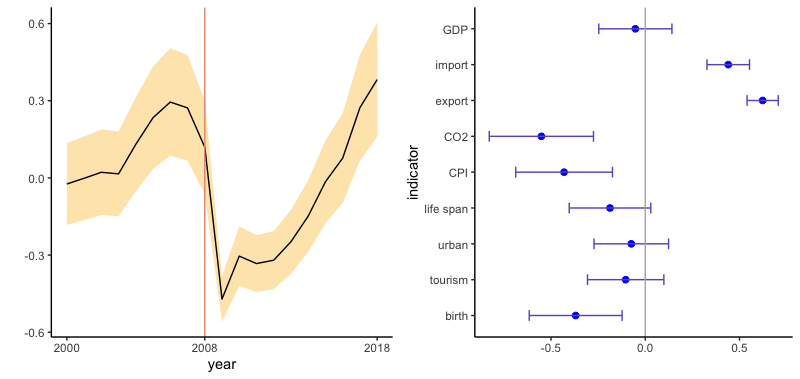}
    \caption{The third PC, international trade: $\bu_3^{(1)}$ and $\bu_3^{(2)}$ plotted with 95\% confidence intervals.}
    \label{WB3}
\end{figure}

Finally, Figure \ref{WB3} shows the third multiway PC. We begin to see much wider confidence intervals as the signal becomes weaker. In fact, only the period around 2008 are significantly different from zero, and likewise, the entries corresponding to life span, urbanization, and tourism are statistically insignificant. This indicates that these patterns likely focus on the impact of the 2008 financial crisis: it caused an immediate economic downturn but recovered not long after.

\subsection{NYC Bike Rental Data}

Another data example we considered is the Citibike trip data\footnote{https://ride.citibikenyc.com/system-data}. In particular, all the Citibike trips from January 1, 2018 to December 31, 2019, on weekdays (522 days in total) that started in Manhattan and lasted for at least 60 seconds were used in our analysis. During this period, there are 35 zip codes in Manhattan with at least one Citibike station.
There are a total of 29,515,527 trips and we form a data tensor $\bY$ of dimension $522 \times 24 \times 35$ where $Y_{kij}$ denotes the number of trips starting during the $i$th hour of the $k$th day from the $j$th zip code.

The number of counts at different zip codes are of drastically different magnitudes, and the total counts during the two years also display a clear seasonal trend. To facilitate our analysis, we standardized the counts from each zip code $j$ at each day $k$ so that they have mean $0$ and mean absolute deviation $1$. As in the previous example, a direct application of the usual PCA can be misleading as the ambient dimension of the daily observation is $24\times 35=840$ and greater than the number of days. Nonetheless, it is helpful to consider multiway PCs of the form
$$
\scrU_k = \bu_k^{(1)} \otimes \bu_k^{(2)} \in \RR^{24 \times 35},
$$
where $\bu_k^{(1)}$ captures the time-of-the-day effect of bike rental, and $\bu_k^{(2)}$ the location pattern.

\begin{figure}[htpb]
    \centering
    \includegraphics[scale=0.4]{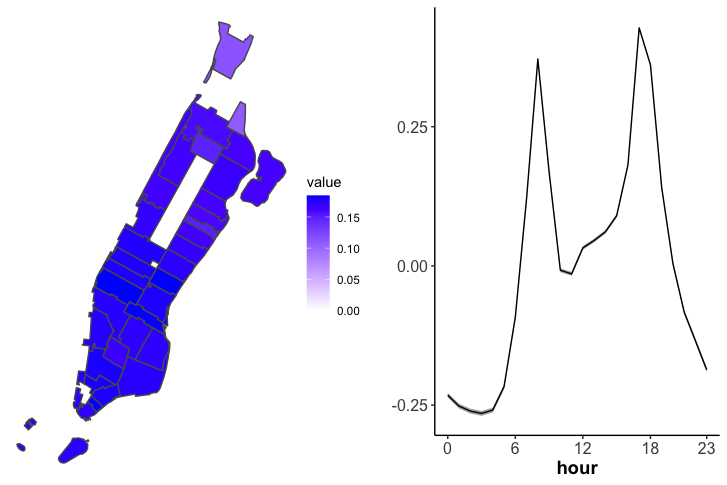}
    \caption{The first PC: overall pattern. }
    \label{citi_start1}
\end{figure}

Figure \ref{citi_start1} plots the first multiway PC. The spatial pattern clearly indicates that this represents an overall pattern across Manhattan with all 35 entries of $\bu_k^{(2)}$ being estimated as positive. The temporal pattern indicates that bike rental strongly coincides with the rush hours with  two peaks during the morning and afternoon rush hours. The blank area downtown is zip code 10006, the big blank rectangular is Central Park, the small blank underneath is zip code 10020, and the blank area to the north of Central Park has zip codes 10030 and 10031. At the time of the recorded period, no Citibike station existed in these areas.

\begin{figure}[htpb]
    \centering
    \includegraphics[scale=0.4]{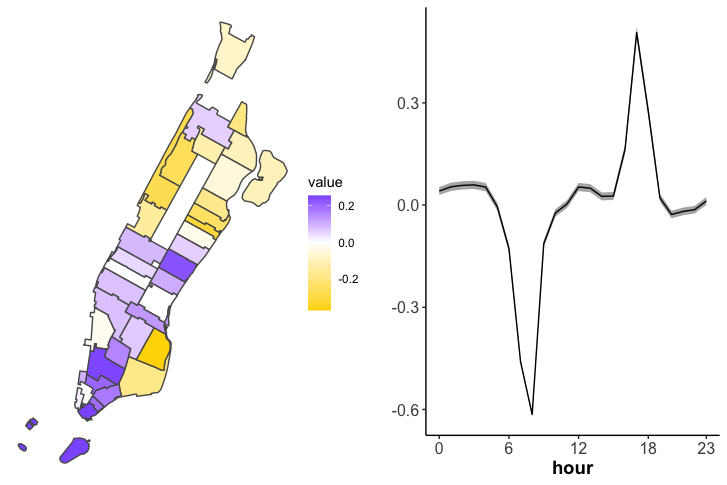}
    \caption{The second PC: rush hour differences.}
    \label{citi_start2}
\end{figure}

The second PC, as shown in Figure \ref{citi_start2}, reveals differences in rental patterns across neighborhoods. While the first PC suggests increased rental activities both in the morning and afternoon rush hours, the second PC captures the difference between morning and evening rental patterns as indicated by the positive peak during the evening rush hours and the negative peak during the morning rush hours. As such, a neighborhood with positive loadings may see more evening rentals than morning rentals. These are the downtown Financial District, Lower Manhattan, and Midtown, largely corresponding to the business area of Manhattan. On the other hand, zip codes corresponding to negative loadings represent mostly residential areas of Manhattan, including the East Village, Upper West Side, and Upper East Side.

\begin{figure}[htpb]
    \centering
    \includegraphics[scale=0.4]{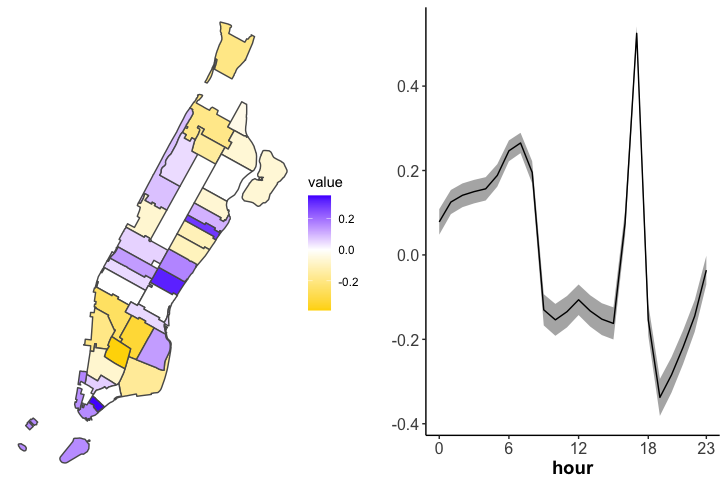}
    \caption{The third PC: afternoon rush hour details.}
    \label{citi_start3}
\end{figure}

Figure \ref{citi_start3} depicts the third PC. The temporal pattern has a narrow and tall peak during the afternoon rush hours suggesting that this PC captures the subtle spatial difference during this time of the day. In particular, the zip codes with large positive values (purple color) are the area around Wall Street (the small purple block in Lower Manhattan), the area around Grand Central Terminal, and an area in Upper East Side. The negative zip codes in this pattern include the areas around SoHo, Greenwich Village, and Harlem.

\section{Summary}\label{sec:summary}

In this paper, we study PCA under the settings that each observation is a matrix or more generally a multiway array. We investigate how to extract multiway PCs and study their statistical properties. In addition to the obvious advantages of increased efficiency and enhanced interpretability, our analysis provides a number of new insights into the operating characteristics of multiway PCA and their methodological implications.

First, we show that multiway PCs can be estimated without the eigengap requirement. Specifically, under a spike covariance model, we establish rates of convergence for the sample multiway PCs. In particular, they are consistent whenever the signal-to-noise ratio $\sigma_k/\sigma_0\gg \max\{d/n,(d/n)^{1/4}\}$ where $d$ is the dimension of one mode. Perhaps more interestingly, we prove that the sample multiway PCs are asymptotically independent of each other, at least when the dimension $d=o(\sqrt{n})$. In higher dimensions, the sample PCs can be biased and the bias can be corrected via sample-splitting to lead to asymptotically normal estimates of the multiway PCs, which enables us to construct confidence intervals or conduct hypothesis testing for linear forms of the PCs.

Our theoretical developments are complemented by numerical experiments, both simulated and real. In particular, meaningful findings can be inferred when applying our methods to two real-world datasets, further demonstrating the merits of our methodology.

\bibliographystyle{plainnat}
\bibliography{ref_factor_model_asymptotics} 

\newpage

\appendix

\section{Notations and Preliminary Bounds}

Write $a \vee b := \max\{a,b\}$ and $a \wedge b := \min\{a,b\}$.
For a positive integer $n$, let $[n]:= \{1,2,\dots,n\}$. For a vector $\bx \in \mathbb{R}^d$, denote $\|\bx \|$ to be its $\ell_2$-norm, $\| \bx \|_1$ to be its $\ell_1$-norm, and $\| \bx \|_{\infty}=\max_i |x_i|$ to be its $\ell_{\infty}$-norm. For two sequences of real numbers $\{a_n\}$ and $\{b_n\}$, write $a_n=O(b_n)$ if $\exists C, \exists M$, such that $\forall n >M$, $|a_n| \le C|b_n|$. Write $a_n=o(b_n)$ if $\lim_{n \to \infty} a_n/b_n =0$. For two sequences of real-valued random variables $X_n$ and $Y_n$, write $X_n=O_p(Y_n)$ if $X_n = R_n Y_n$ and $R_n$ is uniformly tight. Write $X_n=o_p(Y_n)$ if $X_n = R_n Y_n$ and $R_n \overset{p}\to 0$. For linear subspace $U$ of $\mathbb{R}^{d}$, denote $P_U$ and $P_U^{\perp}$ to be the orthogonal projection onto $U$ and its orthogonal complement $U^{\perp}$, respectively. For a non-zero vector $u \in \mathbb{R}^{d}$, denote $P_u := P_{ {\rm span} \{u\}}$ and $P_u^{\perp} := P_{ {\rm span} \{u\}}^{\perp}$.

For an order-$k$ tensor $\scrT \in \mathbb{R}^{d_1 \times d_2 \times \dots \times d_k}$, define its \emph{tensor operator norm} as:
\begin{align}
\| \scrT \| := \sup_{\bu_j \in \mathbb{R}^{d_j}, \| \bu_j \| =1} \scrT(\bu_1,\bu_2,\dots,\bu_k).
\end{align} 
Specifically, when $k=2$ so that $\scrT \in \mathbb{R}^{d_1 \times d_2}$ is a matrix, $\| \scrT \|$ is the matrix spectral norm of $\scrT$. For tensor $\scrT \in \mathbb{R}^{d_1 \times d_2 \times \dots \times d_k}$, write $$\|\scrT\|_{\max} = \max_{i_1,\dots,i_k} \left| \scrT_{i_1,\dots,i_k} \right|$$ to be its $\ell_{\infty}$-norm.

%The $q$-mode product of $\scrT \in \mathbb{R}^{d_1 \times d_2 \times \dots \times d_k}$ with a matrix $A \in \RR^{m \times d_q}$ is an order-$k$ tensor of size $d_1 \times \dots d_{q-1} \times m \times d_{q+1} \dots \times d_k$, and will be denoted by $\scrT \times_q A$, with its elements given by
%$$
%[\scrT \times_q A]_{i_1 \dots i_{q-1} j i_{q+1} \dots i_k} = \sum_{i_q=1}^{d_q} \scrT_{i_1 \dots i_q \dots i_k} A_{j i_q}.
%$$
With a slight abuse of notation, the mode $q$ product of $\scrT \in \mathbb{R}^{d_1 \times d_2 \times \dots \times d_k}$ with a vector $\ba \in \RR^{d_q}$, denoted by $\scrT \times_q \ba$, is defined as an order-$(k-1)$ tensor of size $d_1 \times \dots d_{q-1} \times d_{q+1} \dots \times d_k$, with elements
$$
[\scrT \times_q \ba]_{i_1 \dots i_{q-1} i_{q+1} \dots i_k} = \sum_{i_q=1}^{d_q} \scrT_{i_1 \dots i_q \dots i_k} \ba_{i_q}.
$$

%Recall that for a random variable $X$ and $\alpha>0$, the Orlicz $\psi_{\alpha}$-norm of $X$ is defined as 
%\begin{align*}
%	\|X\|_{\psi_{\alpha}} = \inf\left\{K>0: \mathbb{E} e^{(|X|/K)^2}-1 \le 1 \right\}.
%\end{align*}

%\section{Preliminary Bounds}
Write
$$
\hat{\Sigma}_\theta={1\over n}\sum_{i=1}^n \theta_i\otimes \theta_i,\qquad \hat{\Sigma}_\scrE={1\over n}\sum_{i=1}^n \scrE_i\otimes\scrE_i,
$$
and
$$
\hat{\Sigma}_{\theta,\scrE}={1\over n}\sum_{i=1}^n \theta_i\otimes \scrE_i,
$$
the sample covariance matrices of $\theta$, $\scrE$ and between them respectively. Correspondingly denote by $\Sigma_\theta$, $\Sigma_\scrE$ and $\Sigma_{\theta, \scrE}$ their population counterpart. It is clear $\Sigma_{\theta, \scrE}=0$. Recall also that
$$
\hat{\Sigma}={1\over n}\sum_{i=1}^n \scrX_i\otimes \scrX_i.
$$
and
$$
\Sigma=\sum_{l=1}^r \sigma_l \bu_l^{(1)}\otimes\cdots\otimes \bu_l^{(p)}\otimes \bu_l^{(1)}\otimes\cdots\otimes \bu_l^{(p)}+\sigma_0^2\scrI
$$
are the sample and population covariance matrices of $\scrX$.

The proof relies on the following technical lemmas.

\begin{lemma}
	\label{le:probbd}
There exists a numerical constant $C>0$ such that for any $t\ge 1$,
$$
\|\hat{\Sigma}-\Sigma\|\le C(\sigma_1^2+\sigma_0^2)\max\left\{\sqrt{d\over n}, {d\over n}, \sqrt{t\over n}, {t\over n}\right\},
$$
$$
\|\hat{\Sigma}_{\theta,\scrE}\|\le C\sigma_0\max\left\{\sqrt{d\over n}, {d\over n}, \sqrt{t\over n}, {t\over n}\right\},
$$
$$
\|\hat{\Sigma}_\scrE-\Sigma_\scrE\|\le C\sigma_0^2\max\left\{\sqrt{d\over n}, {d\over n}, \sqrt{t\over n}, {t\over n}\right\},
$$
and
$$
\|\hat{\Sigma}_\theta-\Sigma_\theta\|\le C\max\left\{\sqrt{r\over n}, {r\over n}, \sqrt{t\over n}, {t\over n}\right\},
$$
with probability at least $1-e^{-t}$.
\end{lemma}

Note that we shall use $C$ to denote a constant that may take different values at each appearance. We shall also make use the following bounds:

\begin{lemma}
	\label{le:probbd2}
There exists a numerical constant $C>0$ such that for any $t>0$,
$$
\|\hat{\Sigma}_\theta-\Sigma_\theta\|_{\max}\le C\max\left\{\sqrt{\log r\over n}, {\log r\over n}, \sqrt{t\over n}, {t\over n}\right\},
$$	
$$
\max_{1\le l_1,l_2\le r}\left|\hat{\Sigma}_{\scrE,\theta}(\bu_{l_1}^{(1)},\bu_{l_2}^{(2)},\ldots,\bu_{l_2}^{(p)},{\bf e}_{l_2})\right|\le C\sigma_0\max\left\{\sqrt{\log r\over n}, {\log r\over n}, \sqrt{t\over n}, {t\over n}\right\}
$$
where ${\bf e}_{l_2}$ is the $l_2$th canonical basis of ${\mathbb R}^{r}$, and
$$
\max_{1\le l\le r}\left|(\hat{\Sigma}_\scrE-\Sigma_\scrE)(\bu_l^{(1)},\ldots, \bu_l^{(p)},\bu_l^{(1)}, \ldots,\bu_l^{(p)})\right|\le C\sigma_0^2\max\left\{\sqrt{\log r\over n}, {\log r\over n}, \sqrt{t\over n}, {t\over n}\right\},
$$
with probability at least $1-e^{-t}$.
\end{lemma}

Both Lemmas are well known and follow immediately from an application of union bounds and $\chi^2$ tail bounds. See, e.g., \cite{vershynin2010introduction}.

\section{Proof of Theorems \ref{pr_rate} and \ref{thm_rate}}
Theorem \ref{pr_rate} follows immediately from Theorem \ref{thm_rate} and it suffices to prove the latter. For brevity, we shall focus on the case when $d\le n$ and $r$ diverges with $n$. Denote by $\calE$ the event that
$$
\|\hat{\Sigma}_\theta-\Sigma_\theta\|\le C\sqrt{r\over n},\quad {\rm and}\quad \sigma_0^{-2}\|\hat{\Sigma}_\scrE-\Sigma_\scrE\|,\sigma_0^{-1}\|\hat{\Sigma}_{\theta,\scrE}\|\le C\sqrt{d\over n}
$$
and
$$
\|\hat{\Sigma}_\theta-\Sigma_\theta\|_{\max}, \ \sigma_0^{-1}\max_{1\le l_1,l_2\le r}\left|\hat{\Sigma}_{\scrE,\theta}(\bu_{l_1}^{(1)},\bu_{l_2}^{(2)},\ldots,\bu_{l_2}^{(p)},{\bf e}_{l_2})\right|\le C\sqrt{\log r\over n}
$$
By Lemmas \ref{le:probbd} and \ref{le:probbd2}, $\calE$ holds with probability tending to one. It suffices to proceed conditional on the event $\calE$. 

As noted, the $k$th sample PCs may not correspond to the $k$th population PCs because we do not assume the existence of eigengap and $\sigma_k$s may not even be distinct. Nonetheless, we can match the sample PCs with population PCs as follows. Define
$$
\pi(1)=\argmax_{1\le l\le r} \left\{ \sigma_l^2 \left| \prod_{q=1}^p \langle \bu_l^{(q)}, \hat{\bu}_1^{(q)} \rangle \right|\right\},
$$
and for $k>1$,
$$
\pi(k):=\argmax_{l \notin \pi([k-1])} \left\{ \sigma_l^2 \left| \prod_{q=1}^p \langle \bu_l^{(q)}, \hat{\bu}_k^{(q)} \rangle \right|\right\}.
$$
The goal is to show that with high probability,
\begin{equation}
	\label{eq:rate_main}
	\eta_k:=\max_{1\le q\le p}\sin\angle(\bu_{\pi(k)}^{(q)}, \hat{\bu}_k^{(q)})\le C\left({\sigma_0\over \sigma_{\pi(k)}}+{\sigma_0^2\over \sigma_{\pi(k)}^2}\right)\max\left\{\sqrt{d\over n},{d\over n}\right\}=:\delta_k,
\end{equation}
for $k=1,\ldots, r$. Our proof proceeds by induction over $k$. To facilitate the induction, we shall also prove that
\begin{eqnarray}\nonumber
	\tilde{\eta}_k&:=&\max_{1\le q\le p}\max_{l\notin \pi([k])}\langle\bu_{l}^{(q)}, \hat{\bu}_k^{(q)}\rangle\\\nonumber
	&\le& C\left({\sigma_0\over \sigma_{\pi(k)}}+{\sigma_0^2\over \sigma_{\pi(k)}^2}\right)^2\max\left\{{d\over n},{d^2\over n^2}\right\}+C\left({\sigma_0\over \sigma_{\pi(k)}}+{\sigma_0^2\over \sigma_{\pi(k)}^2}\right)\sqrt{\log r\over n}\\
	&=:&\tilde{\delta}_k.	\label{eq:rate_main1}
\end{eqnarray}

In addition to \eqref{eq:rate_main} and \eqref{eq:rate_main1}, we shall also prove that
\begin{equation}\label{eq:rate_main2}
	\sigma_{\pi(k)}^2\ge \max_{l\notin\pi([k])}\sigma_l^2(1-C\delta_k^2).
\end{equation}
This immediately implies that
$$
\sum_{l=1}^k \tilde{\delta}_k^2\le C\delta_k^2\quad{\rm and}\quad \max_{1\le l\le k} \{\sigma_{\pi(l)}\delta_l\}\le C\sigma_{\pi(k)}\delta_k,
$$
by the taking the constant $c_0$ in \eqref{eq:assumption} small enough. We shall make use of these bounds repeatedly.

As noted, we shall proceed by induction over $k$. In particular, we shall denote by $\delta_0=\tilde{\delta}_0=0$ so that the the base case holds trivially when $k=0$. Now assume the induction hypotheses \eqref{eq:rate_main} and \eqref{eq:rate_main1} holds for $1,\ldots, k-1$. We want to show that they continue to hold for $k$. The general architect of the argument is similar to that for the base case, but additional challenges arise with the need to control the impact of estimation error of $\hat{\bu}_{l}^{(q)}$s for $1\le l<k$.

Denote by $\hat{\calP}^{(q)}$ the projection matrix onto the linear space spanned by $\{ \hat{\bu}_1^{(q)}, \dots, \hat{\bu}_{k-1}^{(q)} \}$, for $q \in [p]$. Note that in the case when $k=1$, $\hat{\calP}^{(q)}={\bf 0}_{d\times d}$. Then
\begin{eqnarray*}
(\hat{\bu}_k^{(1)}, \ldots, \hat{\bu}_k^{(p)})&=& \argmax_{\substack{\| \bw^{(q)} \| \le 1, \langle \bw^{(q)}, \hat{\bu}_l^{(q)} \rangle =0, \\ \forall l\le k-1, q \in [p]}} \hat{\Sigma}(\bw^{(1)},\dots,\bw^{(p)},\bw^{(1)},\dots,\bw^{(p)})\\
&=&\argmax_{\substack{\| \bw^{(q)} \| =1, \forall q \in [p]}} \hat{\Sigma}(\hat{\calP}^{(1)}_\perp\bw^{(1)},\dots,\hat{\calP}^{(p)}_\perp\bw^{(p)},\hat{\calP}^{(1)}_\perp\bw^{(1)},\dots,\hat{\calP}^{(p)}_\perp\bw^{(p)}) 
\end{eqnarray*}
where $\hat{\calP}^{(q)}_\perp=I-\hat{\calP}^{(q)}$. Observe that
$$
\hat{\calP}_\perp^{(q)}\tilde{\Sigma}(\hat{\bu}_k^{(1)},\ldots,\hat{\bu}_k^{(q-1)}, \cdot,\hat{\bu}_k^{(q+1)},\ldots,\hat{\bu}_k^{(p)}, \hat{\bu}_k^{(1)},\ldots, \hat{\bu}_k^{(p)})\propto \hat{\bu}_k^{(q)},
$$
where $\tilde{\Sigma}=\hat{\Sigma}-\sigma_0^2\scrI$. This implies that
$$
\langle \hat{\bu}_k^{(q)},\bw\rangle={\tilde{\Sigma}(\hat{\bu}_k^{(1)},\ldots,\hat{\bu}_k^{(q-1)}, \hat{\calP}_\perp^{(q)}\bw,\hat{\bu}_k^{(q+1)},\ldots,\hat{\bu}_k^{(p)}, \hat{\bu}_k^{(1)},\ldots, \hat{\bu}_k^{(p)})\over \tilde{\Sigma}(\hat{\bu}_k^{(1)},\hat{\bu}_k^{(2)},\ldots,\hat{\bu}_k^{(p)}, \hat{\bu}_k^{(1)},\ldots, \hat{\bu}_k^{(p)})}.
$$
In particular, for $l\neq k$,
$$
\sin\angle(\hat{\bu}_k^{(q)}, \bu_l^{(q)})={\tilde{\Sigma}(\hat{\bu}_k^{(1)},\ldots,\hat{\bu}_k^{(q-1)},\bu_l^{(q)},\hat{\bu}_k^{(q+1)},\ldots,\hat{\bu}_k^{(p)}, \hat{\bu}_k^{(1)},\ldots, \hat{\bu}_k^{(p)})\over \tilde{\Sigma}(\hat{\bu}_k^{(1)},\hat{\bu}_k^{(2)},\ldots,\hat{\bu}_k^{(p)}, \hat{\bu}_k^{(1)},\ldots, \hat{\bu}_k^{(p)})},
$$
and
$$
\sin\angle(\hat{\bu}_k^{(q)}, \bu_k^{(q)})={\tilde{\Sigma}(\hat{\bu}_k^{(1)},\ldots,\hat{\bu}_k^{(q-1)}, \hat{\calP}_\perp^{(q)}(\hat{\bu}_k^{(q)}-\langle\hat{\bu}_k^{(q)},\bu_{\pi(k)}^{(q)}\rangle\bu_{\pi(k)}^{(q)}),\hat{\bu}_k^{(q+1)},\ldots,\hat{\bu}_k^{(p)}, \hat{\bu}_k^{(1)},\ldots, \hat{\bu}_k^{(p)})\over \tilde{\Sigma}(\hat{\bu}_k^{(1)},\hat{\bu}_k^{(2)},\ldots,\hat{\bu}_k^{(p)}, \hat{\bu}_k^{(1)},\ldots, \hat{\bu}_k^{(p)})},
$$
We shall derive lower bounds for the nominators and an upper bound for the denominator. It suffices to consider the case $q=1$. Other indices can be treated in an identical fashion. %To this end, we shall first prove the lower bound for $\sigma_{\pi(k)}^2$ given by \eqref{eq:rate_main2}.

\subsection{Lower Bound for $
	\tilde{\Sigma}(\hat{\bu}_k^{(1)},\hat{\bu}_k^{(2)},\ldots,\hat{\bu}_k^{(p)}, \hat{\bu}_k^{(1)},\ldots, \hat{\bu}_k^{(p)})
	$}
Denote by $\tilde{\Sigma}=\Sigma-\sigma_0^2\scrI$. Observe that
\begin{eqnarray}\nonumber
&&\tilde{\Sigma}(\hat{\bu}_k^{(1)},\hat{\bu}_k^{(2)},\ldots,\hat{\bu}_k^{(p)}, \hat{\bu}_k^{(1)},\ldots, \hat{\bu}_k^{(p)})\\\nonumber
&\ge&\max_{l\notin \pi([k-1])}\tilde{\Sigma}(\hat{\calP}^{(1)}_\perp\bu_l^{(1)},\ldots,\hat{\calP}^{(p)}_\perp\bu_l^{(p)},\hat{\calP}^{(1)}_\perp\bu_l^{(1)},\ldots, \hat{\calP}^{(p)}_\perp\bu_l^{(p)})\\\nonumber
&\ge&\max_{l\notin \pi([k-1])}(\Sigma-\sigma_0^2\scrI)\hat{\calP}^{(1)}_\perp\bu_l^{(1)},\ldots,\hat{\calP}^{(p)}_\perp\bu_l^{(p)},\hat{\calP}^{(1)}_\perp\bu_l^{(1)},\ldots, \hat{\calP}^{(p)}_\perp\bu_l^{(p)})\\
&&-\sup_{\|\bw^{(q)}\|\le 1,1\le q\le p}\left|(\hat{\Sigma}-\Sigma)(\hat{\calP}_\perp^{(1)}\bw^{(1)},\dots,\hat{\calP}_\perp^{(p)}\bw^{(p)},\hat{\calP}_\perp^{(1)}\bw^{(1)},\dots,\hat{\calP}_\perp^{(p)}\bw^{(p)})\right|.\label{eq:denomlower}
\end{eqnarray}
Next we bound the two terms on the rightmost hand side.

Starting with the first term, note that for any $l\notin\pi([k-1])$,
\begin{eqnarray*}
	&&(\Sigma-\sigma_0^2\scrI)(\hat{\calP}^{(1)}_\perp\bu_l^{(1)},\ldots,\hat{\calP}^{(p)}_\perp\bu_l^{(p)},\hat{\calP}^{(1)}_\perp\bu_l^{(1)},\ldots, \hat{\calP}^{(p)}_\perp\bu_l^{(p)})\\
	&=&\sum_{1\le l'\le r}\left[\sigma_{l'}^2\prod_{q=1}^p\langle \hat{\calP}^{(q)}_\perp\bu_l^{(q)}, \bu_{l'}^{(q)}\rangle^2\right]\\
	&\ge&\sigma_l^2\prod_{q=1}^p\langle \hat{\calP}^{(q)}_\perp\bu_l^{(q)}, \bu_l^{(q)}\rangle^2\\
	&=&\sigma_l^2\prod_{q=1}^p\|\hat{\calP}^{(q)}_\perp\bu_l^{(q)}\|^4.
\end{eqnarray*}
By the induction hypothesis \eqref{eq:rate_main1},
$$
\|\hat{\calP}^{(q)}\bu_l^{(q)}\|^2=\sum_{1\le l'<k}\langle \bu_l^{(q)},\hat{\bu}_{l'}^{(q)}\rangle^2\le \sum_{1\le l'<k}\tilde{\delta}^2_{l'}\le C\delta_k^2.
$$
By taking the constant $c_0$ of \eqref{eq:assumption} small enough, we can ensure that
\begin{equation}\label{eq:denomlower1}
\max_{l\notin\pi([k-1])}(\Sigma-\sigma_0^2\scrI)(\hat{\calP}^{(1)}_\perp\bu_l^{(1)},\ldots,\hat{\calP}^{(p)}_\perp\bu_l^{(p)},\hat{\calP}^{(1)}_\perp\bu_l^{(1)},\ldots, \hat{\calP}^{(p)}_\perp\bu_l^{(p)})\ge (1-C\delta_k^4)\tau^2.
\end{equation}
where
$$
\tau^2=\max_{l\notin\pi([k-1])} \sigma_l^2.
$$

Next we derive a bound for 
$$
\sup_{\|\bw^{(q)}\|\le 1, 1\le q\le p}(\Sigma-\hat{\Sigma})(\hat{\calP}^{(1)}_\perp\bw^{(1)},\dots,\hat{\calP}^{(p)}_\perp\bw^{(p)},\hat{\calP}^{(1)}_\perp\bw^{(1)},\dots,\hat{\calP}^{(p)}_\perp\bw^{(p)}).
$$
Note that
\begin{eqnarray}\nonumber
	&&(\Sigma-\hat{\Sigma})(\hat{\calP}^{(1)}_\perp\bw^{(1)},\dots,\hat{\calP}^{(p)}_\perp\bw^{(p)},\hat{\calP}^{(1)}_\perp\bw^{(1)},\dots,\hat{\calP}^{(p)}_\perp\bw^{(p)})\\\nonumber
	&=&\sum_{l_1 =1}^r \sum_{l_2 =1}^r \sigma_{l_1} \sigma_{l_2} \left( \hat{\Sigma}_{\theta,l_1l_2} -\Sigma_{\theta,l_1l_2} \right) \left( \prod_{q=1}^p \langle \bu_{l_1}^{(q)}, \hat{\calP}_\perp^{(q)}\bw^{(q)} \rangle \right) \left( \prod_{q=1}^p \langle \bu_{l_2}^{(q)}, \hat{\calP}_\perp^{(q)}\bw^{(q)} \rangle \right)\\\nonumber
	&&+\frac{2}{n} \sum_{i=1}^n\sum_{l=1}^r \sigma_l \theta_{il} \left( \prod_{q=1}^p \langle \bu_l^{(q)}, \hat{\calP}_\perp^{(q)}\bw^{(q)} \rangle \right) \scrE_i (\hat{\calP}_\perp^{(1)}\bw^{(1)},\dots,\hat{\calP}_\perp^{(p)}\bw^{(p)} )\\
	&&+\left(\frac{1}{n} \sum_{i=1}^n [ \scrE_i (\hat{\calP}_\perp^{(1)}\bw^{(1)},\dots,\hat{\calP}_\perp^{(p)}\bw^{(p)} ) ]^2-\sigma_0^2\prod_{q=1}^p\|\hat{\calP}_\perp^{(q)}\bw^{(q)}\|^2\right).\label{eq:inductemp0}
\end{eqnarray}
We bound the three terms on the right hand side separately.

The first term can be bounded by
\begin{eqnarray*}
	&&\biggl|\sum_{l_1 =1}^r \sum_{l_2 =1}^r \sigma_{l_1} \sigma_{l_2} \left( \frac{1}{n} \sum_{i=1}^n \theta_{il_1 } \theta_{il_2} \right) \left( \prod_{q=1}^p \langle \bu_{l_1}^{(q)}, \hat{\calP}_\perp^{(q)}\bw^{(q)} \rangle \right) \left( \prod_{q=1}^p \langle \bu_{l_2}^{(q)}, \hat{\calP}_\perp^{(q)}\bw^{(q)} \rangle \right)\\
	&&\qquad -  \sum_{l=1}^r \sigma_l^2 \left( \prod_{q=1}^p \langle \bu_l^{(q)}, \hat{\calP}_\perp^{(q)}\bw^{(q)} \rangle \right)^2  \biggr|\\
	&\le&\|\hat{\Sigma}_\theta-I_r\|\left[\sum_{l=1}^r \sigma_l^2 \left( \prod_{q=1}^p \langle \bu_l^{(q)}, \hat{\calP}_\perp^{(q)}\bw^{(q)} \rangle \right)^2\right]\\
	&=&\|\hat{\Sigma}_\theta-I_r\|\left[\sum_{l\in \pi([k-1])} \sigma_l^2 \left( \prod_{q=1}^p \langle \bu_l^{(q)}, \hat{\calP}_\perp^{(q)}\bw^{(q)} \rangle \right)^2+\sum_{l\notin \pi([k-1])} \sigma_l^2 \left( \prod_{q=1}^p \langle \bu_l^{(q)}, \hat{\calP}_\perp^{(q)}\bw^{(q)} \rangle \right)^2\right],
\end{eqnarray*}
Recall that for any $l\in \pi([k-1])$,
$$
|\langle \bu_l^{(q)}, \hat{\calP}_\perp^{(q)}\bw^{(q)} \rangle|=|\langle \hat{\calP}_\perp^{(q)}\bu_l^{(q)}, \bw^{(q)} \rangle|\le\|\hat{\calP}_\perp^{(q)}\bu_l^{(q)}\|\le\delta_l.
$$
Therefore,
\begin{eqnarray*}
\sum_{l\in \pi([k-1])} \sigma_l^2 \left( \prod_{q=1}^p \langle \bu_l^{(q)}, \hat{\calP}_\perp^{(q)}\bw^{(q)} \rangle \right)^2&\le& \max_{l\in \pi([k-1])} \sigma_l^2 \left( \prod_{q=2}^p \langle \bu_l^{(q)}, \hat{\calP}_\perp^{(q)}\bw^{(q)} \rangle \right)^2\\
&\le&\max_{1\le l<k} \{\sigma_{\pi(l)}^2\delta_l^{2p-2}\}\le \max_{1\le l<k} \{\sigma_{\pi(l)}^2\delta_l^{2}\}\le \tau^2,
\end{eqnarray*}
by taking $c_0$ of \eqref{eq:assumption} small enough. On the other hand,
$$
\sum_{l\notin \pi([k-1])} \sigma_l^2 \left( \prod_{q=1}^p \langle \bu_l^{(q)}, \hat{\calP}_\perp^{(q)}\bw^{(q)} \rangle \right)^2\le \max_{l\notin \pi([k-1])} \sigma_l^2=\tau^2.
$$
This implies that
\begin{equation}
\left|\sum_{l_1 =1}^r \sum_{l_2 =1}^r \sigma_{l_1} \sigma_{l_2} \left( \hat{\Sigma}_{\theta,l_1l_2} -\Sigma_{\theta,l_1l_2} \right) \left( \prod_{q=1}^p \langle \bu_{l_1}^{(q)}, \hat{\calP}_\perp^{(q)}\bw^{(q)} \rangle \right) \left( \prod_{q=1}^p \langle \bu_{l_2}^{(q)}, \hat{\calP}_\perp^{(q)}\bw^{(q)} \rangle \right)\right|
\le C\tau^2\sqrt{r\over n}.\label{eq:inductemp1}
\end{equation}

Similarly, the second term can be bounded by
\begin{eqnarray}\nonumber
	&&\left|\frac{1}{n} \sum_{i=1}^n\sum_{l=1}^r \sigma_l \theta_{il} \left( \prod_{q=1}^p \langle \bu_l^{(q)}, \hat{\calP}_\perp^{(q)}\bw^{(q)} \rangle \right) \scrE_i (\hat{\calP}_\perp^{(1)}\bw^{(1)},\dots,\hat{\calP}_\perp^{(p)}\bw^{(p)} )\right|\\\nonumber
	&\le& \|\hat{\Sigma}_{\theta,\scrE}\|\left[\sum_{l=1}^r \sigma_l^2 \left( \prod_{q=1}^p \langle \bu_l^{(q)}, \hat{\calP}_\perp^{(q)}\bw^{(q)} \rangle \right)^2\right]^{1/2}\\
	&\le&C\tau\sigma_0\sqrt{d\over n}.\label{eq:inductemp2}
\end{eqnarray}

Finally, the third term can be bounded by
\begin{equation}\label{eq:inductemp3}
\left|\frac{1}{n} \sum_{i=1}^n [ \scrE_i (\hat{\calP}_\perp^{(1)}\bw^{(1)},\dots,\hat{\calP}_\perp^{(p)}\bw^{(p)} ) ]^2-\sigma_0^2\prod_{q=1}^p\|\hat{\calP}_\perp^{(q)}\bw^{(q)}\|^2\right|\le \|\hat{\Sigma}_{\scrE}-\Sigma_{\scrE}\|\le C\sigma_0^2\sqrt{d\over n}.
\end{equation}
Combing \eqref{eq:inductemp0}-\eqref{eq:inductemp3}, we get
\begin{eqnarray*}\nonumber
&&\sup_{\|\bw^{(q)}\|\le 1,1\le q\le p}\left|(\hat{\Sigma}-\Sigma)(\hat{\calP}_\perp^{(1)}\bw^{(1)},\dots,\hat{\calP}_\perp^{(p)}\bw^{(p)},\hat{\calP}_\perp^{(1)}\bw^{(1)},\dots,\hat{\calP}_\perp^{(p)}\bw^{(p)})\right|\\
&\le& C\tau^2\sqrt{r\over n}+C(\sigma_0\tau+\sigma_0^2)\sqrt{d\over n}.\label{eq:inductemp}
\end{eqnarray*}
Together with \eqref{eq:denomlower}, this implies
\begin{equation}\label{eq:inductlower}
\tilde{\Sigma}(\hat{\bu}_k^{(1)},\hat{\bu}_k^{(2)},\ldots,\hat{\bu}_k^{(p)}, \hat{\bu}_k^{(1)},\ldots, \hat{\bu}_k^{(p)})\ge \tau^2\left(1-C\delta_k^4-C\sqrt{r\over n}\right)-C(\sigma_0\tau+\sigma_0^2)\sqrt{d\over n}-C\sigma_0^2\delta_k^2,
\end{equation}
by taking $c_0$ of \eqref{eq:assumption} small enough.

\subsection{Upper Bounds for $\tilde{\Sigma}(\hat{\calP}_\perp^{(1)}(\hat{\bu}_k^{(1)}-\langle\hat{\bu}_k^{(1)},\bu_{\pi(k)}^{(1)}\rangle\bu_{\pi(k)}^{(1)}),\hat{\bu}_k^{(2)},\ldots,\hat{\bu}_k^{(p)}, \hat{\bu}_k^{(1)},\ldots, \hat{\bu}_k^{(p)})$}

Observe that for any $\bw$ orthogonal to $\bu_{\pi(k)}^{(1)}$, we have
\begin{eqnarray}\nonumber
	&&\tilde{\Sigma}(\hat{\calP}_\perp^{(1)}\bw,\hat{\bu}_k^{(2)},\dots,\hat{\bu}_k^{(p)},\hat{\bu}_k^{(1)},\dots,\hat{\bu}_k^{(p)})  \\ \nonumber
	&=&\sigma_{\pi(k)}^2 \left( \frac{1}{n} \sum_{i=1}^n \theta_{i\pi(k)}^2 \right) \left( \prod_{q=2}^p \langle \bu_{\pi(k)}^{(q)}, \hat{\bu}_k^{(q)} \rangle \right) \left( \prod_{q=1}^p \langle \bu_{\pi(k)}^{(q)}, \hat{\bu}_k^{(q)} \rangle \right) \langle\bu_{\pi(k)}^{(1)},\hat{\calP}_\perp^{(1)}\bw\rangle\\\nonumber
	&&+\sum_{l\neq \pi(k)} \sigma_{\pi(k)} \sigma_{l} \left( \frac{1}{n} \sum_{i=1}^n \theta_{i\pi(k)} \theta_{il} \right) \left( \prod_{q=2}^p \langle \bu_{\pi(k)}^{(q)}, \hat{\bu}_k^{(q)} \rangle \right) \left( \prod_{q=1}^p \langle \bu_{l}^{(q)}, \hat{\bu}_k^{(q)} \rangle \right) \langle\bu_{\pi(k)}^{(1)},\hat{\calP}_\perp^{(1)}\bw\rangle\\\nonumber
	&&+\sum_{l\neq \pi(k)} \sigma_{l} \sigma_{\pi(k)} \left( \frac{1}{n} \sum_{i=1}^n \theta_{il} \theta_{i\pi(k)} \right) \left( \prod_{q=2}^p \langle \bu_{l}^{(q)}, \hat{\bu}_k^{(q)} \rangle \right) \left( \prod_{q=1}^p \langle \bu_{\pi(k)}^{(q)}, \hat{\bu}_k^{(q)} \rangle \right) \langle\bu_{l}^{(1)},\hat{\calP}_\perp^{(1)}\bw\rangle\\\nonumber
	&&+\sum_{l_1,l_2\neq \pi(k)} \sigma_{l_1} \sigma_{l_2} \left( \frac{1}{n} \sum_{i=1}^n \theta_{il_1} \theta_{il_2} \right) \left( \prod_{q=2}^p \langle \bu_{l_1}^{(q)}, \hat{\bu}_k^{(q)} \rangle \right) \left( \prod_{q=1}^p \langle \bu_{l_2}^{(q)}, \hat{\bu}_k^{(q)} \rangle \right) \langle\bu_{l_1}^{(1)},\hat{\calP}_\perp^{(1)}\bw\rangle\\\nonumber
	&&+  \frac{1}{n} \sum_{i=1}^n\left[\sum_{l=1}^r \sigma_l \theta_{il} \left( \prod_{q=2}^p \langle \bu_l^{(q)}, \hat{\bu}_k^{(q)} \rangle \right) \scrE_i (\hat{\bu}_k^{(1)},\dots,\hat{\bu}_k^{(p)} ) \langle\bu_l^{(1)},\hat{\calP}_\perp^{(1)}\bw\rangle\right] \\\nonumber
	&&+  \frac{1}{n} \sum_{i=1}^n\left[\sum_{l=1}^r \scrE_i (\hat{\calP}_\perp^{(1)}\bw,\hat{\bu}_k^{(2)},\dots,\hat{\bu}_k^{(p)} ) \sigma_l \theta_{il} \left( \prod_{q=1}^p \langle \bu_l^{(q)}, \hat{\bu}_k^{(q)} \rangle \right)\right] \\
	&&+ \frac{1}{n} \sum_{i=1}^n  \scrE_i (\hat{\calP}_\perp^{(1)}\bw,\hat{\bu}_k^{(2)},\dots,\hat{\bu}_k^{(p)} )  \scrE_i (\hat{\bu}_k^{(1)},\dots,\hat{\bu}_k^{(p)} )-\sigma_0^2\langle\hat{\calP}_\perp^{(1)}\bw,\hat{\bu}_k^{(p)}\rangle.\label{eq:inductupper0}
\end{eqnarray}
Again each term on the right hand side needs to be bounded carefully.

The first term on the right hand side of \eqref{eq:inductupper0} can be bounded by
\begin{eqnarray*}
\left|\sigma_{\pi(k)}^2 \left( \frac{1}{n} \sum_{i=1}^n \theta_{i\pi(k)}^2 \right) \left( \prod_{q=2}^p \langle \bu_{\pi(k)}^{(q)}, \hat{\bu}_k^{(q)} \rangle \right) \left( \prod_{q=1}^p \langle \bu_{\pi(k)}^{(q)}, \hat{\bu}_k^{(q)} \rangle \right) \langle\bu_{\pi(k)}^{(1)},\hat{\calP}_\perp^{(1)}\bw\rangle\right|\\
\le \tau^2\|\hat{\Sigma}_\theta\|_{\max}|\langle\bu_{\pi(k)}^{(1)},\hat{\calP}_\perp\bw\rangle|.
\end{eqnarray*}
In particular, when $\bw=\hat{\bu}_k^{(1)}-\langle\hat{\bu}_k^{(1)},\bu_{\pi(k)}^{(1)}\rangle\bu_{\pi(k)}^{(1)}$, we have
\begin{eqnarray*}
|\langle\bu_{\pi(k)}^{(1)},\hat{\calP}_\perp^{(1)}\bw\rangle|&=&\left|\langle\hat{\bu}_k^{(1)},\bu_{\pi(k)}^{(1)}\rangle(1-\langle\bu_{\pi(k)}^{(1)},\hat{\calP}_\perp^{(1)}\bu_{\pi(k)}^{(1)}\rangle)\right|\\
&\le&1-\langle\bu_{\pi(k)}^{(1)},\hat{\calP}_\perp^{(1)}\bu_{\pi(k)}^{(1)}\rangle\\
&=&\|\hat{\calP}^{(1)}\bu_{\pi(k)}^{(1)}\|^2\le \sum_{1\le l<k}\tilde{\delta}_l^2\le C\delta_k^2,
\end{eqnarray*}
by taking $c_0$ small enough. Thus,
\begin{equation}\label{eq:inductupper1}
\left|\sigma_{\pi(k)}^2 \left( \frac{1}{n} \sum_{i=1}^n \theta_{i\pi(k)}^2 \right) \left( \prod_{q=2}^p \langle \bu_{\pi(k)}^{(q)}, \hat{\bu}_k^{(q)} \rangle \right) \left( \prod_{q=1}^p \langle \bu_{\pi(k)}^{(q)}, \hat{\bu}_k^{(q)} \rangle \right) \langle\bu_{\pi(k)}^{(1)},\hat{\calP}_\perp^{(1)}\bw\rangle\right|\le C\tau^2\delta_k^2.
\end{equation}

The second term on the right hand side of \eqref{eq:inductupper0} can be bounded as follows:
\begin{eqnarray*}
	&&\left|\sum_{l\neq \pi(k)} \sigma_{\pi(k)} \sigma_{l} \left( \frac{1}{n} \sum_{i=1}^n \theta_{i\pi(k)} \theta_{il} \right) \left( \prod_{q=2}^p \langle \bu_{\pi(k)}^{(q)}, \hat{\bu}_k^{(q)} \rangle \right) \left( \prod_{q=1}^p \langle \bu_{l}^{(q)}, \hat{\bu}_k^{(q)} \rangle \right) \langle\bu_{\pi(k)}^{(1)},\hat{\calP}_\perp^{(1)}\bw\rangle\right|\\
	&\le&\tau\|\hat{\Sigma}_\theta-I\|\left(\sum_{l\neq \pi(k)}\sigma_l^2\prod_{q=1}^p \langle \bu_{l}^{(q)}, \hat{\bu}_k^{(q)} \rangle^2\right)^{1/2}\\
	&\le&\tau\|\hat{\Sigma}_\theta-I\|\left(\sum_{l\neq \pi(k)}\langle \bu_{l}^{(1)}, \hat{\bu}_k^{(1)} \rangle^2\right)^{1/2}\max_{l\neq \pi(k)}\left\{\sigma_l\prod_{q=2}^p |\langle \bu_{l}^{(q)}, \hat{\bu}_k^{(q)} \rangle|\right\}\\
	&=&\tau\delta_k\|\hat{\Sigma}_\theta-I\|\max_{l\neq \pi(k)}\left\{\sigma_l\prod_{q=2}^p |\langle \bu_{l}^{(q)}, \hat{\bu}_k^{(q)} \rangle|\right\}\\
	&\le&\tau\delta_k\|\hat{\Sigma}_\theta-I\|\max\left\{\max_{l\in \pi([k-1])}\left\{\sigma_l\prod_{q=2}^p |\langle \bu_{l}^{(q)}, \hat{\bu}_k^{(q)} \rangle|\right\},\max_{l\notin \pi([k])}\left\{\sigma_l\prod_{q=2}^p |\langle \bu_{l}^{(q)}, \hat{\bu}_k^{(q)} \rangle|\right\}\right\}.
\end{eqnarray*}
Note that
$$
\max_{l\in \pi([k-1])}\left\{\sigma_l\prod_{q=2}^p |\langle \bu_{l}^{(q)}, \hat{\bu}_k^{(q)} \rangle|\right\}\le\max_{l\in \pi([k-1])}\left\{\sigma_l\prod_{q=2}^p |\langle \hat{\calP}_\perp^{(q)}\bu_{l}^{(q)}, \hat{\bu}_k^{(q)} \rangle|\right\}\le\max_{1\le l<k}\{\sigma_{\pi(l)}\delta_l^{p-1}\}\le 2\tau\delta_k^{p-1},
$$
and
$$
\max_{l\notin \pi([k])}\left\{\sigma_l\prod_{q=2}^p |\langle \bu_{l}^{(q)}, \hat{\bu}_k^{(q)} \rangle|\right\}\le \tau\delta_k^{p-1}.
$$
We get
\begin{equation}\label{eq:inductupper2}
\left|\sum_{l\neq \pi(k)} \sigma_{\pi(k)} \sigma_{l} \left( \frac{1}{n} \sum_{i=1}^n \theta_{i\pi(k)} \theta_{il} \right) \left( \prod_{q=2}^p \langle \bu_{\pi(k)}^{(q)}, \hat{\bu}_k^{(q)} \rangle \right) \left( \prod_{q=1}^p \langle \bu_{l}^{(q)}, \hat{\bu}_k^{(q)} \rangle \right)\right|\le C\tau^2\delta_k^p\sqrt{r\over n}.
\end{equation}

And similarly, when $\bw=\hat{\bu}_k^{(1)}-\langle\hat{\bu}_k^{(1)},\bu_{\pi(k)}^{(1)}\rangle\bu_{\pi(k)}^{(1)}$, we bound the third term on the right hand side of \eqref{eq:inductupper0} by
\begin{eqnarray}\nonumber
&&\left|\sum_{l\neq \pi(k)} \sigma_{l} \sigma_{\pi(k)} \left( \frac{1}{n} \sum_{i=1}^n \theta_{il} \theta_{i\pi(k)} \right) \left( \prod_{q=2}^p \langle \bu_{l}^{(q)}, \hat{\bu}_k^{(q)} \rangle \right) \left( \prod_{q=1}^p \langle \bu_{\pi(k)}^{(q)}, \hat{\bu}_k^{(q)} \rangle \right) \langle\bu_{l}^{(1)},\hat{\calP}_\perp^{(1)}\bw\rangle\right|\\\nonumber
&\le&\tau\|\hat{\Sigma}_\theta-I\|\left(\sum_{l\neq \pi(k)}\sigma_l^2\langle\bu_{l}^{(1)},\hat{\calP}_\perp^{(1)}\bw\rangle^2\prod_{q=2}^p \langle \bu_{l}^{(q)}, \hat{\bu}_k^{(q)} \rangle^2\right)^{1/2}\\\nonumber
&\le&\tau\|\hat{\Sigma}_\theta-I\|\left(\sum_{l\neq \pi(k)}\langle\bu_{l}^{(1)},\hat{\calP}_\perp^{(1)}\bw\rangle^2\right)^{1/2}\max_{l\neq \pi(k)}\left\{\sigma_l\prod_{q=2}^p |\langle \bu_{l}^{(q)}, \hat{\bu}_k^{(q)} \rangle\right\}\\
&\le&C\tau^2\delta_k^p\sqrt{r\over n},\label{eq:inductupper3}
\end{eqnarray}
where in the last inequality we used the fact that 
$$
\sum_{l\neq \pi(k)}\langle\bu_{l}^{(1)},\hat{\calP}_\perp^{(1)}\bw\rangle^2\le \|\hat{\calP}_\perp^{(1)}\bw\|^2\le \|\bw\|^2\le \delta_k^2;
$$
the fourth term by
\begin{eqnarray}\nonumber
	&&\left|\sum_{l_1,l_2\neq \pi(k)} \sigma_{l_1} \sigma_{l_2} \left( \frac{1}{n} \sum_{i=1}^n \theta_{il_1} \theta_{il_2} \right) \left( \prod_{q=2}^p \langle \bu_{l_1}^{(q)}, \hat{\bu}_k^{(q)} \rangle \right) \left( \prod_{q=1}^p \langle \bu_{l_2}^{(q)}, \hat{\bu}_k^{(q)} \rangle \right) \langle\bu_{l_1}^{(1)},\hat{\calP}_\perp^{(1)}\bw\rangle\right|\\\nonumber
	&\le&\|\hat{\Sigma}_\theta\|\left(\sum_{l\neq \pi(k)}\sigma_l^2\langle\bu_{l}^{(1)},\hat{\calP}_\perp^{(1)}\bw\rangle^2 \prod_{q=2}^p \langle \bu_{l}^{(q)}, \hat{\bu}_k^{(q)} \rangle^2\right)^{1/2}\left(\sum_{l\neq \pi(k)}\sigma_l^2 \prod_{q=1}^p \langle \bu_{l}^{(q)}, \hat{\bu}_k^{(q)} \rangle^2\right)^{1/2}\\
	&\le&C\tau^2\delta_k^{2(p-1)};\label{eq:inductupper4}
\end{eqnarray}
the fifth term by
\begin{eqnarray}\nonumber
	&&\frac{1}{n} \sum_{i=1}^n\left[\sum_{l\neq \pi(k)} \sigma_l \theta_{il} \left( \prod_{q=2}^p \langle \bu_l^{(q)}, \hat{\bu}_k^{(q)} \rangle \right) \scrE_i (\hat{\bu}_k^{(1)},\dots,\hat{\bu}_k^{(p)} ) \langle\bu_l^{(1)},\hat{\calP}_\perp^{(1)}\bw\rangle\right]\\\nonumber
	&=&\sum_{l\neq \pi(k)} \left\{\sigma_l\left( \prod_{q=2}^p \langle \bu_l^{(q)}, \hat{\bu}_k^{(q)} \rangle \right)\langle\bu_l^{(1)},\hat{\calP}_\perp^{(1)}\bw\rangle \left[{1\over n}\sum_{i=1}^n\theta_{il}\scrE_i (\hat{\bu}_k^{(1)},\dots,\hat{\bu}_k^{(p)} )\right]\right\}\\\nonumber
	&\le&\|\hat{\Sigma}_{\theta,\scrE}\|\cdot\left[\sum_{l\neq \pi(k)} \left(\sigma_l^2\langle\bu_l^{(1)},\hat{\calP}_\perp^{(1)}\bw\rangle^2\prod_{q=2}^p \langle \bu_l^{(q)}, \hat{\bu}_k^{(q)} \rangle^2 \right)\right]^{1/2}\\
	&\le&C\sigma_0\tau\delta_k^{p-1}\sqrt{d\over n};\label{eq:inductupper5}
\end{eqnarray}
and the sixth term by
\begin{eqnarray}\nonumber
	&&\frac{1}{n} \sum_{i=1}^n\left[\sum_{l=1}^r \scrE_i (\hat{\calP}_\perp^{(1)}\bw,\hat{\bu}_k^{(2)},\dots,\hat{\bu}_k^{(p)} ) \sigma_l \theta_{il} \left( \prod_{q=1}^p \langle \bu_l^{(q)}, \hat{\bu}_k^{(q)} \rangle \right)\right]\\\nonumber
	&\le&\frac{1}{n} \sum_{i=1}^n\left[\scrE_i (\hat{\calP}_\perp^{(1)}\bw,\hat{\bu}_k^{(2)},\dots,\hat{\bu}_k^{(p)} ) \sigma_{\pi(k)} \theta_{i\pi(k)} \left( \prod_{q=1}^p \langle \bu_{\pi(k)}^{(q)}, \hat{\bu}_k^{(q)} \rangle \right)\right]\\\nonumber
	&&+\frac{1}{n} \sum_{i=1}^n\left[\sum_{l\neq \pi(k)} \scrE_i (\hat{\calP}_\perp^{(1)}\bw,\hat{\bu}_k^{(2)},\dots,\hat{\bu}_k^{(p)} ) \sigma_l \theta_{il} \left( \prod_{q=1}^p \langle \bu_l^{(q)}, \hat{\bu}_k^{(q)} \rangle \right)\right]\\\nonumber
	&\le&\tau\|\hat{\Sigma}_{\theta,\scrE}\|+\|\hat{\Sigma}_{\theta,\scrE}\|\left(\sum_{l\neq \pi(k)} \sigma_l^2  \prod_{q=1}^p \langle \bu_l^{(q)}, \hat{\bu}_k^{(q)} \rangle^2 \right)^{1/2}\\
	&\le&C\sigma_0\tau\sqrt{d\over n}.\label{eq:inductupper6}
\end{eqnarray}

Finally the last term can be bounded by
$$
\left|\frac{1}{n} \sum_{i=1}^n  \scrE_i (\hat{\calP}_\perp^{(1)}\bw,\hat{\bu}_k^{(2)},\dots,\hat{\bu}_k^{(p)} )  \scrE_i (\hat{\bu}_k^{(1)},\dots,\hat{\bu}_k^{(p)} )-\sigma_0^2\langle \hat{\calP}_\perp^{(1)}\bw, \hat{\bu}_k^{(1)}\rangle\right|\le \|\hat{\Sigma}_{\scrE}-\Sigma_{\scrE}\|\le C\sigma_0^2\sqrt{d\over n}.
$$
Together with \eqref{eq:inductupper1}-\eqref{eq:inductupper6}, we get
\begin{eqnarray}\nonumber
\tilde{\Sigma}(\hat{\calP}_\perp^{(1)}(\hat{\bu}_k^{(1)}-\langle\hat{\bu}_k^{(1)},\bu_{\pi(k)}^{(1)}\rangle\bu_{\pi(k)}^{(1)}),\hat{\bu}_k^{(2)},\ldots,\hat{\bu}_k^{(p)}, \hat{\bu}_k^{(1)},\ldots, \hat{\bu}_k^{(p)})\le C\tau^2\delta_k^2+C(\sigma_0^2+\sigma_0\tau)\sqrt{d\over n}.
\label{eq:inductupper}
\end{eqnarray}

\subsection{Upper Bounds for $\max_{l\notin \pi([k])}\tilde{\Sigma}(\hat{\calP}_\perp^{(1)}\bu_m^{(1)},\hat{\bu}_k^{(2)},\ldots,\hat{\bu}_k^{(p)}, \hat{\bu}_k^{(1)},\ldots, \hat{\bu}_k^{(p)})$}\label{subsec:upper2}

To derive the helper bound \eqref{eq:rate_main1}, we also need an upper bound for 
$$\max_{l\notin \pi([k])}\tilde{\Sigma}(\hat{\calP}_\perp^{(1)}\bu_m^{(1)},\hat{\bu}_k^{(2)},\ldots,\hat{\bu}_k^{(p)}, \hat{\bu}_k^{(1)},\ldots, \hat{\bu}_k^{(p)}).$$
We shall follow a similar step by bound each term on the right hand side of \eqref{eq:inductupper0}, but now with $\bw=\bu_m^{(1)}$  ($m\notin\pi([k])$).

Specifically, the first term can be bounded by
\begin{eqnarray*}
\left|\sigma_{\pi(k)}^2 \left( \frac{1}{n} \sum_{i=1}^n \theta_{i\pi(k)}^2 \right) \left( \prod_{q=2}^p \langle \bu_{\pi(k)}^{(q)}, \hat{\bu}_k^{(q)} \rangle \right) \left( \prod_{q=1}^p \langle \bu_{\pi(k)}^{(q)}, \hat{\bu}_k^{(q)} \rangle \right) \langle\bu_{\pi(k)}^{(1)},\hat{\calP}_\perp^{(1)}\bu_m^{(1)}\rangle\right|\\
\le\tau^2\|\hat{\Sigma}_\theta\|_{\max}|\langle\bu_{\pi(k)}^{(1)},\hat{\calP}_\perp^{(1)}\bu_m^{(1)}\rangle|.
\end{eqnarray*}
Note that
$$
\langle\bu_{\pi(k)}^{(1)},\bu_m^{(1)}\rangle=\langle\bu_{\pi(k)}^{(1)},\hat{\calP}^{(1)}\bu_m^{(1)}\rangle+\langle\bu_{\pi(k)}^{(1)},\hat{\calP}_\perp^{(1)}\bu_m^{(1)}\rangle=0.
$$
We get
$$
|\langle\bu_{\pi(k)}^{(1)},\hat{\calP}_\perp^{(1)}\bu_m^{(1)}\rangle|=|\langle\bu_{\pi(k)}^{(1)},\hat{\calP}^{(1)}\bu_m^{(1)}\rangle|\le \|\hat{\calP}\bu_{\pi(k)}^{(1)}\|\|\hat{\calP}\bu_m^{(1)}\|\le \sum_{1\le l<k}\tilde{\delta}_l^2\le C\delta_k^2,
$$
by Cauchy-Schwartz inequality. This implies that
$$
\left|\sigma_{\pi(k)}^2 \left( \frac{1}{n} \sum_{i=1}^n \theta_{i\pi(k)}^2 \right) \left( \prod_{q=2}^p \langle \bu_{\pi(k)}^{(q)}, \hat{\bu}_k^{(q)} \rangle \right) \left( \prod_{q=1}^p \langle \bu_{\pi(k)}^{(q)}, \hat{\bu}_k^{(q)} \rangle \right) \langle\bu_{\pi(k)}^{(1)},\hat{\calP}_\perp^{(1)}\bu_m^{(1)}\rangle\right|\le C\tau^2\delta_k^2.
$$
by taking $c_0$ small enough. 

The second term can also be bounded by
\begin{eqnarray*}
&&\left|\sum_{l\neq \pi(k)} \sigma_{\pi(k)} \sigma_{l} \left( \frac{1}{n} \sum_{i=1}^n \theta_{i\pi(k)} \theta_{il} \right) \left( \prod_{q=2}^p \langle \bu_{\pi(k)}^{(q)}, \hat{\bu}_k^{(q)} \rangle \right) \left( \prod_{q=1}^p \langle \bu_{l}^{(q)}, \hat{\bu}_k^{(q)} \rangle \right) \langle\bu_{\pi(k)}^{(1)},\hat{\calP}_\perp^{(1)}\bu_m^{(1)}\rangle\right|\\
&\le&\tau\|\hat{\Sigma}_\theta-I\||\langle\bu_{\pi(k)}^{(1)},\hat{\calP}_\perp^{(1)}\bu_m^{(1)}\rangle|\left(\sum_{l\neq \pi(k)}\sigma_l^2\prod_{q=1}^p \langle \bu_{l}^{(q)}, \hat{\bu}_k^{(q)} \rangle^2\right)^{1/2}\\
&\le&\tau^2\delta_k^{p+1}\sqrt{r\over n}.
\end{eqnarray*}

We bound the third term by
\begin{eqnarray*}
&&\left|\sum_{l\neq \pi(k)} \sigma_{l} \sigma_{\pi(k)} \left( \frac{1}{n} \sum_{i=1}^n \theta_{il} \theta_{i\pi(k)} \right) \left( \prod_{q=2}^p \langle \bu_{l}^{(q)}, \hat{\bu}_k^{(q)} \rangle \right) \left( \prod_{q=1}^p \langle \bu_{\pi(k)}^{(q)}, \hat{\bu}_k^{(q)} \rangle \right) \langle\bu_{l}^{(1)},\hat{\calP}_\perp^{(1)}\bu_m^{(1)}\rangle\right|\\
&\le&\left|\sigma_{m} \sigma_{\pi(k)} \left( \frac{1}{n} \sum_{i=1}^n \theta_{im} \theta_{i\pi(k)} \right) \left( \prod_{q=2}^p \langle \bu_{m}^{(q)}, \hat{\bu}_k^{(q)} \rangle \right) \left( \prod_{q=1}^p \langle \bu_{\pi(k)}^{(q)}, \hat{\bu}_k^{(q)} \rangle \right) \langle\bu_{m}^{(1)},\hat{\calP}_\perp^{(1)}\bu_m^{(1)}\rangle\right|\\
&&+\left|\sum_{l\notin \{\pi(k),m\}} \sigma_{l} \sigma_{\pi(k)} \left( \frac{1}{n} \sum_{i=1}^n \theta_{il} \theta_{i\pi(k)} \right) \left( \prod_{q=2}^p \langle \bu_{l}^{(q)}, \hat{\bu}_k^{(q)} \rangle \right) \left( \prod_{q=1}^p \langle \bu_{\pi(k)}^{(q)}, \hat{\bu}_k^{(q)} \rangle \right) \langle\bu_{l}^{(1)},\hat{\calP}_\perp^{(1)}\bu_m^{(1)}\rangle\right|\\
\end{eqnarray*}
The first term on the right hand side can be further bounded by
$$
\tau^2\|\hat{\Sigma}_\theta-\Sigma_\theta\|_{\max}\prod_{q=2}^p \left|\langle \bu_{m}^{(q)}, \hat{\bu}_k^{(q)} \rangle \right| \le C\tau^2\delta_k^{p-1}\sqrt{\log r\over n}.
$$
Now consider the second term:
\begin{eqnarray*}
&&\left|\sum_{l\notin \{\pi(k),m\}} \sigma_{l} \sigma_{\pi(k)} \left( \frac{1}{n} \sum_{i=1}^n \theta_{il} \theta_{i\pi(k)} \right) \left( \prod_{q=2}^p \langle \bu_{l}^{(q)}, \hat{\bu}_k^{(q)} \rangle \right) \left( \prod_{q=1}^p \langle \bu_{\pi(k)}^{(q)}, \hat{\bu}_k^{(q)} \rangle \right) \langle\bu_{l}^{(1)},\hat{\calP}_\perp^{(1)}\bu_m^{(1)}\rangle\right|\\
&\le&\tau\|\hat{\Sigma}_\theta-I\|\left(\sum_{l\notin \{\pi(k),m\}}\sigma_l^2\langle\bu_{l}^{(1)},\hat{\calP}_\perp^{(1)}\bu_m^{(1)}\rangle^2\prod_{q=2}^p \langle \bu_{l}^{(q)}, \hat{\bu}_k^{(q)} \rangle^2\right)^{1/2}\\
&\le&\tau\|\hat{\Sigma}_\theta-I\|\Biggl(\sum_{l\in \pi([k-1])}\sigma_l^2\langle\bu_{l}^{(1)},\hat{\calP}_\perp^{(1)}\bu_m^{(1)}\rangle^2\prod_{q=2}^p \langle \bu_{l}^{(q)}, \hat{\bu}_k^{(q)} \rangle^2\\
&&\hskip100pt+\sum_{l\notin \pi([k])\cup\{m\}}\sigma_l^2\langle\bu_{l}^{(1)},\hat{\calP}_\perp^{(1)}\bu_m^{(1)}\rangle^2\prod_{q=2}^p \langle \bu_{l}^{(q)}, \hat{\bu}_k^{(q)} \rangle^2\Biggr)^{1/2}.
%	&\le&\tau\delta_k\|\hat{\Sigma}_\theta-I\|\left(\max_{1\le l<k}\{\sigma_{\pi(l)}\delta_l^{p-1}\}+\tau\delta_k^{p-2}\left( \sum_{1\le l<k}\tilde{\delta}_l^2\right)+\tau\delta_k^{p-1}\right),
\end{eqnarray*}
The first term in the bracket on the rightmost hand side can be bounded by
\begin{eqnarray*}
&&\sum_{l\in \pi([k-1])}\sigma_l^2\|\hat{\calP}_\perp^{(1)}\bu_{l}^{(1)}\|^2\prod_{q=2}^p \langle \bu_{l}^{(q)}, \hat{\bu}_k^{(q)} \rangle^2\\
&\le& \left(\sum_{l\in \pi([k-1])}\langle \bu_{l}^{(2)}, \hat{\bu}_k^{(2)} \rangle^2\right)\left(\max_{l\in \pi([k-1])}\sigma_l^2\|\hat{\calP}_\perp^{(1)}\bu_{l}^{(1)}\|^2\prod_{q=3}^p \langle \bu_{l}^{(q)}, \hat{\bu}_k^{(q)} \rangle^2\right)\\
&\le&\delta_k^2\left(\max_{l\in \pi([k-1])}\sigma_l^2\delta_l^2\delta_k^{2(p-2)}\right)\le C\tau^2\delta_k^{2p};
\end{eqnarray*}
the second term by
$$
\|\hat{\calP}^{(1)}\bu_{m}^{(1)}\|^2\left(\max_{l\notin \pi([k])\cup\{m\}}\sigma_l^2\prod_{q=2}^p \langle \bu_{l}^{(q)}, \hat{\bu}_k^{(q)} \rangle^2\right)\le C\tau^2\delta_k^{2p},
$$
so that
\begin{eqnarray*}
&&\left|\sum_{l\neq \pi(k)} \sigma_{l} \sigma_{\pi(k)} \left( \frac{1}{n} \sum_{i=1}^n \theta_{il} \theta_{i\pi(k)} \right) \left( \prod_{q=2}^p \langle \bu_{l}^{(q)}, \hat{\bu}_k^{(q)} \rangle \right) \left( \prod_{q=1}^p \langle \bu_{\pi(k)}^{(q)}, \hat{\bu}_k^{(q)} \rangle \right) \langle\bu_{l}^{(1)},\hat{\calP}_\perp^{(1)}\bu_m^{(1)}\rangle\right|\\
&\le& C\left(\tau^2\delta_k^{p-1}\sqrt{\log r\over n}+\tau^2\delta_k^p\sqrt{r\over n}\right).
\end{eqnarray*}

Similar to before, the fourth term on the right hand side of \eqref{eq:inductupper0} can be bounded by
\begin{eqnarray*}
	&&\left|\sum_{l_1,l_2\neq \pi(k)} \sigma_{l_1} \sigma_{l_2} \left( \frac{1}{n} \sum_{i=1}^n \theta_{il_1} \theta_{il_2} \right) \left( \prod_{q=2}^p \langle \bu_{l_1}^{(q)}, \hat{\bu}_k^{(q)} \rangle \right) \left( \prod_{q=1}^p \langle \bu_{l_2}^{(q)}, \hat{\bu}_k^{(q)} \rangle \right) \langle\bu_{l_1}^{(1)},\hat{\calP}_\perp^{(1)}\bu_m^{(1)}\rangle\right|\\
	&\le&\|\hat{\Sigma}_\theta\|\left(\sum_{l\neq \pi(k)}\sigma_l^2\langle\bu_{l}^{(1)},\hat{\calP}_\perp^{(1)}\bu_m^{(1)}\rangle^2 \prod_{q=2}^p \langle \bu_{l}^{(q)}, \hat{\bu}_k^{(q)} \rangle^2\right)^{1/2}\left(\sum_{l\neq \pi(k)}\sigma_l^2 \prod_{q=1}^p \langle \bu_{l}^{(q)}, \hat{\bu}_k^{(q)} \rangle^2\right)^{1/2}\\
	&\le&C\tau^2\delta_k^{2(p-1)};
\end{eqnarray*}
and the fifth term by
\begin{eqnarray*}
&&\left|\frac{1}{n} \sum_{i=1}^n\left[\sum_{l\neq \pi(k)} \sigma_l \theta_{il} \left( \prod_{q=2}^p \langle \bu_l^{(q)}, \hat{\bu}_k^{(q)} \rangle \right) \scrE_i (\hat{\bu}_k^{(1)},\dots,\hat{\bu}_k^{(p)} ) \langle\bu_l^{(1)},\hat{\calP}_\perp^{(1)}\bu_m^{(1)}\rangle\right]\right|\\
&=&\left|\sum_{l\neq\pi(k)} \left\{\sigma_l\left( \prod_{q=2}^p \langle \bu_l^{(q)}, \hat{\bu}_k^{(q)} \rangle \right)\langle\bu_l^{(1)},\hat{\calP}_\perp^{(1)}\bu_m^{(1)}\rangle \left[{1\over n}\sum_{i=1}^n\theta_{il}\scrE_i (\hat{\bu}_k^{(1)},\dots,\hat{\bu}_k^{(p)} )\right]\right\}\right|\\
&\le&\|\hat{\Sigma}_{\theta,\scrE}\|\cdot\left(\sum_{l\neq\pi(k)} \sigma_l^2\langle\bu_l^{(1)},\hat{\calP}_\perp^{(1)}\bu_m^{(1)}\rangle^2\prod_{q=2}^p \langle \bu_l^{(q)}, \hat{\bu}_k^{(q)} \rangle^2\right)^{1/2}\\
&\le&C\sigma_0\tau\delta_k^{p-1}\sqrt{d\over n}.
\end{eqnarray*}

We now turn to the sixth term on the right hand side of \eqref{eq:inductupper0}. Write
\begin{eqnarray*}
&&\frac{1}{n} \sum_{i=1}^n\left[\sum_{l=1}^r \scrE_i (\hat{\calP}_\perp^{(1)}\bu_m^{(1)},\hat{\bu}_k^{(2)},\dots,\hat{\bu}_k^{(p)} ) \sigma_l \theta_{il} \left( \prod_{q=1}^p \langle \bu_l^{(q)}, \hat{\bu}_k^{(q)} \rangle \right)\right]\\
&\le&\frac{1}{n} \sum_{i=1}^n\left[\sum_{l\neq \pi(k)} \scrE_i (\hat{\calP}_\perp^{(1)}\bu_m^{(1)},\hat{\bu}_k^{(2)},\dots,\hat{\bu}_k^{(p)} ) \sigma_l \theta_{il} \left( \prod_{q=1}^p \langle \bu_l^{(q)}, \hat{\bu}_k^{(q)} \rangle \right)\right]\\
&&+\frac{1}{n} \sum_{i=1}^n\left[\scrE_i (\hat{\calP}_\perp^{(1)}\bu_m^{(1)},\hat{\bu}_k^{(2)},\dots,\hat{\bu}_k^{(p)} ) \sigma_{\pi(k)} \theta_{i\pi(k)} \left( \prod_{q=1}^p \langle \bu_{\pi(k)}^{(q)}, \hat{\bu}_k^{(q)} \rangle \right)\right].%\\
%	&\le&\left(\tau+\tau\delta_k^{p-1}+\max_{l\in \pi([k-1])}\{\sigma_l\delta_l^{p-1}\}\right)\|\hat{\Sigma}_{\theta,\scrE}\|,
\end{eqnarray*}
The first term again can be bounded by
\begin{eqnarray*}
&&\left|\frac{1}{n} \sum_{i=1}^n\left[\sum_{l\neq \pi(k)} \scrE_i (\hat{\calP}_\perp^{(1)}\bu_m^{(1)},\hat{\bu}_k^{(2)},\dots,\hat{\bu}_k^{(p)} ) \sigma_l \theta_{il} \left( \prod_{q=1}^p \langle \bu_l^{(q)}, \hat{\bu}_k^{(q)} \rangle \right)\right]\right|\\
&\le&\|\hat{\Sigma}_{\theta,\scrE}\|\left(\sum_{l\neq \pi(k)}\sigma_l^2\prod_{q=1}^p \langle \bu_l^{(q)}, \hat{\bu}_k^{(q)} \rangle^2\right)^{1/2}\le C\sigma_0\tau\delta_k^{p-1}\sqrt{d\over n}.
\end{eqnarray*}
For the second term, note that
\begin{eqnarray}\nonumber
&&\left|\frac{1}{n} \sum_{i=1}^n\left[\scrE_i (\hat{\calP}_\perp^{(1)}\bu_m^{(1)},\hat{\bu}_k^{(2)},\dots,\hat{\bu}_k^{(p)} ) \sigma_{\pi(k)} \theta_{i\pi(k)} \left( \prod_{q=1}^p \langle \bu_{\pi(k)}^{(q)}, \hat{\bu}_k^{(q)} \rangle \right)\right]\right|\\\nonumber
&\le&\tau\left|\frac{1}{n} \sum_{i=1}^n\left[\scrE_i (\hat{\calP}_\perp^{(1)}\bu_m^{(1)},\hat{\bu}_k^{(2)},\dots,\hat{\bu}_k^{(p)} ) \theta_{i\pi(k)} \right]\right|\\
&\le&\tau\left|\frac{1}{n} \sum_{i=1}^n\left[\scrE_i (\bu_m^{(1)},\hat{\bu}_k^{(2)},\dots,\hat{\bu}_k^{(p)} ) \theta_{i\pi(k)} \right]\right|+\tau\left|\frac{1}{n} \sum_{i=1}^n\left[\scrE_i (\hat{\calP}^{(1)}\bu_m^{(1)},\hat{\bu}_k^{(2)},\dots,\hat{\bu}_k^{(p)} ) \theta_{i\pi(k)} \right]\right|,\label{eq:bd62}
\end{eqnarray}
where the second inequality follows from triangular inequality. As before,
$$
\left|\frac{1}{n} \sum_{i=1}^n\left[\scrE_i (\hat{\calP}^{(1)}\bu_m^{(1)},\hat{\bu}_k^{(2)},\dots,\hat{\bu}_k^{(p)} ) \theta_{i\pi(k)} \right]\right|\le \|\hat{\Sigma}_{\theta,\scrE}\|\|\hat{\calP}^{(1)}\bu_m^{(1)}\|\le C\sigma_0\delta_k\sqrt{d\over n}.
$$
To bound the first term on the rightmost hand side of \eqref{eq:bd62}, write
$$
\hat{\bu}_k^{(q)}=\alpha_q\bu_{\pi(k)}+\bv^{(q)}
$$
where $\alpha_q=\langle\bu_{\pi(k)},\hat{\bu}_k\notin \{\pi(k),m\}\notin \{\pi(k),m\}\notin \{\pi(k),m\}^{(q)}\rangle$ and $\|\bv^{(q)}\|=\sqrt{1-\alpha_q^2}\le \delta_k$.
Then
\begin{eqnarray*}
&&\scrE_i (\bu_m^{(1)},\hat{\bu}_k^{(2)},\dots,\hat{\bu}_k^{(p)})\\
&=&\alpha_p\scrE_i (\bu_m^{(1)},\hat{\bu}_k^{(2)},\dots,\hat{\bu}_k^{(p-1)},\bu_{\pi(k)}^{(p)})+\scrE_i (\bu_m^{(1)},\hat{\bu}_k^{(2)},\dots,\hat{\bu}_k^{(p-1)},\bv^{(p)})\\
&=&\alpha_p\alpha_{p-1}\scrE_i (\bu_m^{(1)},\hat{\bu}_k^{(2)},\dots,\hat{\bu}_k^{(p-2)},\bu_{\pi(k)}^{(p-1)},\bu_{\pi(k)}^{(p)})\\
&&+\alpha_p\scrE_i (\bu_m^{(1)},\hat{\bu}_k^{(2)},\dots,\hat{\bu}_k^{(p-2)},\bv^{(p-1)},\bu_{\pi(k)}^{(p)})\\
&&+\scrE_i (\bu_m^{(1)},\hat{\bu}_k^{(2)},\dots,\hat{\bu}_k^{(p-1)},\bv^{(p)})\\
&=&\cdots\cdots\\
&=&\left(\prod_{q=2}^p\alpha_q\right)\scrE_i (\bu_m^{(1)},\bu_{\pi(k)}^{(2)},\dots,\bu_{\pi(k)}^{(p)})\\
&&+\cdots+\scrE_i (\bu_m^{(1)},\hat{\bu}_k^{(2)},\dots,\hat{\bu}_k^{(p-1)},\bv^{(p)}).
\end{eqnarray*}
Therefore,
\begin{eqnarray*}
&&\left|\frac{1}{n} \sum_{i=1}^n\left[\scrE_i (\bu_m^{(1)},\hat{\bu}_k^{(2)},\dots,\hat{\bu}_k^{(p)} ) \theta_{i\pi(k)} \right]\right|\\
&\le&\left|{1\over n}\sum_{i=1}^n\left[\scrE_i (\bu_m^{(1)},\bu_{\pi(k)}^{(2)},\dots,\bu_{\pi(k)}^{(p)} )\theta_{i\pi(k)} \right]\right|+\cdots+\left|\frac{1}{n} \sum_{i=1}^n\left[\scrE_i (\bu_m^{(1)},\hat{\bu}_k^{(2)},\dots,\hat{\bu}_k^{(p-1)},\bv^{(p)})\theta_{i\pi(k)} \right]\right|\\
&\le&\left|{1\over n}\sum_{i=1}^n\left[\scrE_i (\bu_m^{(1)},\bu_{\pi(k)}^{(2)},\dots,\bu_{\pi(k)}^{(p)} )\theta_{i\pi(k)} \right]\right|+C_p\delta_k\|\hat{\Sigma}_{\theta,\scrE}\|\\
&\le&C\left(\sigma_0\sqrt{\log r\over n}+\sigma_0\delta_k\sqrt{d\over n}\right),
\end{eqnarray*}
so that the six term on the rightmost hand side of \eqref{eq:inductupper0} can be upper bounded by
$$
\left|\frac{1}{n} \sum_{i=1}^n\left[\sum_{l=1}^r \scrE_i (\hat{\calP}_\perp^{(1)}\bu_m^{(1)},\hat{\bu}_k^{(2)},\dots,\hat{\bu}_k^{(p)} ) \sigma_l \theta_{il} \left( \prod_{q=1}^p \langle \bu_l^{(q)}, \hat{\bu}_k^{(q)} \rangle \right)\right]\right|\le C\left(\sigma_0\tau\sqrt{\log r\over n}+\sigma_0\tau\delta_k\sqrt{d\over n}\right).
$$

Finally consider the seventh term:
\begin{eqnarray*}
&&\left|\frac{1}{n} \sum_{i=1}^n  \scrE_i (\hat{\calP}_\perp^{(1)}\bu_m^{(1)},\hat{\bu}_k^{(2)},\dots,\hat{\bu}_k^{(p)} )  \scrE_i (\hat{\bu}_k^{(1)},\dots,\hat{\bu}_k^{(p)} )-\langle \hat{\calP}_\perp^{(1)}\bu_m^{(1)}, \hat{\bu}_k^{(1)}\rangle\right|\\%\le \|\hat{\Sigma}_{\scrE}-I\|.
&\le&\left|\frac{1}{n} \sum_{i=1}^n  \scrE_i (\bu_m^{(1)},\hat{\bu}_k^{(2)},\dots,\hat{\bu}_k^{(p)} )  \scrE_i (\hat{\bu}_k^{(1)},\dots,\hat{\bu}_k^{(p)} )-\langle \bu_m^{(1)}, \hat{\bu}_k^{(1)}\rangle\right|\\
&&+\left|\frac{1}{n} \sum_{i=1}^n  \scrE_i (\hat{\calP}^{(1)}\bu_m^{(1)},\hat{\bu}_k^{(2)},\dots,\hat{\bu}_k^{(p)} )  \scrE_i (\hat{\bu}_k^{(1)},\dots,\hat{\bu}_k^{(p)} )-\langle \hat{\calP}^{(1)}\bu_m^{(1)}, \hat{\bu}_k^{(1)}\rangle\right|.
\end{eqnarray*}
Similar to before, the second term can be bounded by
\begin{eqnarray*}
&&\left|\frac{1}{n} \sum_{i=1}^n  \scrE_i (\hat{\calP}^{(1)}\bu_m^{(1)},\hat{\bu}_k^{(2)},\dots,\hat{\bu}_k^{(p)} )  \scrE_i (\hat{\bu}_k^{(1)},\dots,\hat{\bu}_k^{(p)} )-\sigma_0^2\langle \hat{\calP}^{(1)}\bu_m^{(1)}, \hat{\bu}_k^{(1)}\rangle\right|\\
&\le&\|\hat{\calP}^{(1)}\bu_m^{(1)}\|\|\hat{\Sigma}_{\scrE}-\Sigma_{\scrE}\|\le C\sigma_0^2\delta_k\sqrt{d\over n};
\end{eqnarray*}
the first term by
\begin{eqnarray*}
&&\left|\frac{1}{n} \sum_{i=1}^n  \scrE_i (\bu_m^{(1)},\hat{\bu}_k^{(2)},\dots,\hat{\bu}_k^{(p)} )  \scrE_i (\hat{\bu}_k^{(1)},\dots,\hat{\bu}_k^{(p)} )-\sigma_0^2\langle \bu_m^{(1)}, \hat{\bu}_k^{(1)}\rangle\right|\\
&\le&\left|\frac{1}{n} \sum_{i=1}^n  \scrE_i (\bu_m^{(1)},\hat{\bu}_k^{(2)},\dots,\hat{\bu}_k^{(p)} )  \scrE_i (\bv^{(1)},\dots,\hat{\bu}_k^{(p)} )-\sigma_0^2\langle \bu_m^{(1)}, \bv^{(1)}\rangle\right|\\
&&+\left|\frac{1}{n} \sum_{i=1}^n  \scrE_i (\bu_m^{(1)},\hat{\bu}_k^{(2)},\dots,\hat{\bu}_k^{(p)} )  \scrE_i (\bu_{\pi(k)}^{(1)},\dots,\hat{\bu}_k^{(p)} )\right|\\
&\le&C\delta_k\|\hat{\Sigma}_{\scrE}-\Sigma_{\scrE}\|+\left|\frac{1}{n} \sum_{i=1}^n  \scrE_i (\bu_m^{(1)},\bu_{\pi(k)}^{(2)},\cdots,\bu_{\pi(k)}^{(p)} )  \scrE_i (\bu_{\pi(k)}^{(1)},\cdots,\bu_{\pi(k)}^{(p)} )\right|\\
&&+\cdots+\left|\frac{1}{n} \sum_{i=1}^n  \scrE_i (\bu_m^{(1)},\hat{\bu}_k^{(2)},\dots,\hat{\bu}_k^{(p)} )  \scrE_i (\bu_{\pi(k)}^{(1)},\hat{\bu}_k^{(2)},\dots,\hat{\bu}_k^{(p-1)},\bv^{(p)} )\right|\\
&\le&C\left(\sigma_0^2\delta_k\sqrt{d\over n}+\sigma_0^2\sqrt{\log r\over n}\right).
\end{eqnarray*}

Putting all seven upper bounds together, we have
$$
\max_{l\notin \pi([k])}\tilde{\Sigma}(\hat{\calP}_\perp^{(1)}\bu_m^{(1)},\hat{\bu}_k^{(2)},\ldots,\hat{\bu}_k^{(p)}, \hat{\bu}_k^{(1)},\ldots, \hat{\bu}_k^{(p)})\le C\left(\tau^2\delta_k^2+(\tau^2\delta_k^{p-1}+\sigma_0\tau+\sigma_0^2)\sqrt{\log r\over n}\right).
$$

\subsection{Finishing Up}
We first verify \eqref{eq:rate_main2}. Note that
\begin{eqnarray*}
	&&(\Sigma-\Sigma_{\scrE})(\hat{\bu}_k^{(1)},\hat{\bu}_k^{(2)},\ldots,\hat{\bu}_k^{(p)}, \hat{\bu}_k^{(1)},\ldots, \hat{\bu}_k^{(p)})\\
	&=&\sum_{l=1}^r\left(\sigma_l^2\prod_{q=1}^p \langle\bu_l^{(q)},\hat{\bu}_k^{(q)}\rangle^2\right)\\
	&\le& \max_{1\le l \le r} \left\{ \sigma_l^2 \left| \prod_{q=1}^p \langle \bu_l^{(q)}, \hat{\bu}_k^{(q)} \rangle \right| \right\} \cdot \left(\sum_{l=1}^r \left| \prod_{q=1}^p \langle \bu_l^{(q)}, \hat{\bu}_k^{(q)} \rangle \right|\right) \\
	&\le&\max_{1\le l \le r} \left\{ \sigma_l^2 \left| \prod_{q=1}^p \langle \bu_l^{(q)}, \hat{\bu}_k^{(q)} \rangle \right| \right\}.
\end{eqnarray*}
where the last inequality follows from Cauchy-Schwartz inequality. Therefore, by definition,
\begin{equation}\label{eq:taubd}
\sigma_{\pi(k)}^2\ge (\Sigma-\Sigma_{\scrE})(\hat{\bu}_k^{(1)},\hat{\bu}_k^{(2)},\ldots,\hat{\bu}_k^{(p)}, \hat{\bu}_k^{(1)},\ldots, \hat{\bu}_k^{(p)})
\end{equation}

On the other hand,
\begin{eqnarray*}
	&&(\Sigma-\Sigma_{\scrE})(\hat{\bu}_k^{(1)},\hat{\bu}_k^{(2)},\ldots,\hat{\bu}_k^{(p)}, \hat{\bu}_k^{(1)},\ldots, \hat{\bu}_k^{(p)})\\
	&\ge&\tilde{\Sigma}(\hat{\bu}_k^{(1)},\hat{\bu}_k^{(2)},\ldots,\hat{\bu}_k^{(p)}, \hat{\bu}_k^{(1)},\ldots, \hat{\bu}_k^{(p)})\\
	&&-\sup_{\|\bw^{(q)}\|\le 1,1\le q\le p}\left|(\hat{\Sigma}-\Sigma)(\hat{\calP}_\perp^{(1)}\bw^{(1)},\dots,\hat{\calP}_\perp^{(p)}\bw^{(p)},\hat{\calP}_\perp^{(1)}\bw^{(1)},\dots,\hat{\calP}_\perp^{(p)}\bw^{(p)})\right|\\
	&\ge&\tilde{\Sigma}(\hat{\calP}_\perp^{(1)}\bu_{l_\ast}^{(1)},\ldots,\hat{\calP}_\perp^{(p)}\bu_{l_\ast}^{(p)},\hat{\calP}_\perp^{(1)}\bu_{l_\ast}^{(1)},\ldots, \hat{\calP}_\perp^{(p)}\bu_{l_\ast}^{(p)})\\
	&&-\sup_{\|\bw^{(q)}\|\le 1,1\le q\le p}\left|(\hat{\Sigma}-\Sigma)(\hat{\calP}_\perp^{(1)}\bw^{(1)},\dots,\hat{\calP}_\perp^{(p)}\bw^{(p)},\hat{\calP}_\perp^{(1)}\bw^{(1)},\dots,\hat{\calP}_\perp^{(p)}\bw^{(p)})\right|\\
	&\ge&(\Sigma-\Sigma_{\scrE})(\hat{\calP}_\perp^{(1)}\bu_{l_\ast}^{(1)},\ldots,\hat{\calP}_\perp^{(p)}\bu_{l_\ast}^{(p)},\hat{\calP}_\perp^{(1)}\bu_{l_\ast}^{(1)},\ldots, \hat{\calP}_\perp^{(p)}\bu_{l_\ast}^{(p)})\\
	&&-2\sup_{\|\bw^{(q)}\|\le 1,1\le q\le p}\left|(\hat{\Sigma}-\Sigma)(\hat{\calP}_\perp^{(1)}\bw^{(1)},\dots,\hat{\calP}_\perp^{(p)}\bw^{(p)},\hat{\calP}_\perp^{(1)}\bw^{(1)},\dots,\hat{\calP}_\perp^{(p)}\bw^{(p)})\right|
\end{eqnarray*}
where $l_\ast$ is the index such that $\sigma_{l_\ast}^2=\tau^2$. Following the same derivation as before, we have
\begin{eqnarray*}
(\Sigma-\Sigma_{\scrE})(\hat{\bu}_k^{(1)},\hat{\bu}_k^{(2)},\ldots,\hat{\bu}_k^{(p)}, \hat{\bu}_k^{(1)},\ldots, \hat{\bu}_k^{(p)})\ge \tau^2\left(1-C\sqrt{r\over n}-C\delta_k\right).
\end{eqnarray*}
Together with \eqref{eq:taubd}, we get
$$
\sigma_{\pi(k)}^2\ge\tau^2\left(1-C\sqrt{r\over n}-C\delta_k\right).
$$

Combing the lower bound for $
\tilde{\Sigma}(\hat{\bu}_k^{(1)},\hat{\bu}_k^{(2)},\ldots,\hat{\bu}_k^{(p)}, \hat{\bu}_k^{(1)},\ldots, \hat{\bu}_k^{(p)})
$ and upper bound for $\tilde{\Sigma}(\hat{\calP}_\perp^{(1)}(\hat{\bu}_k^{(1)}-\langle\hat{\bu}_k^{(1)},\bu_{\pi(k)}^{(1)}\rangle\bu_{\pi(k)}^{(1)}),\hat{\bu}_k^{(2)},\ldots,\hat{\bu}_k^{(p)}, \hat{\bu}_k^{(1)},\ldots, \hat{\bu}_k^{(p)})$, we have
\begin{eqnarray*}
\eta_k&\le& {C\tau^2\delta_k^2+C(\sigma_0^2+\sigma_0\tau)\sqrt{d\over n}\over \tau^2\left(1-C\delta_k^4-C\sqrt{r\over n}\right)-C(\sigma_0\tau+\sigma_0^2)\sqrt{d\over n}-C\sigma_0^2\delta_k^2}\\
&\le&C\delta_k^2+C\left({\sigma_0\over \tau}+{\sigma_0^2\over \tau^2}\right)\sqrt{d\over n}\\
&\le&C\left({\sigma_0\over \sigma_{\pi(k)}}+{\sigma_0^2\over \sigma_{\pi(k)}^2}\right)\sqrt{d\over n}=\delta_k.
\end{eqnarray*}

Similarly, Combing the lower bound for $
\tilde{\Sigma}(\hat{\bu}_k^{(1)},\hat{\bu}_k^{(2)},\ldots,\hat{\bu}_k^{(p)}, \hat{\bu}_k^{(1)},\ldots, \hat{\bu}_k^{(p)})
$ and upper bound for $\max_{l\notin \pi([k])}\tilde{\Sigma}(\hat{\calP}_\perp^{(1)}\bu_m^{(1)},\hat{\bu}_k^{(2)},\ldots,\hat{\bu}_k^{(p)}, \hat{\bu}_k^{(1)},\ldots, \hat{\bu}_k^{(p)})$, we get
\begin{eqnarray*}
\tilde{\eta}_k&\le& {C\left(\tau^2\delta_k^2+(\tau^2\delta_k^{p-1}+\sigma_0\tau+\sigma_0^2)\sqrt{\log r\over n}\right)\over \tau^2\left(1-C\delta_k^4-C\sqrt{r\over n}\right)-C(\sigma_0\tau+\sigma_0^2)\sqrt{d\over n}-C\sigma_0^2\delta_k^2}\\
&\le&C\delta_k^2+C\left(\delta_k+{\sigma_0\over \tau}+{\sigma_0^2\over \tau^2}\right)\sqrt{\log r\over n}\\
&\le& C\delta_k^2+C\left({\sigma_0\over \sigma_{\pi(k)}}+{\sigma_0^2\over \sigma_{\pi(k)}^2}\right)\sqrt{\log r\over n}=\tilde{\delta}_k.
\end{eqnarray*}
 
\section{Proof of Theorems \ref{thm:low_dim} and \ref{thm:linear_form}}

Note that Theorem \ref{thm:low_dim} can be viewed as special case of Theorem \ref{thm:linear_form} and it suffices to prove Theorem \ref{thm:linear_form}. As before, we only need to consider the case when $q=1$. Write
$$
\bw=\langle \bw,\bu_{\pi(k)}^{(1)}\rangle\bu_{\pi(k)}^{(1)}+\tilde{\bw}.
$$
Then
$$
\langle\tilde{\bu}_k^{(1)},\bw\rangle-\langle \bu_{\pi(k)}^{(1)}, \bw\rangle=\langle \bw,\bu_{\pi(k)}^{(1)}\rangle\left(\langle\tilde{\bu}_k^{(1)},\bu_{\pi(k)}^{(1)}\rangle-1\right)+\langle\tilde{\bu}_k^{(1)},\tilde{\bw}\rangle.
$$

Under the assumption $d = o(n^{1/2})$, by Lemma \ref{le:probbd} and Theorem \ref{thm_rate}, it is not hard to see that
\begin{align}
\sin \angle (\tilde{\bu}_k^{(1)}, \bu_{\pi(k)}^{(1)} ) = O_p\left( \sqrt{\frac{d}{n}} \right), \label{eq:low_start}
\end{align}
so
$$
1-\langle \tilde{\bu}_k^{(1)},\bu_{\pi(k)}^{(1)}\rangle\le 1-\langle \tilde{\bu}_k^{(1)},\bu_{\pi(k)}^{(1)}\rangle^2=\sin^2 \angle (\tilde{\bu}_k^{(1)}, \bu_{\pi(k)}^{(1)} )=O_p\left( \frac{d}{n} \right) = o_p(n^{-1/2}).
$$
Therefore it suffices to prove that% the asymptotic normality of
$$
\sqrt{n}\left(\langle\tilde{\bu}_k^{(1)},\tilde{\bw}\rangle-\langle\bu_{\pi(k)}^{(1)},\tilde{\bw}\rangle\right)\to_d N\left(0,\|\tilde{\bw}\|^2\left({\sigma_0^2\over \sigma_{\pi(k)}^2}+{\sigma_0^4\over \sigma_{\pi(k)}^4}\right)\right).
$$

Recall that $\tilde{\bu}_k^{(1)}$ is the leading eigenvector of 
$$
\hat{\Sigma}(\cdot,\hat{\bu}_k^{(2)},\ldots,\hat{\bu}_k^{(p)}, \cdot ,\hat{\bu}_k^{(2)},\ldots, \hat{\bu}_k^{(p)}),
$$
which is the same as the leading eigenvector of
$$
\tilde{\Sigma}(\cdot,\hat{\bu}_k^{(2)},\ldots,\hat{\bu}_k^{(p)}, \cdot ,\hat{\bu}_k^{(2)},\ldots, \hat{\bu}_k^{(p)}),
$$
so
\begin{align*}
\tilde{\bu}_k^{(1)} = \frac{ \tilde{\Sigma}(\cdot,\hat{\bu}_k^{(2)},\ldots,\hat{\bu}_k^{(p)}, \tilde{\bu}_k^{(1)},\hat{\bu}_k^{(2)},\ldots, \hat{\bu}_k^{(p)}) }{ \tilde{\Sigma}(\tilde{\bu}_k^{(1)},\hat{\bu}_k^{(2)},\ldots,\hat{\bu}_k^{(p)}, \tilde{\bu}_k^{(1)},\hat{\bu}_k^{(2)},\ldots, \hat{\bu}_k^{(p)}) },
\end{align*}
which implies
\begin{align*}
\langle \tilde{\bu}_k^{(1)}, \tilde{\bw} \rangle
= \frac{ \tilde{\Sigma}(\tilde{\bw},\hat{\bu}_k^{(2)},\ldots,\hat{\bu}_k^{(p)}, \tilde{\bu}_k^{(1)},\hat{\bu}_k^{(2)},\ldots, \hat{\bu}_k^{(p)}) }{ \tilde{\Sigma}(\tilde{\bu}_k^{(1)},\hat{\bu}_k^{(2)},\ldots,\hat{\bu}_k^{(p)}, \tilde{\bu}_k^{(1)},\hat{\bu}_k^{(2)},\ldots, \hat{\bu}_k^{(p)}) }.
\end{align*}

We start with the nominator. Following an identical argument as that for \eqref{eq:inductupper0} in Subsection \ref{subsec:upper2}, we have
\begin{align*}
& \tilde{\Sigma}(\tilde{\bw},\hat{\bu}_k^{(2)},\ldots,\hat{\bu}_k^{(p)}, \tilde{\bu}_k^{(1)},\hat{\bu}_k^{(2)},\ldots, \hat{\bu}_k^{(p)})-\tilde{\Sigma}(\tilde{\bw},\bu_{\pi(k)}^{(2)},\dots,\bu_{\pi(k)}^{(p)},\bu_{\pi(k)}^{(1)},\dots,\bu_{\pi(k)}^{(p)}) \\
\le \ & C\left(\tau^2\delta_k^2+\tau^2\delta^{p+1}\sqrt{r\over n}+\tau^2\delta^{p-1}\sqrt{r\over n}+\tau^2\delta_k^{2(p-1)}+\sigma_0\tau\delta_k\sqrt{d\over n}\right)=o(n^{-1/2})
\end{align*}
with probability tending to one.
So
\begin{align}
& \tilde{\Sigma}(\tilde{\bw},\hat{\bu}_k^{(2)},\ldots,\hat{\bu}_k^{(p)}, \tilde{\bu}_k^{(1)},\hat{\bu}_k^{(2)},\ldots, \hat{\bu}_k^{(p)}) \nonumber \\
=&\tilde{\Sigma}(\tilde{\bw},\bu_{\pi(k)}^{(2)},\dots,\bu_{\pi(k)}^{(p)},\bu_{\pi(k)}^{(1)},\dots,\bu_{\pi(k)}^{(p)}) \nonumber + o_p(n^{-1/2}) \\ 
=& \sigma_{\pi(k)} \frac{1}{n} \sum_{i=1}^n \theta_{\pi(k) i} \scrE_i (\tilde{\bw},\bu_{\pi(k)}^{(2)},\dots,\bu_{\pi(k)}^{(p)}) \nonumber \\
& + \frac{1}{n} \sum_{i=1}^n  \scrE_i (\bu_{\pi(k)}^{(1)},\dots,\bu_{\pi(k)}^{(p)}) \scrE_i (\tilde{\bw},\bu_{\pi(k)}^{(2)},\dots,\bu_{\pi(k)}^{(p)}) - \bu_{\pi(k)}^{(1)} + o_p(n^{-1/2}) \nonumber\\
\to_d & N(0,(\sigma_0^4+\sigma_0^2\sigma_{\pi(k)}^2)\|\tilde{\bw}\|^2). \nonumber
\end{align}
On the other hand, similar to the previous section,
$$
\tilde{\Sigma}(\tilde{\bu}_k^{(1)},\hat{\bu}_k^{(2)},\ldots,\hat{\bu}_k^{(p)}, \tilde{\bu}_k^{(1)},\hat{\bu}_k^{(2)},\ldots, \hat{\bu}_k^{(p)}) = \sigma_{\pi(k)}^2(1+o_p(1)).
$$
Theorem \ref{thm:linear_form} then follows from Slutsky's Theorem.

The claim in Theorem \ref{thm:low_dim} that 
\begin{eqnarray*}
&\sqrt{n} \left[{\rm vec}(\tilde{\bU}^{(q)})-{\rm vec}(\bU^{(q)}_\pi)\right]\\
&\overset{d}{\to} N \left(0,   {\rm diag}\left(\left(\frac{\sigma_0^2}{\sigma_{\pi(1)}^2} + 
\frac{\sigma_0^4}{\sigma_{\pi(1)}^4}\right)\calP_{\bu_{\pi(1)}^{(q)}}^\perp,\ldots, \left(\frac{\sigma_0^2}{\sigma_{\pi(r)}^2} + 
\frac{\sigma_0^4}{\sigma_{\pi(r)}^4}\right)\calP_{\bu_{\pi(r)}^{(q)}}^\perp\right)\right)
\end{eqnarray*}
follows from Theorem \ref{thm:linear_form} and the fact that
$$
\sigma_{\pi(k)} \frac{1}{n} \sum_{i=1}^n \theta_{\pi(k) i} \scrE_i (\cdot,\bu_{\pi(k)}^{(2)},\dots,\bu_{\pi(k)}^{(p)}) + \frac{1}{n} \sum_{i=1}^n  \scrE_i (\bu_{\pi(k)}^{(1)},\dots,\bu_{\pi(k)}^{(p)}) \scrE_i (\cdot,\bu_{\pi(k)}^{(2)},\dots,\bu_{\pi(k)}^{(p)}) - \bu_{\pi(k)}^{(1)}
$$
are independent for any $k_1 \neq k_2$.

\section{Proof of Theorems \ref{thm:bias_simple} and \ref{thm_b}}  

Theorem \ref{thm:bias_simple} is a special case of Theorem \ref{thm_b} and it suffices to prove the latter. We first need to introduce a number of notations. Denote
\begin{align}
	\bar{\mathbb{P}}(\cdot):=\mathbb{P}(\cdot|\scrX_{n/2+1},\scrX_{n/2+2},\dots,\scrX_{n}), \label{eq:def_barP}
\end{align} 
and 
\begin{align}
	\bar{\mathbb{E}}(\cdot):=\mathbb{E}(\cdot|\scrX_{n/2+1},\scrX_{n/2+2},\dots,\scrX_{n}), \label{eq:def_barE}
\end{align} 
the conditional probability and expectation given $\{ \scrX_{n/2+1},\scrX_{n/2+2},\dots,\scrX_{n} \},$ respectively. 

Write
$$\hat{\calP}_j^{(q)[1]}:= \check{\bu}_j^{(q)[1]} \otimes \check{\bu}_j^{(q)[1]},\qquad \calP_j^{(q)}=\bu_j^{(q)} \otimes \bu_j^{(q)}=\calP_{\bu_j^{(q)}},$$
$$\calC_j^{(q)}:= \frac{1}{\sigma_j^2} \left( I_{d_q} - \bu_j^{(q)} \otimes \bu_j^{(q)} \right ),$$
\begin{align*}
	z_{jk} := \scrX_k \times_1 \hat{\bu}_j^{(1)[2]} \dots \times_{q-1} \hat{\bu}_j^{(q-1)[2]} \times_{q+1} \hat{\bu}_j^{(q+1)[2]} \dots \times_p \hat{\bu}_j^{(p)[2]}, \qquad k =1,\dots,n/2,
\end{align*}
and
\begin{align*}
	M_j^{(q)[1]}:=\sum_{l=1}^r \sigma_l^2 \left( \prod_{q' \neq q} \langle \bu_l^{(q')}, \hat{\bu}_j^{(q')[2]} \rangle \right)^2 \bu_l^{(1)} \otimes \bu_l^{(1)} + I_{d_q}= \sum_{l=1}^r \tilde{\sigma}_l^2 \bu_l^{(1)} \otimes \bu_l^{(1)} + I_{d_q},
\end{align*}
where
\begin{align*}
	\tilde{\sigma}_l^2=\sigma_l^2 \left( \prod_{q' \neq q} \langle \bu_l^{(q')}, \hat{\bu}_j^{(q')[2]} \rangle \right)^2.
\end{align*}
Furthermore, let
\begin{align*}
	\calC_j^{(q)[1]}=\sum_{l \neq j} \frac{1}{\tilde{\sigma}_j^2 - \tilde{\sigma}_l^2} \bu_l^{(q)} \otimes \bu_l^{(q)} + \frac{1}{\tilde{\sigma}_j^2} \left( I_{d_q} - \sum_{l=1}^r \bu_l^{(q)} \otimes \bu_l^{(q)} \right),
\end{align*}
\begin{align*}
	\calL_j^{(q)[1]}= \frac{2}{n} \sum_{k=1}^{n/2} \left( \calC^{(q)[1]} z_{jk} \otimes \calP_j^{(q)} z_{jk} + \calP_j^{(q)} z_{jk} \otimes \calC^{(q)[1]} z_{jk} \right) ,
\end{align*}
\begin{align*}
	\calS_j^{(q)[1]}= \hat{\calP}_j^{(q)[1]} - \calP_j^{(q)} - \calL_j^{(q)[1]}.
\end{align*}
We use calligraphic capital letters for $\calP_j^{(q)}$, $\hat{\calP}_j^{(q)[1]}$, $\calC_j^{(q)}$, $\calC_j^{(q)[1]}$, $\calL_j^{(q)[1]}$ and $\calS_j^{(q)[1]}$ to remind the readers that they are matrices with specific definitions.

Define
\begin{align}
	b_j^{(q)[1]} := \left\langle \bar{\mathbb{E}}(\calS_j^{(q)[1]}) \bu_j^{(q)}, \bu_j^{(q)} \right\rangle.  \label{bjq}
\end{align}
and $b_j^{(q)[2]}$ is similarly defined. Finally, we define
%
%Then, denote 
%\begin{align}
%	y_{jk}^{(q)} := \scrX_k \times_2 \bu_j^{(2)} \dots \times_{q-1} \bu_j^{(q-1)} \times_{q+1} \bu_j^{(q+1)} \dots \times_p \bu_j^{(p)}, k =1,\dots,n/2
%\end{align}
%
%Define random variables $b_j^{(q)[1]}$ and $b_j^{(q)[2]}$ as in Lemma \ref{linear_basic}, then define
\begin{align}
b_j^{(q)} = \frac{ \left\|\check{\bu}_j^{(q),[1]}+\check{\bu}_j^{(q),[2]}\right\| }{ \sqrt{1+b_j^{(q)[1]}}+\sqrt{1+b_j^{(q)[2]}} } -1.
\label{defbj1}.
\end{align}
The proof is rather involved and we shall break it into several steps.

\paragraph{Step 1.} We shall represent linear forms of 
$\check{\bu}_j^{(q),[1]}$ as bilinear forms of $\hat{\calP}_j^{(q)[1]}$, and prove that 
\begin{align}
\sqrt{n}\left((1+b_j^{(q)})\langle \check{\bu}^{(q)}_j, \bv\rangle-\langle \bu^{(q)}_{\pi(j)}, \bv\rangle\right)\to_d N\left(0,\left(\frac{\sigma_0^2}{\sigma_{\pi(j)}^2} + 
\frac{\sigma_0^4}{\sigma_{\pi(j)}^4}\right)\|\calP_{\bu_{\pi(j)}^{(q)}}^\perp\bv\|^2\right). \label{eq:bias_part1}
\end{align}
%The asymptotics of the bilinear forms are available in Lemma \ref{linear_basic}.

\paragraph{Step 2.} We prove that
\begin{align}
\langle \check{\bu}_j^{(q)}, \bu_{\pi(j)}^{(q)} \rangle 
= \frac{1}{\sqrt{ 1+ \frac{d_q}{n} \left(\frac{\sigma_0^2}{\sigma_{\pi(j)}^2} + \frac{\sigma_0^4}{\sigma_{\pi(j)}^4}\right) } } + O_p\left( \frac{d^{3/2}}{n^{3/2}} \right).
\label{eq:uu}
\end{align}
Notice that by letting $\bv = \bu^{(q)}_{\pi(j)}$, \eqref{eq:bias_part1} implies
\begin{align*}
(1+b_j^{(q)}) \langle \check{\bu}_j^{(q)}, \bu_{\pi(j)}^{(q)} \rangle - 1
= o_p(n^{-1/2}).
\end{align*}
Combine it with \eqref{eq:uu}, we immediately have 
\begin{align*}
b_j^{(q)}=\sqrt{ 1+ \frac{d_q}{n} \left(\frac{\sigma_0^2}{\sigma_{\pi(j)}^2} + \frac{\sigma_0^4}{\sigma_{\pi(j)}^4}\right) }-1 + O_p\left( \frac{d^{3/2}}{n^{3/2}} \right) + o_p\left( \frac{1}{\sqrt{n}} \right),
\end{align*}

\paragraph{Step 3.} Finally, we show that
\begin{align}
\sqrt{n} \left( \hat{b}_j^{(q)} - b_j^{(q)} \right) \overset{p} \to 0. \label{eq:bb}
\end{align}

For simplicity, in the rest of the proof we shall assume without loss of generality that the permutation $\pi$ that matches $\hat{\bu}_j^{(q)}$ to $\bu_{\pi(j)}^{(1)}$ is the identity. We shall also make repeated use of the following facts, oftentimes without explicit mentioning.

Similar to the proof of Theorem \ref{thm_rate}, write
$$
\hat{\Sigma}_\scrE^{[1]}={2\over n}\sum_{i=1}^{n/2} \scrE_i\otimes\scrE_i, \qquad
\hat{\Sigma}_{\theta,\scrE}^{[1]}={2\over n}\sum_{i=1}^{n/2} \theta_i\otimes \scrE_i.
$$
By Lemmas \ref{le:probbd} and \ref{le:probbd2}, with probability tending to one,
\begin{align}
\sigma_0^{-2}\|\hat{\Sigma}_\scrE^{[1]}-\scrI\| \le C\sqrt{d\over n} ,\ 
\sigma_0^{-1}\|\hat{\Sigma}_{\theta,\scrE}^{[1]}\|\le C\sqrt{d\over n}. \label{eq:SS}
\end{align}
Let
$$
	\Delta^{(q),[2]} = \hat{\bu}_j^{(q),[2]} - \bu_j^{(q)}.
$$
By Theorem \ref{thm_rate}, we have 
\begin{align*}
	\max_{ q \in [p] } \|\Delta^{(q),[2]}\| = O_p \left(\sqrt{\frac{d}{n}}\right) 
\end{align*}
Moreover, under the assumption $d = o(n)$, by Lemma \ref{le:probbd} and Theorem \ref{thm_rate}, it is not hard to see that
\begin{align*}
\sin \angle (\check{\bu}_j^{(1)[1]}, \bu_{j}^{(1)} ) = O_p\left( \sqrt{\frac{d}{n}} \right).
\end{align*}
Denote
$$
\delta^{(1)[2]}:= \max\left\{ \max_{ q \in [p] } \|\Delta^{(q),[2]}\|, \|\check{\bu}_j^{(1)[1]}-\bu_{j}^{(1)}\| \right\},
$$
combine the two bounds above, we have that under the assumptions for Theorem \ref{thm_b},
\begin{align}
	\delta^{(1)[2]} = O_p \left(\sqrt{\frac{d}{n}}\right). \label{eq:bias_fact}
\end{align}

\subsection{Step 1.}

Without loss of generality, for this step we assume $\sigma_0=1$. We only need to prove for the case $q=1$, so within this step, we shall also write $\calP_j = \calP_j^{(1)} = \calP_{\bu_j^{(1)}}$ and $\calC_j = \calC_j^{(1)}=\frac{1}{\sigma_j^2} \left( I-\calP_j\right)$ for simplicity.

Define 
\begin{align*}
\eta_{jk}(\bv) := \langle y_{jk},\calP_j \bu_j^{(1)} \rangle \langle y_{jk},\calC_j \bv \rangle ,
\end{align*}
where
\begin{align*}
y_{jk} := \scrX_k \times_2 \bu_j^{(2)} \dots \times_p \bu_j^{(p)} = \sigma_j \theta_{jk} \bu_j^{(1)} + \scrE_k \times_2 \bu_j^{(2)} \dots \times_p \bu_j^{(p)} .
\end{align*}
Recall that $\hat{\calP}_j^{(1)[1]}:= \check{\bu}_j^{(1),[1]} \otimes \check{\bu}_j^{(1),[1]}$ and $\hat{\calP}_j^{(1)[2]}:= \check{\bu}_j^{(1),[2]} \otimes \check{\bu}_j^{(1),[2]}$. Define
\begin{align}
\rho_j(\bv)^{[1]}&:= \left\langle ((\hat{\calP}_j^{(1)[1]}-(1+b_j^{(1)[1]})\calP_j) \bu_j^{(1)}, \bv \right\rangle, \bv \in \mathbb{H} .  \\
\rho_j(\bv)^{[2]}&:= \left\langle ((\hat{\calP}_j^{[2]}-(1+b_j^{(1)[2]})\calP_j) \bu_j^{(1)}, \bv \right\rangle, \bv \in \mathbb{H} .
\end{align}
Equation (6.6) in \cite{kolt14_2} provides the representation of linear forms of $\check{\bu}_j^{(1),[1]}$ in terms of $\rho_j(\bv)^{[1]}$ and $\rho_j(\bv)^{[2]}$:
\begin{align}
& \sqrt{n} \left\langle \frac{1}{2}\left[\check{\bu}_j^{(1),[1]}+ \check{\bu}_j^{(1),[2]} - \left( \sqrt{1+b_j^{(1)[1]}} + \sqrt{1+b_j^{(1)[2]}} \right)  \bu_j^{(1)} \right], \bv \right\rangle  \nonumber \\
=&\sum_{h=1}^2 \sqrt{n} \bigg[ \frac{\rho_j(\bv)^{[h]}}{\sqrt{1+b_j^{(1)[h]}+\rho_j(\bv)^{[h]}}}   \nonumber \\
&+ \frac{\sqrt{1+b_j^{(1)[h]}}}{\sqrt{1+b_j^{(1)[h]}+\rho_j(\bv)^{[h]}}(\sqrt{1+b_j^{(1)[h]}+\rho_j(\bv)^{[h]}}+\sqrt{1+b_j^{(1)[h]}} ) } \rho_j(u_j^{(1)})^{[h]} \langle \bu_j^{(1)}, \bv \rangle \bigg]  . \label{bitoli}
\end{align}

We shall make use of the following lemma:

\begin{lemma} \label{linear_basic}
Under the same assumptions in Theorem \ref{thm_b}, 
for any $\bu, \bv \in \RR^{d_q}$,
\begin{align*}
& \sqrt{\frac{n}{2}} \bigg\langle \left[\hat{\calP}_j^{(q)[1]} - (1+b_j^{(q)[1]}) \calP_j^{(q)}  \right] \bu,\bv \bigg\rangle \\
& \quad- {\sqrt{2\over n}} \sum_{k=1}^{n/2} \left[ \langle y_{jk}^{(q)},\calP_j^{(q)} \bv \rangle \langle y_{jk}^{(q)},\calC_j^{(q)} \bu \rangle + \langle y_{jk}^{(q)},\calP_j^{(q)} \bu \rangle \langle y_{jk}^{(q)},\calC_j^{(q)} \bv \rangle \right]
=o_p\left(\|\bu\|\|\bv\|\right).  
\end{align*}
Further more, there exists universal constant $C$ such that $ \mathbb{P} ( | b_j^{(q)[1]} | \le C {d}/{n} ) \to 1 $ as $n \to \infty$.
\end{lemma}

Following from Lemma \ref{linear_basic}, observe that $\calC_j \bu_j^{(1)} = 0$, we have
\begin{align}
\sqrt{\frac{n}{2}} \rho_j(\bv)^{[1]} - \sqrt{\frac{2}{n}} \sum_{k=1}^{\frac{n}{2}} \eta_{jk}(\bv) \overset{p}{\to} 0,  \label{eq:rhoeta1}  \\
\sqrt{\frac{n}{2}} \rho_j(\bv)^{[2]} - \sqrt{\frac{2}{n}} \sum_{k=\frac{n}{2}+1}^{n} \eta_{jk}(\bv) \overset{p}{\to} 0.  \label{eq:rhoeta2}
\end{align}
If we can show that for any $\bv \in \RR^{d_1}$,
\begin{align}
\frac{1}{\sqrt{n}} \sum_{k=1}^n \eta_{jk}(\bv)
\overset{d}{\to}
N\left(0,
\frac{1+\sigma_j^2}{\sigma_j^4} \|\calP_j^\perp\bv\|^2 \right),
\label{eq:etaN}
\end{align}
then combining with the facts that $b_j^{(1)[h]} \overset{p}{\to} 0$ and \eqref{bitoli}, \eqref{eq:rhoeta1} and \eqref{eq:rhoeta2}, we have:
\begin{align*}
& \sqrt{n}\left\langle \frac{1}{2}\left[\check{\bu}_j^{(1),[1]}+ \check{\bu}_j^{(1),[2]} - \left( \sqrt{1+b_j^{(1)[1]}} + \sqrt{1+b_j^{(1)[2]}} \right)  \bu_j^{(1)} \right], \bv \right\rangle  \nonumber \\
=& \frac{1}{\sqrt{n}} \sum_{k=1}^n \eta_{jk}(\bv) + o_p(1) \\
\overset{d}{\to} &
N\left(0,
\frac{1+\sigma_j^2}{\sigma_j^4} \|\calP_j^\perp\bv\|^2 \right).
\end{align*}
Recall that $b_j^{(1)[h]} \overset{p}{\to} 0$. By Slutsky's Theorem,
\begin{align*}
& \sqrt{n}\left((1+b_j^{(1)})\langle \check{\bu}^{(1)}_j, \bv\rangle-\langle \bu^{(1)}_{j}, \bv\rangle\right) \\
=& \frac{1}{\sqrt{1+b_j^{(1)[1]}} + \sqrt{1+b_j^{(1)[2]}}} \cdot \sqrt{n}\left\langle \frac{1}{2}\left[\check{\bu}_j^{(1),[1]}+ \check{\bu}_j^{(1),[2]} - \left( \sqrt{1+b_j^{(1)[1]}} + \sqrt{1+b_j^{(1)[2]}} \right)  \bu_j^{(1)} \right], \bv \right\rangle \\
\overset{d}{\to} &
N\left(0,
\frac{1+\sigma_j^2}{\sigma_j^4} \|\calP_j^\perp\bv\|^2 \right).
\end{align*}
The claim \eqref{eq:bias_part1} then follows.

We shall now prove \eqref{eq:etaN}. Note that $\calP_j \bu_j^{(1)}= \bu_j^{(1)}$, and $\calC_j = \frac{1}{\sigma_j^2} \calP_j^{\perp}$, we have
\begin{align*}
\eta_{jk}(\bv) = \frac{1}{\sigma_j^2} \left( \sigma_j \theta_{jk} + \scrE_k \times_1 \bu_j^{(1)} \times_2 \bu_j^{(2)} \dots \times_p \bu_j^{(p)} \right) \left( \scrE_k \times_1 (\calP_j^{\perp} \bv ) \times_2 \bu_j^{(2)} \dots \times_p \bu_j^{(p)} \right) .
\end{align*}
Both $\sigma_j \theta_{jk} + \scrE_k \times_1 \bu_j^{(1)} \times_2 \bu_j^{(2)} \dots \times_p \bu_j^{(p)}$ and $\scrE_k \times_1 ( \mathbb{I}_{d_1} \calP_j^{\perp} \bv ) \times_2 \bu_j^{(2)} \dots \times_p \bu_j^{(p)}$ are centered Gaussian random variables. Moreover, they are independent since $\theta_{jk}$ is independent with $\scrE_k$, and $\bu_j^{(1)} \perp \calP_j^{\perp} \bv$. So
\begin{align*}
\mathbb{E} \eta_{jk}(\bv) = 0.
\end{align*}
More generally, Gaussian variable $\sigma_j \theta_{jk} + \scrE_k \times_1 \bu_j^{(1)} \times_2 \bu_j^{(2)} \dots \times_p \bu_j^{(p)}$ and Gaussian vector $\scrE_k \times_1 (\calP_j^{\perp} ) \times_2 \bu_j^{(2)} \dots \times_p \bu_j^{(p)}$ are independent for the exact same reason. So direct calculation gives:
\begin{align*}
& {\rm var}(\eta_{jk}(\bv))  \\
=& \frac{1}{\sigma_j^4} \mathbb{E} \left( \sigma_j \theta_{jk} + \scrE_k \times_1 \bu_j^{(1)} \times_2 \bu_j^{(2)} \dots \times_p \bu_j^{(p)} \right)^2   \\
&\mathbb{E} \bigg[ \left( \scrE_k \times_1 (\calP_j^{\perp} \bv ) \times_2 \bu_j^{(2)} \dots \times_p \bu_j^{(p)} \right)^2 \bigg]  \\
= & \frac{1+\sigma_j^2}{\sigma_j^4} \| \calP_j^{\perp} \bv \|^2. 
%= \Gamma(\bv_1,v_2).
\end{align*}

With the fact that $\eta_{jk}(\bv)$ are i.i.d for $k=1,2,\dots,n$, to finish this part of  the proof with CLT, it remains to check the Lindeberg condition for CLT, which reduced to 
\begin{align*}
\frac{ \mathbb{E} \eta_{jk}(\bv)^2 \mathbb{I} \left( |\eta_{jk}(\bv)| \ge \tau \sqrt{n} \mathbb{E}^{1/2} \eta_{jk}(\bv)^2  \right) }{ \mathbb{E} \eta_{jk}(\bv)^2 } \to 0 \ {\rm as } \ n \to \infty
\end{align*}
for all $\tau > 0$. Note that
\begin{align*}
\frac{ \mathbb{E} \eta_{jk}(\bv)^2 \mathbb{I} \left( |\eta_{jk}(\bv)| \ge \tau \sqrt{n} \mathbb{E}^{1/2} \eta_{jk}(\bv)^2  \right) }{ \mathbb{E} \eta_{jk}(\bv)^2 } \le \frac{ \mathbb{E} \eta_{jk}(\bv)^4 }{ \tau^2 n \left( \mathbb{E} \eta_{jk}(\bv)^2 \right)^2 } .
\end{align*}
Since
\begin{align*}
&\mathbb{E} \eta_{jk}(\bv)^4  \\
=& \mathbb{E} \left( \sigma_j \theta_{jk} + \scrE_k \times_1 \bu_j^{(1)} \times_2 \bu_j^{(2)} \dots \times_p \bu_j^{(p)} \right)^4 \mathbb{E} \left( \scrE_k \times_1 (\calP_j^{\perp} \bv ) \times_2 \bu_j^{(2)} \dots \times_p \bu_j^{(p)} \right)^4 ,
\end{align*}
and
\begin{align*}
&\left( \mathbb{E} \eta_{jk}(\bv)^2 \right)^2 \\
=& \mathbb{E}^2 \left( \sigma_j \theta_{jk} + \scrE_k \times_1 \bu_j^{(1)} \times_2 \bu_j^{(2)} \dots \times_p \bu_j^{(p)} \right)^2 \mathbb{E}^2 \left( \scrE_k \times_1 (\calP_j^{\perp} \bv ) \times_2 \bu_j^{(2)} \dots \times_p \bu_j^{(p)} \right)^2 ,
\end{align*}
with the fact that for a centered normal random variable $\xi$, $\mathbb{E} \xi^4 = 3 \mathbb{E}^2 \xi^2$, we get
\begin{align*}
\frac{ \mathbb{E} \eta_{jk}(\bv)^4 }{ \tau^2 n \left( \mathbb{E} \eta_{jk}(\bv)^2 \right)^2 } = \frac{1}{\tau^2 n} \to 0,
\end{align*}
and \eqref{eq:etaN} follows.

\subsection{Step 2.} 

Again, it suffices to consider the case $q=1$. We shall now argue that
\begin{align}
\langle \check{\bu}_j^{(1)}, \bu_j^{(1)} \rangle 
= \frac{1}{\sqrt{ 1+ \frac{d_1}{n} \left(\frac{\sigma_0^2}{\sigma_j^2} + \frac{\sigma_0^4}{\sigma_j^4}\right) } } + O_p\left( \frac{d^{3/2}}{n^{3/2}} \right).
\label{eq:uu1}
\end{align}

Write $\tilde{\Sigma}^{[1]} = \hat{\Sigma}^{[1]} - \scrI$, $\tilde{\Sigma}^{[2]} = \hat{\Sigma}^{[2]} - \scrI$, and let
\begin{align*}
\grave{\bu}_j^{(1),[1]} = \tilde{\Sigma}^{[1]}(\cdot,\hat{\bu}_j^{(2)[2]},\dots,\hat{\bu}_j^{(p)[2]},\check{\bu}_j^{(1),[1]},\hat{\bu}_j^{(2)[2]},\dots,\hat{\bu}_j^{(p)[2]}),
\end{align*}
and
\begin{align*}
\grave{\bu}_j^{(1),[2]} = \tilde{\Sigma}^{[2]}(\cdot,\hat{\bu}_j^{(2)[1]},\dots,\hat{\bu}_j^{(p)[1]},\check{\bu}_j^{(1),[2]},\hat{\bu}_j^{(2)[1]},\dots,\hat{\bu}_j^{(p)[1]}).
\end{align*}
Since $\check{\bu}_j^{(1),[1]}$ is the leading eigenvector of 
$$
\hat{\Sigma}^{[1]}(\cdot,\hat{\bu}_j^{(2)[2]},\dots,\hat{\bu}_j^{(p)[2]},\cdot,\hat{\bu}_j^{(2)[2]},\dots,\hat{\bu}_j^{(p)[2]}),
$$
it is also the leading eigenvector of
$$
\tilde{\Sigma}^{[1]}(\cdot,\hat{\bu}_j^{(2)[2]},\dots,\hat{\bu}_j^{(p)[2]},\cdot,\hat{\bu}_j^{(2)[2]},\dots,\hat{\bu}_j^{(p)[2]}),
$$
so
$$
\check{\bu}_j^{(1),[1]} = \frac{\grave{\bu}_j^{(1),[1]}}{\|\grave{\bu}_j^{(1),[1]}\|},
$$
and similarly 
$$
\check{\bu}_j^{(1),[2]} = \frac{\grave{\bu}_j^{(1),[2]}}{\|\grave{\bu}_j^{(1),[2]}\|}.
$$

Write
$$
\bz_j^{(1)[1]}:= \grave{\bu}_j^{(1),[1]} - \langle \grave{\bu}_j^{(1),[1]}, \bu_j^{(1)} \rangle \bu_j^{(1)} - \by_j^{(1)[1]} ,
$$
where
\begin{equation}
\by_j^{(1)[1]}:= \calP_{\bu_j^{(1)}}^\perp \left[ \frac{2}{n} \sum_{k=1}^{n/2} \scrE_k (\cdot,\bu_j^{(2)},\dots,\bu_j^{(p)} ) \sigma_j \theta_{jk} + \frac{2}{n} \sum_{k=1}^{n/2}   \scrE_k (\cdot,\bu_j^{(2)},\dots,\bu_j^{(p)} )  \scrE_k (\bu_j^{(1)},\dots,\bu_j^{(p)} ) \right]. \label{def:y}
\end{equation}
$\bz_j^{(1)[2]}$ and $\by_j^{(1)[2]}$ are defined similarly so that
$$
\grave{\bu}_j^{(1),[1]} = \langle \grave{\bu}_j^{(1),[1]}, \bu_j^{(1)} \rangle \bu_j^{(1)} + \by_j^{(1)[1]} + \bz_j^{(1)[1]}, 
$$
$$
\grave{\bu}_j^{(1),[2]} = \langle \grave{\bu}_j^{(1),[2]}, \bu_j^{(1)} \rangle \bu_j^{(1)} + \by_j^{(1)[2]} + \bz_j^{(1)[2]}.
$$

We shall treat the three terms in $\grave{\bu}_j^{(1),[1]}$ and $\grave{\bu}_j^{(1),[1]}$ separately to show that
\begin{align}
\langle \grave{\bu}_j^{(1),[1]}, \bu_j^{(1)} \rangle = \sigma_j^2 + O_p\left(\frac{d}{n}\right), \ \langle \grave{\bu}_j^{(1),[2]}, \bu_j^{(1)} \rangle = \sigma_j^2 + O_p\left(\frac{d}{n}\right), \label{eq:grave_inner}
\end{align}
\begin{align}
\left\| \by_j^{(1)[1]} + \by_j^{(1)[2]} \right\| = 2 \sqrt{\frac{d_1}{n}} \sqrt{\sigma_0^2 \sigma_j^2 + \sigma_0^4} + O_p \left(\frac{\sqrt{d}}{n}\right) + O_p \left(\frac{1}{\sqrt{n}}\right)
\label{eq:yy12},
\end{align}
\begin{align}
\label{eq:yy1}
\left\| \by_j^{(1)[1]} \right\| =& \sqrt{\frac{2d_1}{n}} \sqrt{\sigma_0^2 \sigma_j^2 + \sigma_0^4} + O_p \left(\frac{\sqrt{d}}{n}\right) + O_p \left(\frac{1}{\sqrt{n}}\right) ,\\
\left\| \by_j^{(1)[2]} \right\| =& \sqrt{\frac{2d_1}{n}} \sqrt{\sigma_0^2 \sigma_j^2 + \sigma_0^4} + O_p \left(\frac{\sqrt{d}}{n}\right) + O_p \left(\frac{1}{\sqrt{n}}\right) , \nonumber
\end{align}
and
\begin{align}
\bz_j^{(1)[1]} = O_p\left(\frac{d}{n}\right), \ \bz_j^{(1)[2]} = O_p\left(\frac{d}{n}\right). \label{eq:zz}
\end{align}

We first show that equation \eqref{eq:uu1} follows from the bounds given in \eqref{eq:grave_inner}-\eqref{eq:zz}. We start with $\|\grave{\bu}_j^{(1),[1]}\|$. Combining \eqref{eq:grave_inner}-\eqref{eq:zz}, we have
\begin{align}
& \|\grave{\bu}_j^{(1),[1]}\| \nonumber\\
=& \sqrt{\left\langle \grave{\bu}_j^{(1),[1]}, \bu_j^{(1)} \right\rangle^2 + \left\| \by_j^{(1)[1]} + \bz_j^{(1)[1]}\right\|^2} \nonumber\\
=& \sqrt{ \left[\sigma_j^2 + O_p\left(\frac{d}{n}\right)\right]^2 + \left [ O_p\left(\sqrt{\frac{d}{n}}\right) + O_p\left(\frac{d}{n}\right) \right]^2 } \nonumber\\ 
=& \sigma_j^2 + O_p\left(\frac{d}{n}\right), \label{eq:uu1norm}
\end{align}
where the second equality follows from \eqref{eq:grave_inner}, \eqref{eq:yy1}, \eqref{eq:zz}, and the last equality holds because $d=o(n)$. Similarly we can derive that
$$\|\grave{\bu}_j^{(1),[2]}\| =\sigma_j^2 + O_p\left({d}/{n}\right).$$
Hence,
\begin{align}
&\left\|\frac{\by_j^{(1)[1]}}{\|\grave{\bu}_j^{(1),[1]}\|} + \frac{\by_j^{(1)[2]}}{\|\grave{\bu}_j^{(1),[2]}\|} \right\| \nonumber\\
=&\left\| \frac{1}{\sigma_j^2} (\by_j^{(1)[1]} + \by_j^{(1)[2]}) + \left(\frac{1}{\|\grave{\bu}_j^{(1),[1]}\|} - \frac{1}{\sigma_j^2}\right) \by_j^{(1)[1]} + \left(\frac{1}{\|\grave{\bu}_j^{(1),[2]}\|} - \frac{1}{\sigma_j^2}\right) \by_j^{(1)[2]} \right\| \nonumber\\
=& 2 \sqrt{\frac{d_1}{n} \left( \frac{\sigma_0^2}{\sigma_j^2} + \frac{\sigma_0^4}{\sigma_j^4} \right)} + O_p\left(\frac{\sqrt{d}}{n}\right) + O_p\left(\frac{1}{\sqrt{n}}\right) + O_p\left(\frac{\sqrt{d}}{n}\right) \cdot O_p\left(\frac{1}{\sqrt{n}}\right) \nonumber\\
=& 2 \sqrt{\frac{d_1}{n} \left( \frac{\sigma_0^2}{\sigma_j^2} + \frac{\sigma_0^4}{\sigma_j^4} \right)} + O_p\left(\frac{\sqrt{d}}{n}\right) + O_p\left(\frac{1}{\sqrt{n}}\right), \label{eq:yuyu}
\end{align}
where the second equality follows from \eqref{eq:yy12}, \eqref{eq:yy1} and \eqref{eq:uu1norm}.

Moreover, \eqref{eq:zz}, \eqref{eq:uu1norm} imply that
\begin{align}
\left\|\frac{\bz_j^{(1)[1]}}{\|\grave{\bu}_j^{(1),[1]}\|} + \frac{\bz_j^{(1)[2]}}{\|\grave{\bu}_j^{(1),[2]}\|} \right\| = O_p\left(\frac{d}{n}\right), \label{eq:zuzu}
\end{align}
and \eqref{eq:grave_inner} and \eqref{eq:uu1norm} imply that
\begin{align}
\frac{\langle \grave{\bu}_j^{(1),[1]}, \bu_j^{(1)} \rangle}{\|\grave{\bu}_j^{(1),[1]}\|} + \frac{\langle \grave{\bu}_j^{(1),[2]}, \bu_j^{(1)} \rangle}{\|\grave{\bu}_j^{(1),[2]}\|} = 2 + O_p\left(\frac{d}{n}\right). \label{eq:uuuu}
\end{align}
Therefore,
\begin{align*}
&\langle \check{\bu}_j^{(1)}, \bu_j^{(1)} \rangle \\
=& { \frac{\langle \grave{\bu}_j^{(1),[1]}, \bu_j^{(1)} \rangle}{\|\grave{\bu}_j^{(1),[1]}\|} + \frac{\langle \grave{\bu}_j^{(1),[2]}, \bu_j^{(1)} \rangle}{\|\grave{\bu}_j^{(1),[2]}\|} } 
\over 
\sqrt{ \left( \frac{\langle \grave{\bu}_j^{(1),[1]}, \bu_j^{(1)} \rangle}{\|\grave{\bu}_j^{(1),[1]}\|} + \frac{\langle \grave{\bu}_j^{(1),[2]}, \bu_j^{(1)} \rangle}{\|\grave{\bu}_j^{(1),[2]}\|} \right)^2 + \left( \frac{\by_j^{(1)[1]}}{\|\grave{\bu}_j^{(1),[1]}\|} + \frac{\by_j^{(1)[2]}}{\|\grave{\bu}_j^{(1),[2]}\|} + \frac{\bz_j^{(1)[1]}}{\|\grave{\bu}_j^{(1),[1]}\|} + \frac{\bz_j^{(1)[2]}}{\|\grave{\bu}_j^{(1),[2]}\|} \right)^2 } \\
=& 1 \Bigg/ \sqrt{1+ \left( \frac{ \frac{\by_j^{(1)[1]}}{\|\grave{\bu}_j^{(1),[1]}\|} + \frac{\by_j^{(1)[2]}}{\|\grave{\bu}_j^{(1),[2]}\|} + \frac{\bz_j^{(1)[1]}}{\|\grave{\bu}_j^{(1),[1]}\|} + \frac{\bz_j^{(1)[2]}}{\|\grave{\bu}_j^{(1),[2]}\|} }{ \frac{\langle \grave{\bu}_j^{(1),[1]}, \bu_j^{(1)} \rangle}{\|\grave{\bu}_j^{(1),[1]}\|} + \frac{\langle \grave{\bu}_j^{(1),[2]}, \bu_j^{(1)} \rangle}{\|\grave{\bu}_j^{(1),[2]}\|} } \right)^2 } \\
=& 1 \Bigg/ \sqrt{1+ \left( \frac{ 2 \sqrt{\frac{d_1}{n} \left( \frac{\sigma_0^2}{\sigma_j^2} + \frac{\sigma_0^4}{\sigma_j^4} \right)} + O_p\left(\frac{\sqrt{d}}{n}\right) + O_p\left(\frac{1}{\sqrt{n}}\right) + O_p\left(\frac{d}{n}\right) }{ 2 + O_p\left(\frac{d}{n}\right) } \right)^2 } \\
=& 1 \Bigg/ \sqrt{ 1 + \left\{ \left[ \sqrt{\frac{d_1}{n} \left( \frac{\sigma_0^2}{\sigma_j^2} + \frac{\sigma_0^4}{\sigma_j^4} \right)} + O_p\left(\frac{1}{\sqrt{n}}\right) + O_p\left(\frac{d}{n}\right) \right] \cdot \left[ 1+O_p\left(\frac{d}{n}\right) \right]\right\}^2 }  \\
=& 1 \over \sqrt{ 1 +  \left[ \sqrt{\frac{d_1}{n} \left( \frac{\sigma_0^2}{\sigma_j^2} + \frac{\sigma_0^4}{\sigma_j^4} \right)} + O_p\left(\frac{d}{n}\right) \right]^2 } \\
=& \frac{1}{\sqrt{ 1+ \frac{d_1}{n} \left(\frac{\sigma_0^2}{\sigma_j^2} + \frac{\sigma_0^4}{\sigma_j^4}\right) } } + O_p\left( \frac{d^{3/2}}{n^{3/2}} \right) ,
\end{align*}
where the third equation follows from \eqref{eq:yuyu}-\eqref{eq:uuuu}, and the last equation follows from the simple fact that
$$
\left| \frac{1}{\sqrt{1+(x+y)^2}} - \frac{1}{\sqrt{1+x^2}} \right| \le \left| xy+\frac{1}{2} y^2 \right|.
$$

It now remains to show \eqref{eq:grave_inner}-\eqref{eq:zz}.

\subsubsection{Equations \eqref{eq:yy12} and \eqref{eq:yy1}.}
Write
$$
\bx_k :=\scrE_k (\cdot,\bu_j^{(2)},\dots,\bu_j^{(p)} ) \in \RR^{d_1}.
$$
If we choose an orthogonal basis of $\RR^{d_1}$ with $\be_1=\bu_j^{(1)}$, then
\begin{align*}
	\left\| \frac{1}{2} ( \by_j^{(1)[1]} + \by_j^{(1)[2]} ) \right\| =& \left\| \calP_{\bu_j^{(1)}}^\perp \left[ \frac{1}{n} \sum_{k=1}^{n} \bx_k \big( \sigma_j \theta_{jk} + \bx_{1,k} \big) \right] \right\|=\left\| \frac{1}{n} \sum_{k=1}^{n} \bx_{(-1),k} \big( \sigma_j \theta_{jk} + \bx_{1,k} \big) \right\|, 
\end{align*}
in which $\bx_{1,k}$ means the first entry of $\bx_k$, while $\bx_{(-1),k}$ stands for all the other entries (having a dimension of $d_1-1$).

Observe that $\bx_{(-1),k}$ and $\big( \sigma_j \theta_{jk} + \bx_{1,k} \big)$ two independent group of i.i.d. random variables, $\bx_{(-1),k}$ follows distribution $N(0,\sigma_0^2 I_{d_1-1})$, and $\big( \sigma_j \theta_{jk} + \bx_{1,k} \big)$ follows distribution $N(0,\sigma_0^2+\sigma_j^2)$. Thus,
\begin{align*}
	\left\| \frac{1}{2} ( \by_j^{(1)[1]} + \by_j^{(1)[2]} ) \right\| =& \sqrt{\frac{d_1-1}{n}} \sqrt{\sigma_0^2 \sigma_j^2 + \sigma_0^4} + O_p \left(\frac{\sqrt{d_1-1}}{n}\right) + O_p \left(\frac{1}{\sqrt{n}}\right) .
\end{align*} 
\eqref{eq:yy1} can be proven in a similar way.

\subsubsection{Equation \eqref{eq:zz}.} 

Recall that
\begin{align}
&\tilde{\Sigma}(\cdot,\hat{\bu}_j^{(2)[2]},\dots,\hat{\bu}_j^{(p)[2]}, \grave{\bu}_j^{(1),[1]} ,\hat{\bu}_j^{(2)[2]},\dots,\hat{\bu}_j^{(p)[2]}) \nonumber \\ 
=& \sum_{l=1}^r \sigma_l^2 \left( \frac{2}{n} \sum_{k=1}^{n/2}  \theta_{lk}^2 \right) \left( \prod_{q=2}^p \langle \bu_l^{(q)}, \hat{\bu}_j^{(q)[2]} \rangle \right)^2 \langle \bu_l^{(1)}, \grave{\bu}_j^{(1),[1]} \rangle \bu_l^{(1)}   \nonumber \\
&+ \sum_{l_1 \neq l_2} \sigma_{l_1} \sigma_{l_2} \left( \frac{2}{n} \sum_{k=1}^{n/2}  \theta_{l_1 k} \theta_{l_2 k} \right) \left( \prod_{q=2}^p \langle \bu_{l_1}^{(q)}, \hat{\bu}_j^{(q)[2]} \rangle \right) \left( \prod_{q=2}^p \langle \bu_{l_2}^{(q)}, \hat{\bu}_j^{(q)[2]} \rangle \right) \langle \bu_{l_2}^{(1)}, \grave{\bu}_j^{(1),[1]} \rangle \bu_{l_1}^{(1)} \nonumber \\
&+  \frac{2}{n} \sum_{k=1}^{n/2}  \sum_{l=1}^r \sigma_l \theta_{lk} \left( \prod_{q=2}^p \langle \bu_l^{(q)}, \hat{\bu}_j^{(q)[2]} \rangle \right) \scrE_k (\grave{\bu}_j^{(1),[1]}, \hat{\bu}_j^{(2)[2]},\dots,\hat{\bu}_j^{(p)[2]} ) \bu_l^{(1)}   \nonumber \\
&+  \frac{2}{n} \sum_{k=1}^{n/2}  \sum_{l=1}^r \scrE_k (\cdot, \hat{\bu}_j^{(2)[2]},\dots,\hat{\bu}_j^{(p)[2]} ) \sigma_l \theta_{lk} \left( \prod_{q=2}^p \langle \bu_l^{(q)}, \hat{\bu}_j^{(q)[2]} \rangle \right) \langle \bu_l^{(1)}, \grave{\bu}_j^{(1),[1]} \rangle    \nonumber \\
&+ \frac{2}{n} \sum_{k=1}^{n/2}   \scrE_k (\cdot, \hat{\bu}_j^{(2)[2]},\dots,\hat{\bu}_j^{(p)[2]} )  \scrE_k ( \grave{\bu}_j^{(1),[1]} ,\hat{\bu}_j^{(2)[2]},\dots,\hat{\bu}_j^{(p)[2]})  - \grave{\bu}_j^{(1),[1]} , \nonumber
\end{align}
so by definition,
\begin{align}
\bz_j^{(1)[1]}
=&  \sum_{l\neq j, l\in [r]} \sigma_l^2 \left( \frac{2}{n} \sum_{k=1}^{n/2}  \theta_{lk}^2 \right) \left( \prod_{q=2}^p \langle \bu_l^{(q)}, \hat{\bu}_j^{(q)[2]} \rangle \right)^2 \langle \bu_l^{(1)}, \grave{\bu}_j^{(1),[1]} \rangle \bu_l^{(1)}   \nonumber \\
&+ \sum_{l_1 \neq l_2, l_1 \neq j} \sigma_{l_1} \sigma_{l_2} \left( \frac{2}{n} \sum_{k=1}^{n/2}  \theta_{l_1 k} \theta_{l_2 k} \right) \left( \prod_{q=2}^p \langle \bu_{l_1}^{(q)}, \hat{\bu}_j^{(q)[2]} \rangle \right) \left( \prod_{q=2}^p \langle \bu_{l_2}^{(q)}, \hat{\bu}_j^{(q)[2]} \rangle \right) \langle \bu_{l_2}^{(1)}, \grave{\bu}_j^{(1),[1]} \rangle \bu_{l_1}^{(1)} \nonumber \\
&+  \frac{2}{n} \sum_{k=1}^{n/2}  \sum_{l\neq j, l\in [r]} \sigma_l \theta_{lk} \left( \prod_{q=2}^p \langle \bu_l^{(q)}, \hat{\bu}_j^{(q)[2]} \rangle \right) \scrE_k (\grave{\bu}_j^{(1),[1]}, \hat{\bu}_j^{(2)[2]},\dots,\hat{\bu}_j^{(p)[2]} ) \bu_l^{(1)}   \nonumber \\
&+  \frac{2}{n} \sum_{k=1}^{n/2}  \sum_{l\neq j, l\in [r]} \scrE_k (\cdot, \hat{\bu}_j^{(2)[2]},\dots,\hat{\bu}_j^{(p)[2]} ) \sigma_l \theta_{lk} \left( \prod_{q=2}^p \langle \bu_l^{(q)}, \hat{\bu}_j^{(q)[2]} \rangle \right) \langle \bu_l^{(1)}, \grave{\bu}_j^{(1),[1]} \rangle    \nonumber \\
&+  \calP_{\bu_j^{(1)}}^\perp \bigg[ \frac{2}{n} \sum_{k=1}^{n/2} \scrE_k (\cdot,\hat{\bu}_j^{(2)[2]},\dots,\hat{\bu}_j^{(p)[2]} ) \sigma_j \theta_{jk} \left( \prod_{q=2}^p \langle \bu_j^{(q)}, \hat{\bu}_j^{(q)[2]} \rangle \right) \langle \bu_j^{(1)}, \grave{\bu}_j^{(1),[1]} \rangle  \nonumber \\
& \qquad \qquad - \frac{2}{n} \sum_{k=1}^{n/2} \scrE_k (\cdot,\bu_j^{(2)},\dots,\bu_j^{(p)} ) \sigma_j \theta_{jk} \bigg] \nonumber \\
&+ \calP_{\bu_j^{(1)}}^\perp \bigg[ \frac{2}{n} \sum_{k=1}^{n/2}   \scrE_k (\cdot,\hat{\bu}_j^{(2)[2]},\dots,\hat{\bu}_j^{(p)[2]} )  \scrE_k (\grave{\bu}_j^{(1),[1]}, \hat{\bu}_j^{(2)[2]},\dots,\hat{\bu}_j^{(p)[2]} )  - \grave{\bu}_j^{(1),[1]} \nonumber \\
& \qquad \qquad - \frac{2}{n} \sum_{k=1}^{n/2}   \scrE_k (\cdot,\bu_j^{(2)},\dots,\bu_j^{(p)} )  \scrE_k (\bu_j^{(1)},\dots,\bu_j^{(p)} ) + \bu_j^{(1)} \bigg],
\end{align}
and we will deal with those six terms one by one.

\paragraph{The first term.}
\begin{align*}
	\bz_1^{[1]}:=\sum_{l\neq j, l\in [r]} \sigma_l^2 \left( \frac{2}{n} \sum_{k=1}^{n/2}  \theta_{lk}^2 \right) \left( \prod_{q=2}^p \langle \bu_l^{(q)}, \hat{\bu}_j^{(q)[2]} \rangle \right)^2 \langle \bu_l^{(1)}, \grave{\bu}_j^{(1),[1]} \rangle \bu_l^{(1)} .
\end{align*}
Observe that
\begin{align}
	\left\| \bz_1^{[1]} \right\| \le 
	\big(\delta^{(1)[2]}\big)^{2p-1} \sum_{l\neq j, l\in [r]} \sigma_l^2 \left| \frac{2}{n} \sum_{k=1}^{n/2}  \theta_{lk}^2 \right|
	=O_p \left[ \left( \sqrt{\frac{d}{n}} \right)^{2p-1} \right] 
	= O_p \left[ \left( \frac{d}{n} \right)^{3/2} \right]. \label{eq:bias_second_noise_1}
\end{align}
The last two equality's follow from \eqref{eq:bias_fact}, and $p \ge 2$. Because of the assumption $d=o(n)$, it implies
$$
\left\| \bz_1^{[1]} \right\| \le O_p \left( \frac{d}{n} \right).
$$

\paragraph{The second term.}
\begin{align*}
	\bz_2^{[1]}:=&\sum_{l_1 \neq l_2, l_1 \neq j} \sigma_{l_1} \sigma_{l_2} \left( \frac{2}{n} \sum_{k=1}^{n/2}  \theta_{l_1 k} \theta_{l_2 k} \right) \left( \prod_{q=2}^p \langle \bu_{l_1}^{(q)}, \hat{\bu}_j^{(q)[2]} \rangle \right) \left( \prod_{q=2}^p \langle \bu_{l_2}^{(q)}, \hat{\bu}_j^{(q)[2]} \rangle \right) \langle \bu_{l_2}^{(1)}, \grave{\bu}_j^{(1),[1]} \rangle \bu_{l_1}^{(1)} \nonumber\\
	=& \sum_{l_1 \neq l_2, l_1 \neq j, l_2 \neq j} \sigma_{l_1} \sigma_{l_2} \left( \frac{2}{n} \sum_{k=1}^{n/2}  \theta_{l_1 k} \theta_{l_2 k} \right) \left( \prod_{q=2}^p \langle \bu_{l_1}^{(q)}, \hat{\bu}_j^{(q)[2]} \rangle \right) \left( \prod_{q=2}^p \langle \bu_{l_2}^{(q)}, \hat{\bu}_j^{(q)[2]} \rangle \right) \langle \bu_{l_2}^{(1)}, \grave{\bu}_j^{(1),[1]} \rangle \bu_{l_1}^{(1)} \nonumber\\
	&+\sum_{l_1 \neq j} \sigma_{l_1} \sigma_j \left( \frac{2}{n} \sum_{k=1}^{n/2}  \theta_{l_1 k} \theta_{j k} \right) \left( \prod_{q=2}^p \langle \bu_{l_1}^{(q)}, \hat{\bu}_j^{(q)[2]} \rangle \right) \left( \prod_{q=2}^p \langle \bu_j^{(q)}, \hat{\bu}_j^{(q)[2]} \rangle \right) \langle \bu_j^{(1)}, \grave{\bu}_j^{(1),[1]} \rangle \bu_{l_1}^{(1)}. 
\end{align*}
Because
$$
\left| \frac{2}{n} \sum_{k=1}^{n/2}  \theta_{l_1 k} \theta_{l_2 k} \right|=O_p(\frac{1}{\sqrt{n}}), \ \forall l_1 \neq l_2,
$$ 
we immediately have
\begin{align}
	&\left\| \bz_2^{[1]} \right\| \nonumber \\ 
	\le &
	\big( \delta^{(1)[2]} \big)^{2p-1} \sum_{l_1 \neq l_2, l_1 \neq j, l_2 \neq j} \sigma_{l_1} \sigma_{l_2} \left| \frac{2}{n} \sum_{k=1}^{n/2}  \theta_{l_1 k} \theta_{l_2 k} \right| 
	+ \big( \delta^{(1)[2]} \big)^{p-1} \sum_{l_1 \neq j} \sigma_{l_1} \sigma_{j} \left| \frac{2}{n} \sum_{k=1}^{n/2}  \theta_{l_1 k} \theta_{j k} \right| \nonumber \\
	=& O_p(\frac{1}{\sqrt{n}}) \cdot O_p \left[ \left( \sqrt{\frac{d}{n}} \right)^{2p-1} \right] + O_p(\frac{1}{\sqrt{n}}) \cdot \left[ \left( \sqrt{\frac{d}{n}} \right)^{p-1} \right] \nonumber\\
	=& O_p \left( \frac{\sqrt{d}}{n} \right).
	\label{eq:bias_second_noise_2}
\end{align}

\paragraph{The third term.}
\begin{align*}
	\bz_3^{[1]}:=\frac{2}{n} \sum_{k=1}^{n/2}  \sum_{l\neq j, l\in [r]} \sigma_l \theta_{lk} \left( \prod_{q=2}^p \langle \bu_l^{(q)}, \hat{\bu}_j^{(q)[2]} \rangle \right) \scrE_k (\grave{\bu}_j^{(1),[1]}, \hat{\bu}_j^{(2)[2]},\dots,\hat{\bu}_j^{(p)[2]} ) \bu_l^{(1)}.
\end{align*}
Observe that
\begin{align}
\left\| \bz_3^{[1]} \right\|\le& \sum_{l\neq j, l\in [r]} \sigma_l \left| \prod_{q=2}^p \langle \bu_l^{(q)}, \hat{\bu}_j^{(q)[2]} \rangle \right| \cdot \left| \frac{2}{n} \sum_{k=1}^{n/2} \theta_{lk} \scrE_k (\grave{\bu}_j^{(1),[1]}, \hat{\bu}_j^{(2)[2]},\dots,\hat{\bu}_j^{(p)[2]} ) \right| \nonumber\\
\le& r \sigma_1 (\delta^{(1)[2]})^{p-1} \cdot \|\hat{\Sigma}_{\theta,\scrE}^{[1]}\| \nonumber\\
=& O_p \left(\sqrt{\frac{d}{n}}\right)^{p-1} \cdot O_p \left(\sqrt{\frac{d}{n}}\right) \nonumber\\
=& O_p \left(\frac{d}{n}\right),
\label{eq:bias_second_noise_3}
\end{align}
in which the second to last equality follows from \eqref{eq:bias_fact} and \eqref{eq:SS}.

\paragraph{The fourth term.}
\begin{align*}
	\bz_4^{[1]}:= \frac{2}{n} \sum_{k=1}^{n/2}  \sum_{l\neq j, l\in [r]} \scrE_k (\cdot, \hat{\bu}_j^{(2)[2]},\dots,\hat{\bu}_j^{(p)[2]} ) \sigma_l \theta_{lk} \left( \prod_{q=2}^p \langle \bu_l^{(q)}, \hat{\bu}_j^{(q)[2]} \rangle \right) \langle \bu_l^{(1)}, \grave{\bu}_j^{(1),[1]} \rangle .
\end{align*}
Similar to $\bz_3^{[1]}$,
\begin{align*}
	\big\| \bz_4^{[1]}\big\| \le \|\hat{\Sigma}_{\theta,\scrE}\| \cdot  \left(\delta^{(1)[2]}\right)^p = O_p \left( \frac{d^{3/2}}{n^{3/2}}\right)=O_p\left(d \over n \right).
\end{align*}

\paragraph{The fifth term.}
\begin{align*}
	\bz_5^{[1]}:= &   \calP_{\bu_j^{(1)}}^\perp \bigg[ \frac{2}{n} \sum_{k=1}^{n/2} \scrE_k (\cdot,\hat{\bu}_j^{(2)[2]},\dots,\hat{\bu}_j^{(p)[2]} ) \sigma_j \theta_{jk} \left( \prod_{q=2}^p \langle \bu_j^{(q)}, \hat{\bu}_j^{(q)[2]} \rangle \right) \langle \bu_j^{(1)}, \grave{\bu}_j^{(1),[1]} \rangle  \nonumber \\
& \qquad \qquad - \frac{2}{n} \sum_{k=1}^{n/2} \scrE_k (\cdot,\bu_j^{(2)},\dots,\bu_j^{(p)} ) \sigma_j \theta_{jk} \bigg] .
\end{align*}
we shall introduce the notation $\biguplus$, as the operation of expanding all the 
$$
\hat{\bu}_j^{(q)[2]}=\bu_j^{(q)}+\Delta_j^{(q)[2]},
$$
and 
$$
\grave{\bu}_j^{(1),[1]}=\bu_j^{(1)}+\grave{\Delta}_j^{(1)[1]},
$$
and then keep all the terms with at least one $\Delta_j^{(\cdot)[2]}$ or $\grave{\Delta}_j^{(1)[1]}$ in it. For instance, here, expanding 
$$
\frac{2}{n} \sum_{k=1}^{n/2} \scrE_k (\cdot,\hat{\bu}_j^{(2)[2]},\dots,\hat{\bu}_j^{(p)[2]} ) \sigma_j \theta_{jk} \left( \prod_{q=2}^p \langle \bu_j^{(q)}, \hat{\bu}_j^{(q)[2]} \rangle \right) \langle \bu_j^{(1)}, \grave{\bu}_j^{(1),[1]} \rangle
$$
would result in $2^{2p-1}$ terms, and we keep everything other than the term 
$$
\frac{2}{n} \sum_{k=1}^{n/2} \scrE_k (\cdot,\bu_j^{(2)},\dots,\bu_j^{(p)} ) \sigma_j \theta_{jk} \left( \prod_{q=1}^p \langle \bu_j^{(q)}, \bu_j^{(q)} \rangle \right).
$$
Then we have:
\begin{align*}
	\bz_5^{[1]}=& \calP_{\bu_j^{(1)}}^\perp \left\{ \biguplus \Bigg[ \frac{2}{n} \sum_{k=1}^{n/2} \scrE_k (\cdot,\hat{\bu}_j^{(2)[2]},\dots,\hat{\bu}_j^{(p)[2]} ) \sigma_j \theta_{jk} \left( \prod_{q=2}^p \langle \bu_j^{(q)}, \hat{\bu}_j^{(q)[2]} \rangle \right) \langle \bu_j^{(1)}, \grave{\bu}_j^{(1),[1]} \rangle \Bigg] \right\}.
	%\\ =& \biguplus \Bigg[ \frac{2}{n} \sum_{k=1}^{n/2} \scrE_k (\cdot,\bu_j^{(2)}+\Delta_j^{(2)[2]},\dots,\bu_j^{(p)}+\Delta_j^{(p)[2]} ) \sigma_j \theta_{jk} \left( \prod_{q=1}^p \langle \bu_j^{(q)}, \bu_j^{(q)}+\Delta_j^{(q)[2]} \rangle \right) \Bigg]
\end{align*}
Then,
\begin{align*}
	\big\| \bz_5^{[1]}\big\|\le \left\| \biguplus \Bigg[ \frac{2}{n} \sum_{k=1}^{n/2} \scrE_k (\cdot,\hat{\bu}_j^{(2)[2]},\dots,\hat{\bu}_j^{(p)[2]} ) \sigma_j \theta_{jk} \left( \prod_{q=1}^p \langle \bu_j^{(q)}, \hat{\bu}_j^{(q)[2]} \rangle \right) \Bigg] \right\|,
	%\\ =& \biguplus \Bigg[ \frac{2}{n} \sum_{k=1}^{n/2} \scrE_k (\cdot,\bu_j^{(2)}+\Delta_j^{(2)[2]},\dots,\bu_j^{(p)}+\Delta_j^{(p)[2]} ) \sigma_j \theta_{jk} \left( \prod_{q=1}^p \langle \bu_j^{(q)}, \bu_j^{(q)}+\Delta_j^{(q)[2]} \rangle \right) \Bigg]
\end{align*}
and for every term in $\biguplus$, its norm is bounded by 
$$ \|\hat{\Sigma}_{\theta,\scrE}\| \cdot \delta^{(1)[2]}=O_p\big(\frac{d}{n}\big),$$
so that
\begin{align*}
	\big\| \bz_5^{[1]}\big\| = O_p \left( \frac{d}{n}\right).
\end{align*}

\paragraph{The sixth term.}
This term can be treated in a similar fashion as the last term.
\begin{align*}
\bz_6^{[1]}:= & \calP_{\bu_j^{(1)}}^\perp \bigg[ \frac{2}{n} \sum_{k=1}^{n/2}   \scrE_k (\cdot,\hat{\bu}_j^{(2)[2]},\dots,\hat{\bu}_j^{(p)[2]} )  \scrE_k (\grave{\bu}_j^{(1),[1]}, \hat{\bu}_j^{(2)[2]},\dots,\hat{\bu}_j^{(p)[2]} )  - \grave{\bu}_j^{(1),[1]} \nonumber \\
& \qquad \qquad - \frac{2}{n} \sum_{k=1}^{n/2}   \scrE_k (\cdot,\bu_j^{(2)},\dots,\bu_j^{(p)} )  \scrE_k (\bu_j^{(1)},\dots,\bu_j^{(p)} ) + \bu_j^{(1)} \bigg] \nonumber \\
=& \calP_{\bu_j^{(1)}}^\perp \Bigg[  \bigg(\frac{2}{n} \sum_{k=1}^{n/2} \scrE_k \otimes \scrE_k -\scrI \bigg) (\cdot,\hat{\bu}_j^{(2)[2]},\dots,\hat{\bu}_j^{(p)[2]}, \grave{\bu}_j^{(1),[1]}, \hat{\bu}_j^{(2)[2]},\dots,\hat{\bu}_j^{(p)[2]} ) \nonumber \\
& \qquad \qquad - \bigg(\frac{2}{n} \sum_{k=1}^{n/2} \scrE_k \otimes \scrE_k -\scrI \bigg) (\cdot,\bu_j^{(2)},\dots,\bu_j^{(p)}, \bu_j^{(1)},\dots,\bu_j^{(p)} ) \Bigg]\\
=& \calP_{\bu_j^{(1)}}^\perp \left\{ \biguplus \Bigg[ \bigg(\frac{2}{n} \sum_{k=1}^{n/2} \scrE_k \otimes \scrE_k -\scrI \bigg) (\cdot,\hat{\bu}_j^{(2)[2]},\dots,\hat{\bu}_j^{(p)[2]}, \grave{\bu}_j^{(1),[1]}, \hat{\bu}_j^{(2)[2]},\dots,\hat{\bu}_j^{(p)[2]} ) \Bigg] \right\},
\end{align*}
then,
\begin{align*}
	\big\| \bz_6^{[1]}\big\|\le \left\| \biguplus \Bigg[ \bigg(\frac{2}{n} \sum_{k=1}^{n/2} \scrE_k \otimes \scrE_k -\scrI \bigg) (\cdot,\hat{\bu}_j^{(2)[2]},\dots,\hat{\bu}_j^{(p)[2]}, \hat{\bu}_j^{(1)[2]},\dots,\hat{\bu}_j^{(p)[2]} ) \Bigg] \right\|,
\end{align*}
and for every term in $\biguplus$, its norm is bounded by
$$\| \hat{\Sigma}_{\scrE} - \scrI \| \cdot \delta^{(1)[2]}=O_p\big(\frac{d}{n}\big),$$
so that
\begin{align*}
	\big\| \bz_6^{[1]}\big\| = O_p \left( \frac{d}{n}\right).
\end{align*}

\subsubsection{Equation \eqref{eq:grave_inner}}

We now show that 
$$
\langle \grave{\bu}_j^{(1),[1]}, \bu_j^{(1)} \rangle = \sigma_j^2 + O_p\left(\frac{d}{n}\right).
$$
It follows immediately, by symmetry, that
\begin{align*}
\langle \grave{\bu}_j^{(1),[2]}, \bu_j^{(1)} \rangle = \sigma_j^2 + O_p\left(\frac{d}{n}\right).
\end{align*}

By definition,
\begin{align}
&  \langle \grave{\bu}_j^{(1),[1]}, \bu_j^{(1)} \rangle \nonumber \\
=& \sigma_j^2 \left( \frac{2}{n} \sum_{k=1}^{n/2}  \theta_{jk}^2 \right) \left( \prod_{q=2}^p \langle \bu_j^{(q)}, \hat{\bu}_j^{(q)[2]} \rangle \right)^2 \langle \bu_j^{(1)}, \grave{\bu}_j^{(1),[1]} \rangle  \nonumber \\
&+ \sum_{l \neq j} \sigma_{l} \left( \frac{2}{n} \sum_{k=1}^{n/2}  \theta_{j k} \theta_{l k} \right) \left( \prod_{q=2}^p \langle \bu_{j}^{(q)}, \hat{\bu}_j^{(q)[2]} \rangle \right) \left( \prod_{q=2}^p \langle \bu_{l}^{(q)}, \hat{\bu}_j^{(q)[2]} \rangle \right) \langle \bu_l^{(1)}, \grave{\bu}_j^{(1),[1]} \rangle  \nonumber \\
&+  \frac{2}{n} \sum_{k=1}^{n/2} \sigma_j \theta_{jk} \left( \prod_{q=2}^p \langle \bu_j^{(q)}, \hat{\bu}_j^{(q)[2]} \rangle \right) \scrE_k ( \grave{\bu}_j^{(1),[1]} ,\hat{\bu}_j^{(2)[2]},\dots,\hat{\bu}_j^{(p)[2]})   \nonumber \\
&+  \frac{2}{n} \sum_{k=1}^{n/2}  \sum_{l=1}^r \scrE_k (\bu_j^{(1)},\hat{\bu}_j^{(2)[2]},\dots,\hat{\bu}_j^{(p)[2]} ) \sigma_l \theta_{lk} \left( \prod_{q=2}^p \langle \bu_l^{(q)}, \hat{\bu}_j^{(q)[2]} \rangle \right)  \langle \bu_j^{(1)}, \grave{\bu}_j^{(1),[1]} \rangle  \nonumber \\
&+ \frac{2}{n} \sum_{k=1}^{n/2}   (\scrE_k \otimes \scrE_k - \scrI) (\bu_j^{(1)},\hat{\bu}_j^{(2)[2]},\dots,\hat{\bu}_j^{(p)[2]}, \grave{\bu}_j^{(1),[1]} ,\hat{\bu}_j^{(2)[2]},\dots,\hat{\bu}_j^{(p)[2]} ) .
\end{align}
All the terms except the first one will be bounded with similar techniques as we bound the six terms in  $\bz_j^{(1)[1]}$, and we omit some of the details.

\paragraph{The first term.} Observe that
$$
1- \langle \bu_j^{(q)}, \hat{\bu}_j^{(q)[2]} \rangle \le \big( \delta^{(1)[2]} \big)^2, \ 1- \langle \bu_j^{(1)}, \grave{\bu}_j^{(1)[1]} \rangle \le \big( \delta^{(1)[2]} \big)^2.
$$
Thus,
$$
\sigma_j^2 \left( \frac{2}{n} \sum_{k=1}^{n/2}  \theta_{jk}^2 \right) \left( \prod_{q=2}^p \langle \bu_j^{(q)}, \hat{\bu}_j^{(q)[2]} \rangle \right)^2 \langle \bu_j^{(1)}, \grave{\bu}_j^{(1),[1]} \rangle = \sigma_j^2 +O_p\left(\frac{d}{n} \right) .
$$

\paragraph{The second term.} 
\begin{align*}
	&\sum_{l \neq j} \sigma_{l} \left( \frac{2}{n} \sum_{k=1}^{n/2}  \theta_{j k} \theta_{l k} \right) \left( \prod_{q=2}^p \langle \bu_{j}^{(q)}, \hat{\bu}_j^{(q)[2]} \rangle \right) \left( \prod_{q=2}^p \langle \bu_{l}^{(q)}, \hat{\bu}_j^{(q)[2]} \rangle \right) \langle \bu_l^{(1)}, \grave{\bu}_j^{(1),[1]} \rangle \\
	=& O_p(\frac{1}{\sqrt{n}}) \cdot O_p \left[ \left( \sqrt{\frac{d}{n}} \right)^{2p-1} \right] \\=& o_p(\frac{1}{\sqrt{n}}).
\end{align*}

\paragraph{The third term.} 
Observe that
\begin{align*}
& \frac{2}{n} \sum_{k=1}^{n/2} \sigma_j \theta_{jk} \left( \prod_{q=2}^p \langle \bu_j^{(q)}, \hat{\bu}_j^{(q)[2]} \rangle \right) \scrE_k ( \grave{\bu}_j^{(1),[1]} ,\hat{\bu}_j^{(2)[2]},\dots,\hat{\bu}_j^{(p)[2]}) \\
=& \frac{2}{n} \sum_{k=1}^{n/2} \sigma_j \theta_{jk} \left( \prod_{q=2}^p \langle \bu_j^{(q)}, \hat{\bu}_j^{(q)[2]} \rangle \right) \scrE_k ( \bu_j^{(1),[1]} ,\hat{\bu}_j^{(2)[2]},\dots,\hat{\bu}_j^{(p)[2]}) \\
&+ \frac{2}{n} \sum_{k=1}^{n/2} \sigma_j \theta_{jk} \left( \prod_{q=2}^p \langle \bu_j^{(q)}, \hat{\bu}_j^{(q)[2]} \rangle \right) \scrE_k ( \grave{\bu}_j^{(1),[1]} - \bu_j^{(1),[1]},\hat{\bu}_j^{(2)[2]},\dots,\hat{\bu}_j^{(p)[2]}) ,
\end{align*}
in which the first term is $O_p(1/\sqrt{n})$ because $\theta_{jk} \scrE_k, k \in [n/2]$ are independent with $\hat{\bu}_j^{(2)[2]},\dots,\hat{\bu}_j^{(p)[2]}$, and the second term is bounded by
$$ \|\hat{\Sigma}_{\theta,\scrE}\| \cdot \delta^{(1)[2]}=O_p\big(\frac{d}{n}\big),$$
so
$$
\frac{2}{n} \sum_{k=1}^{n/2} \sigma_j \theta_{jk} \left( \prod_{q=2}^p \langle \bu_j^{(q)}, \hat{\bu}_j^{(q)[2]} \rangle \right) \scrE_k ( \grave{\bu}_j^{(1),[1]} ,\hat{\bu}_j^{(2)[2]},\dots,\hat{\bu}_j^{(p)[2]}) = O_p \left(d \over n\right) + O_p\left(1\over\sqrt{n}\right) = O_p \left(d \over n\right) .
$$
The last equality follows from the assumption $d=o(n)$.

\paragraph{The fourth term.}
Recall that $\hat{\bu}_j^{(q)[2]}$ is a function of the second half of the data, which is independent of the first half of the data, i.e., all the random variables with index $k \le n/2$, so conditional on the second half of the data, for any $l\in [r]$,
$$
\scrE_k (\bu_j^{(1)},\hat{\bu}_j^{(2)[2]},\dots,\hat{\bu}_j^{(p)[2]} ) \theta_{lk}
$$
are a product of two independent $N(0,\sigma_0^2)$ and $N(0,1)$ variables, and they are i.i.d across $k \le n/2$.
So we have
$$
\frac{2}{n} \sum_{k=1}^{n/2}  \sum_{l=1}^r \scrE_k (\bu_j^{(1)},\hat{\bu}_j^{(2)[2]},\dots,\hat{\bu}_j^{(p)[2]} ) \sigma_l \theta_{lk} \left( \prod_{q=2}^p \langle \bu_l^{(q)}, \hat{\bu}_j^{(q)[2]} \rangle \right) \langle \bu_j^{(1)}, \grave{\bu}_j^{(1),[1]} \rangle =O_p(\frac{1}{\sqrt{n}}).
$$

\paragraph{The fifth term.} Similar to the sixth term in $\bz_j^{(1)[1]}$,
$$
\frac{2}{n} \sum_{k=1}^{n/2}   (\scrE_k \otimes \scrE_k - \scrI) (\bu_j^{(1)},\hat{\bu}_j^{(2)[2]},\dots,\hat{\bu}_j^{(p)[2]}, \grave{\bu}_j^{(1),[1]} ,\hat{\bu}_j^{(2)[2]},\dots,\hat{\bu}_j^{(p)[2]} ) .
$$

Combining the five bounds above, we get
$$
\langle \grave{\bu}_j^{(1),[1]}, \bu_j^{(1)} \rangle = \sigma_j^2 + O_p\left(\frac{d}{n}\right) + O_p\left(1\over\sqrt{n}\right) = O_p \left(d \over n\right) .
$$
The last equality follows from the assumption $d=o(n)$.

\subsection{Step 3.} 

Recall that
$$
\hat{b}_j^{(q)[1]} := \left\langle \hat{\bu}_j^{(q),[1][1]}, \hat{\bu}_j^{(q),[1][2]} \right \rangle -1,
$$
$$
\hat{b}_j^{(q)[2]} := \left\langle \hat{\bu}_j^{(q),[2][1]}, \hat{\bu}_j^{(q),[2][2]} \right \rangle -1.
$$

In this part, we use Lemma \ref{p3kolt} to show that 
\begin{align}
\sqrt{n} \left( \hat{b}_j^{(1)[1]} - b_j^{(1)[1]} \right) \overset{p} \to 0 , \label{eq:hatb_b}
\end{align}
in which $b_j^{(1)[1]}$ is defined in Lemma \ref{linear_basic}.
Similarly $\sqrt{n} \left( \hat{b}_j^{(1)[2]} - b_j^{(1)[2]} \right) \overset{p} \to 0$. Then, recall that by definition \eqref{defbj1},
\begin{align*}
b_j^{(1)} = \frac{ \left\|\check{\bu}_j^{(1),[1]}+\check{\bu}_j^{(1),[2]}\right\| }{ \sqrt{1+b_j^{(1)[1]}}+\sqrt{1+b_j^{(1)[2]}} } -1,
\end{align*}
and by definition \eqref{eq:bias_general},
\begin{align*}
\hat{b}_j^{(1)} = 
\frac{ \left\|\check{\bu}_j^{(q),[1]}+\check{\bu}_j^{(q),[2]}\right\| }{ \sqrt{1+\hat{b}_j^{(1)[1]}} + \sqrt{1+\hat{b}_j^{(1)[2]}} } -1,
\end{align*}
we have that \eqref{eq:bb} follows from \eqref{eq:hatb_b} and the fact that $\left\|\check{\bu}_j^{(1),[1]}+\check{\bu}_j^{(1),[2]}\right\| \le 2$. Now we turn our attention to \eqref{eq:hatb_b}.

By Lemma \ref{thm1kolt},
\begin{align*}
&  \bar{\mathbb{E}} \| \hat{M}_j^{[1]} - M_j^{[1]} \|  \\
\le & C \| M_j^{[1]} \| \left( \sqrt{ \frac{r(M_j^{[1]})}{n} } \bigvee \frac{r(M_j^{[1]})}{n} \right)   \\
\le & \frac{C (\sigma_j^2 +1) (\sigma_j^2 + d_1 +1)}{(\frac{1}{2} \sigma_j^2 +1) n}  ,
\end{align*}
on events $\mathcal{B}_n$. ($M_j^{[1]}$, $\hat{M}_j^{[1]}$ and events $\mathcal{B}_n$ are defined at the beginning of the proof for Lemma \ref{linear_basic}.) The last inequality holds because of \eqref{ineqr} and \eqref{ineqnorm}. Remember that $\bar{g}_1 \ge \frac{1}{2} \sigma_j^2$ (inequality \eqref{ineqg1}), so with assumption $d=o(n)$, we have that for large enough $n$, $\bar{\mathbb{E}} \| \hat{M}_j^{[1]} - M_j^{[1]} \| \le \frac{1}{4} \bar{g}_1$, i.e., if we let $\gamma= \frac{1}{4}$, then
\begin{align*}
\bar{\mathbb{E}} \| \hat{M}_j^{[1]} - M_j^{[1]} \| \le \frac{(1-2\gamma)\bar{g}_1}{2}.
\end{align*}
$1+b_j^{(1)[1]} \ge 2 \gamma$ is also satisfied for large enough $n$ by inequality \eqref{ineqb1}.

Let $t= \left( \frac{n}{2r(M_j^{[1]})} \right)^{\frac{1}{3}}$. First note that on the event $\mathcal{B}_n$,
\begin{align*}
t \ge \left( \frac{n(\frac{1}{2} \sigma_j^2 +1)}{2(\sigma_j^2 + d_1 +1)} \right)^{\frac{1}{3}},
\end{align*}
by inequality \eqref{ineqr}, with assumption $d=o(n)$, we have $t \ge 1$ for large enough $n$.

Let $D:=D_{\frac{1}{4}}$ as in Lemma \ref{p3kolt}. We have
\begin{align*}
 D \| M_j^{[1]} \| \left( \sqrt{\frac{t}{n}} \bigvee \sqrt{\frac{\log n}{n}} \right) 
\le  D(\sigma_j^2 +1) \left( n^{-\frac{2}{3}} \bigvee \sqrt{\frac{\log n}{n}} \right) \le \frac{1}{8} \bar{g}_1
\end{align*} 
for large enough $n$.

So all the conditions in Lemma \ref{p3kolt} are satisfied with $\hat{M}_j^{[1]}$ and $M_j^{[1]}$, conditional on $\{ \scrX_k, k=n/2+1,\dots,n \}$ under events $\mathcal{B}_n$, for large enough $n$.

Observe that conditional on $\{ \scrX_k, k=n/2+1,\dots,n \}$, 
$\hat{b}_j^{(1)[1]}$ and $b_j^{(1)[1]}$ are just defined as the $\hat{b}_1$ and $b_1$ [defined in \eqref{b1} and \eqref{hatb1}] corresponding to $M_j^{[1]}$.  Now we can apply Lemma \ref{p3kolt}:
\begin{align}
& \sqrt{n} \left| \hat{b}_j^{(1)[1]} - b_j^{(1)[1]} \right| \nonumber \\
\le & D \sqrt{n} \frac{\| M_j^{[1]} \|^2}{\bar{g}_1^2} \left( \sqrt{\frac{r(M_j^{[1]})}{n}} \bigvee \sqrt{ \frac{t}{n} } \bigvee \sqrt{\frac{\log n}{n}} \right) \sqrt{ \frac{t}{n} } \nonumber \\
\le & D \frac{\| M_j^{[1]} \|^2}{\bar{g}_1^2} \left( \left(\frac{r(M_j^{[1]})}{n}\right)^{\frac{1}{3}} \bigvee \frac{1}{r(M_j^{[1]})^{\frac{1}{3}} n^{\frac{1}{6}} } \bigvee \frac{ \sqrt{\log n} }{ r(M_j^{[1]})^{\frac{1}{6}} n^{\frac{1}{3}}  } \right)  \nonumber \\
\le & D \frac{\| M_j^{[1]} \|^2}{\bar{g}_1^2} \left( \left(\frac{r(M_j^{[1]})}{n}\right)^{\frac{1}{3}} \bigvee \frac{1}{ n^{\frac{1}{6}} } \right)  \nonumber \\
\le & D \left( \frac{\sigma_j^2+1}{\frac{1}{2}\sigma_j^2} \right)^2 \left[ \left( \frac{\sigma_j^2 + d_1 +1}{n(\frac{1}{2} \sigma_j^2 +1)} \right)^{\frac{1}{3}} \bigvee \frac{1}{n^{\frac{1}{6}} } \ \right]  \label{032992}
\end{align}
conditional on $\{ \scrX_k, k=n/2+1,\dots,n \}$ under events $\mathcal{B}_n$, for large enough $n$, with probability at least $1-e^{-t}$. The last inequality holds because of \eqref{ineqnorm}, \eqref{ineqg1} and \eqref{ineqr}.

Since on the event $\mathcal{B}_n$, conditional on $\{ \scrX_{n/2+1},\dots,\scrX_{n} \}$, \eqref{032992} holds with probability at least $1-e^{-t} \ge 1 - \exp \left[ - \left( \frac{n(\frac{1}{2} \sigma_j^2 +1)}{2(\sigma_j^2 + d_1 +1)} \right)^{\frac{1}{3}} \right]$, we have that on the event $\mathcal{B}_n$, \eqref{032992} also holds with probability at least $1 - \exp \left[ - \left( \frac{n(\frac{1}{2} \sigma_j^2 +1)}{2(\sigma_j^2 + d_1 +1)} \right)^{\frac{1}{3}} \right]$. Combine with assumption $\frac{d}{n} \to 0$, we have that
\begin{align*}
\sqrt{n} \left( \hat{b}_j^{(1)[1]} - b_j^{(1)[1]} \right) \cdot \mathbb{I}(\mathcal{B}_n)
\overset{p}{\to} 0,
\end{align*}
which leads to
\begin{align*}
\sqrt{n} \left( \hat{b}_j^{(1)[1]} - b_j^{(1)[1]} \right)  \overset{p}{\to} 0.  
\end{align*}

\section{Proof of Lemma \ref{linear_basic}}

The proof of Lemma \ref{linear_basic} relies heavily on the techniques and results from \cite{kolt14_2} which we will review first.

\subsection{Preliminaries}

Let $\mathbb{H}$ be a Hilbert space and $M: \mathbb{H} \to \mathbb{H}$ be a compact symmetric nonnegative definite operator. It is well known that the following spectral representation holds
\begin{align*}
M = \sum_{r>1} \mu_r \calP_r
\end{align*}
with distinct non-zero eigenvalues $\mu_r$ arranged in decreasing order $\mu_1 > \mu_2 > \dots \ge 0$, and $\calP_r$ are the corresponding spectral projectors. The effective rank of $M$ is defined as
\begin{align*}
r(M) := \frac{{\rm tr}(M)}{ \|M\| }.
\end{align*}
We will use in particular the results from \cite{kolt14_2} for the estimation of $\calP_1$, in the case where 
\begin{align*}
\calP_1 = \bu_1 \otimes \bu_1,
\end{align*}
i.e., estimating the leading eigenvector in the case that the leading eigenvalue is an isolated simple eigenvalue. 
Let $Y_1, Y_2, \dots, Y_n$ be i.i.d. centered Gaussian random vectors in $\mathbb{R}^m$ with covariance $M = \mathbb{E}(Y \otimes Y)$. Let 
\begin{align*}
\hat{M} := \frac{1}{n} \sum_{k=1}^n Y_k \otimes Y_k
\end{align*}
be the sample covariance matrix based on the observations $(Y_1, Y_2, \dots, Y_n)$. The following lemma is a restatement of Theorem 1 from \cite{kolt14_2}.

\begin{lemma}
	\label{thm1kolt}
	\begin{align*}
		\mathbb{E} \| \hat{M} - M \| \asymp \| M \| \left( \sqrt{ \frac{r(M)}{n} } \bigvee \frac{r(M)}{n} \right) ,
	\end{align*}
	and
	\begin{align*}
		\mathbb{E} \| \hat{M} - M \|^2 \asymp \| M \|^2 \left( \sqrt{ \frac{r(M)}{n} } \bigvee \frac{r(M)}{n} \right)^2 .
	\end{align*}
\end{lemma}

Let $\hat{\bu}_1$ be the leading eigenvector of $\hat{M}$. Without loss of generality, to make the linear form of $\hat{\bu}_1$ well-defined, we always assume that $\langle \hat{\bu}_1, \bu_1 \rangle \ge 0$. Denote $\hat{\calP}_1 := \hat{\bu}_1 \otimes \hat{\bu}_1$. Define $\bar{g}_1:=\mu_1 - \mu_2$, the spectral gap of $\mu_1$, and write
\begin{align}
\calC_1 = \sum_{s \neq 1} \frac{1}{\mu_1 - \mu_s} \calP_s,
\label{koltdefc1}
\end{align}
\begin{align}
\calL_1 &:= \calC_1(\hat{M} - M)\calP_1 + \calP_1(\hat{M} - M)\calC_1 = \frac{1}{n} \sum_{j=1}^n (\calC_1 Y_j \otimes \calP_1 Y_j + \calP_1 Y_j \otimes \calC_1 Y_j) , \label{koltdefl1} \\
\calS_1 &:= \hat{\calP}_1 - \calP_1 - \calL_1 , \label{koltdefs1}
\end{align}
and the remainder in terms of operator
\begin{align}
\calR_1 := \hat{\calP}_1 - \mathbb{E} \hat{\calP}_1 - \calL_1. \label{koltdefr1}
\end{align}
Note that $\mathbb{E} \calL_1 =0$, so $\calR_1 = \calS_1 -\mathbb{E} \calS_1$. As in the proof of Theorem \ref{thm_b}, we use calligraphic capital letters on $\hat{\calP}_1$, $\calP_1$, $\calC_1$, $\calL_1$, $\calS_1$ and $\calR_1$ to signify that they are matrices. Rephrasing Lemma 1 in \cite{kolt14_2}, we have

\begin{lemma}
	\label{lemma1kolt}
	\begin{align}
		\| \calS_1 \| \le 14 \left( \frac{ \| \hat{M} - M \| }{\bar{g}_1} \right)^2
	\end{align}
\end{lemma}

Combine Lemma \ref{lemma1kolt} and \ref{thm1kolt}, we have
\begin{align}
	\mathbb{E} \| \calS_1 \| \le C  \frac{\| M \|^2}{ \bar{g}_1^2 } \left( \sqrt{ \frac{r(M)}{n} } \bigvee \frac{r(M)}{n} \right)^2 ,
	\label{ineqs1}
\end{align}
where $C$ is a universal constant. Restating Theorems 3 and 4 of \cite{kolt14_2}, we get

\begin{lemma}
	\label{thm3kolt}
Let $t>1$ and suppose that, for some $\gamma \in (0,1)$ and a sufficiently large constant $C>0$,
	\begin{align}
		\mathbb{E} \| \hat{M} - M \| + C \| M \| \sqrt{\frac{t}{n}} \le \frac{1-\gamma}{1+\gamma} \frac{\bar{g}_1}{2}.
	\end{align}
Then there exists a constant $D_{\gamma} > 0$ such that, for all $u,v \in \mathbb{H}$, the following bound holds with probability at least $1-e^{-t}$ :
	\begin{align}
		\left| \big\langle (\hat{\calP}_1 - \mathbb{E} \hat{\calP}_1 - \calL_1) u,v \big\rangle \right| \le D_{\gamma} \frac{\| M \|^2}{\bar{g}_1^2} \left( \sqrt{\frac{r(M)}{n}} \bigvee \sqrt{\frac{t}{n}} \right) \sqrt{\frac{t}{n}}  \|u\| \|v\|.
	\end{align}
\end{lemma}

\begin{lemma}
	\label{thm4kolt}
Suppose that for some $\gamma \in (0,1)$ and a sufficiently large constant $C>0$,
	\begin{align}
		\mathbb{E} \| \hat{M} - M \| + C \|M\| \frac{\log n}{n} \le (1-\gamma) \frac{\bar{g}_1}{2}.
	\end{align}
	Then, there exists a constant $D_{\gamma} > 0$ such that
	\begin{align}
		\| \mathbb{E} \hat{\calP}_1 - \calP_1 - \calP_1 \mathbb{E}(\calS_1) \calP_1 \|  \le D_{\gamma} \frac{\| M \|^2}{\bar{g}_1^2} \frac{1}{\sqrt{n}} \left( \sqrt{\frac{r(M)}{n}} \bigvee \sqrt{\frac{\log n}{n}} \right).
	\end{align}
\end{lemma}

With $\calS_1$ defined, we can define a critical quantity that characterizes the bias of $\hat{\calP}_1$:
\begin{align}
b_1 := \left\langle \mathbb{E}(\calS_1) \bu_1, \bu_1 \right\rangle.  \label{b1}
\end{align}
Note that $\mathbb{E} \calL_1 =0$ and $\hat{\calP}_1 := \hat{\bu}_1 \otimes \hat{\bu}_1$, in which $\langle \hat{\bu}_1, \bu_1 \rangle \ge 0$, we have
\begin{align*}
b_1 = \left\langle \mathbb{E}(\hat{\bu}_1 \otimes \hat{\bu}_1) \bu_1, \bu_1 \right\rangle -1,
\end{align*}
so $-1 \le b_1 \le 0$.

% In the context of our proof, it will be a random variable that is $o_p(1)$.

There is a way to estimate $b_1$. Suppose we divide the sample $(Y_1, Y_2, \dots, Y_n)$ into two subsamples of sample size $\lfloor \frac{n}{2} \rfloor$ each. Let $\check{M}$ be the sample covariance based on the first subsample and $\check{M}'$ be the sample covariance based on the second subsample. Denote by $\check{\bu}_1$ the leading eigenvector of $\check{M}$ and by $\check{\bu}_1'$ the leading eigenvector of $\check{M}'$. Assume that their signs are chosen in such a way that $\langle \check{\bu}_1, \check{\bu}_1' \rangle \ge 0$. Define
\begin{align}
\hat{b}_1 := \langle \check{\bu}_1, \check{\bu}_1' \rangle -1. \label{hatb1}
\end{align}
The following lemma, a restatement of Proposition 3 from \cite{kolt14_2}, provides a concentration inequality of $|\hat{b}_1-b_1|$. 

\begin{lemma}[]
	\label{p3kolt}
Let $t\ge 1$ and $\gamma \in (0,1/2)$. There exists a constant $D_{\gamma} > 0$ such that, if
	\begin{align}
		\mathbb{E} \| \hat{M} - M \| \le \frac{(1-2\gamma)\bar{g}_1}{2},\ 1+b_1 \ge 2 \gamma 
	\end{align}
	and
	\begin{align}
		D_{\gamma} \| M \| \left( \sqrt{\frac{t}{n}} \bigvee \sqrt{\frac{\log n}{n}} \right) \le \frac{\gamma \bar{g}_1}{2},\end{align}
	then with probability at least $1-e^{-t}$,
	\begin{align}
		| \hat{b}_1 - b_1 | \le D_{\gamma} \frac{\| M \|^2}{\bar{g}_1^2} \left( \sqrt{\frac{r(M)}{n}} \bigvee \sqrt{ \frac{t}{n} } \bigvee \sqrt{\frac{\log n}{n}} \right) \sqrt{ \frac{t}{n} } .
	\end{align}
\end{lemma}

\paragraph{Proof outline.}

Without loss of generality, we assume $ \|\bu\|, \|\bv\| = 1$ throughout the proof. Write
\begin{align}
\calL_j := \frac{2}{n} \sum_{k=1}^{n/2} \left( \calC_j y_{jk} \otimes \calP_j y_{jk} + \calP_j y_{jk} \otimes \calC_j y_{jk} \right) ,
\end{align}
then
\begin{align}
\langle \calL_j \bu, \bv \rangle = \frac{2}{n} \sum_{k=1}^{n/2} \left[ \langle y_{jk},\calP_j \bv \rangle \langle y_{jk},\calC_j \bu \rangle + \langle y_{jk},\calP_j \bu \rangle \langle y_{jk},\calC_j \bv \rangle \right].  \label{luv}
\end{align}
Now Lemma \ref{linear_basic} is equivalent to: there exists universal constant $C$ such that
\begin{align}
\mathbb{P} ( | b_j^{(q)[1]} | \le C \frac{d}{n} ) \to 1,   \label{eq:b_upper}
\end{align}
and for $\forall \bu,\bv \in \mathbb{H}$, $\|\bu\|, \|\bv\| \le 1$,
\begin{align}
\sqrt{\frac{n}{2}} \bigg\langle \left[\hat{\calP}_j - \calP_j - \calP_j \bar{\mathbb{E}}(\calS^{[1]}) \calP_j  - \calL_j \right] \bu,\bv \bigg\rangle   \overset{p}{\to} 0.   \label{eq:linear_basic_proof}
\end{align}

We separate the proof for \eqref{eq:linear_basic_proof} into three parts:
\begin{align}
\sqrt{\frac{n}{2}}  \big\langle (\hat{\calP}_j - \bar{\mathbb{E}} \hat{\calP}_j - \calL^{[1]}) u,v \big\rangle  \overset{p}{\to} 0,  \label{part1finished}
\end{align}
\begin{align}
\sqrt{\frac{n}{2}} \left\| \bar{\mathbb{E}} \hat{\calP}_j - \calP_j - \calP_j \bar{\mathbb{E}}(S^{[1]}) \calP_j \right\| \overset{p}{\to} 0, \label{part2finished}
\end{align}
\begin{align}
\sqrt{\frac{n}{2}} \big\langle ( \calL^{[1]} - \calL_j) u,v \big\rangle \overset{p}{\to} 0.  \label{part3finished}
\end{align}

We will first prove some preliminary bounds and \eqref{eq:b_upper}, and then come back to \eqref{part1finished}, \eqref{part2finished} and \eqref{part3finished} to complete the proof.

%\paragraph{Notations and preliminary bounds}
\subsection{Proof for \eqref{eq:b_upper}.}

Upper bounds \eqref{eq:rate_main} and \eqref{eq:rate_main1} imply that
$\hat{\bu}_j^{(q)[2]}$ satisfies the following conditions:
\begin{align}
\lim_{n \to \infty} \mathbb{P} \left( \| \hat{\bu}_j^{(q)[2]} - \bu_j^{(q)} \| \le a_n, \ \forall q \in [p] \right) = 1, \label{eq:big_lemma_a1}
\end{align}
where $a_n$ is a numeric sequence such that $ \lim_{n \to \infty} a_n =0$, and
\begin{align}
\lim_{n \to \infty} \mathbb{P} \left( \max_{l \neq j} \left| \sigma_l^2 \langle \bu_l^{(q)}, \hat{\bu}_j^{(q)[2]} \rangle \right| \le A, \ \forall q \in [p] \right) = 1, \label{eq:big_lemma_a2}
\end{align}
where $A$ is a numeric constant. Define events 
$$\mathcal{B}_n:= \left\{ \| \hat{\bu}_j^{(q)[2]} - \bu_j^{(q)} \| \le a_n, \ \| \hat{\bu}_j^{(q)[2]} - \bu_j^{(q)} \| \le a_n, \ \forall q \in [p] \right\}.$$
By \eqref{eq:big_lemma_a1} and \eqref{eq:big_lemma_a2}, $\lim_{n \to \infty} \mathbb{P} (\mathcal{B}_n) =1$.
Note that event $\mathcal{B}_n$ belongs to the sigma field of $\{ \scrX_{n/2+1},\scrX_{n/2+2},\dots,\scrX_{n} \}$, so it suffices to treat $\bar{\mathbb{P}}(\cdot)$ and $\bar{\mathbb{E}}(\cdot)$ as conditional on the event $\mathcal{B}_n$. Since events $\mathcal{B}_n$ satisfy $\lim_{n \to \infty} \mathbb{P} (\mathcal{B}_n) =1$, for any sequence of random variables $Z_n$, to prove $ Z_n \overset{p}{\to} 0$, we only need to show $ Z_n \cdot \mathbb{I}(\mathcal{B}_n) \overset{p}{\to} 0$, where $\mathbb{I}(\mathcal{B}_n)$ is the indicator function of event $\mathcal{B}_n$. We will use this technique extensively.

We only need to prove for the case $q=1$. From now on till the end of this proof, for simplicity of notations, we denote $\calP_j = \calP_j^{(1)}$, $\calC_j = \calC_j^{(1)}$, $ \hat{\calP}_j =\hat{\calP}_j^{(1)}$, 
$M_j^{[1]} = M^{(1)[1]}$, $\calC_j^{[1]} = \calC_j^{(1)[1]}$, $\calL_j^{[1]} = \calL_j^{(1)[1]}$, $\calS_j^{[1]} = \calS_j^{(1)[1]}$. 
Furthermore, write \begin{align}
 \calP_{-} = I_{d_q} - \sum_{l=1}^r \bu_l^{(q)} \otimes \bu_l^{(q)}
\end{align}
Note that $\calC_j^{[1]}, \calL_j^{[1]}, \calS_j^{[1]}, R^{[1]}$ are the $\calC_1, \calL_1, \calS_1, \calR_1$ [defined in \eqref{koltdefc1}, \eqref{koltdefl1}, \eqref{koltdefs1}, \eqref{koltdefr1}] corresponding to our covariance matrix $M_j^{[1]}$. Conditional on $\{ \scrX_k, k=n/2+1,\dots,n \}$, 
\begin{align*}
z_{jk} := \scrX_k \times_2 \hat{\bu}_j^{(1)[2]} \dots \times_p \hat{\bu}_j^{(p)[2]}, k =1,\dots,n/2
\end{align*}
has covariance matrix $M_j^{[1]}$, and $ \hat{\bu}_j^{(q)[1]}$ is the leading eigenvector of the sample covariance matrix
\begin{align}
\hat{M}_j^{[1]} = \frac{2}{n} \sum_{k=1}^{n/2} z_{jk} \otimes z_{jk}.
\end{align}
We first prove some inequalities for $\|M_j^{[1]}\|$, $\bar{g}_1$ the first spectral gap of $M_j^{[1]}$, and $r(M_j^{[1]})$ the effective rank of $M_j^{[1]}$, under event $\mathcal{B}_n$. These inequalities will be used extensively throughout the proof. 

Under $\mathcal{B}_n$, since $\| \hat{\bu}_j^{(q)[2]} - \bu_j^{(q)} \| \le a_n $, we have $ (1-a_n^2)^{\frac{p}{2}} \sigma_j^2 \le \tilde{\sigma}_j^2 \le \sigma_j^2$. Moreover, because $ \max_{l \neq j} \left| \sigma_l^2 \langle \bu_l^{(2)}, \hat{\bu}_j^{(2)[2]} \rangle \right| \le A $ is bounded, and  $ \left| \langle \bu_l^{(2)}, \hat{\bu}_j^{(2)[2]} \rangle \right| \le \sqrt{1- \langle\hat{\bu}_j^{(2)[2]} , \bu_j^{(2)} \rangle^2 } \le a_n $, we have $\max_{l \neq j} \{ \tilde{\sigma}_l^2 \} \le A a_n $. Since $\lim_{n\to \infty} a_n =0$, for large enough $n$, we have $ (1-a_n^2)^{\frac{p}{2}} \sigma_j^2 > A a_n$. So the leading eigenvector of $M_j^{[1]}$ is $\bu_j^{(1)}$, with corresponding eigenvalue $ \tilde{\sigma}_j^2 + 1$. So for large enough $n$,
\begin{align}
\frac{1}{2} \sigma_j^2 +1 \le (1-a_n^2)^{\frac{p}{2}} \sigma_j^2 +1 \le \| M_j^{[1]} \| \le \sigma_j^2 +1,  \label{ineqnorm}
\end{align}
and the spectral gap for the leading eigenvalue of $M_j^{[1]}$ 
\begin{align}
\bar{g}_1 \ge (1-a_n^2)^{\frac{p}{2}} \sigma_j^2 - A a_n.
\end{align}
Since $\lim_{n\to \infty} a_n =0$, we have for large enough $n$,
\begin{align}
\bar{g}_1 \ge \frac{1}{2} \sigma_j^2.  \label{ineqg1}
\end{align}
By definition, 
\begin{align}
r(M_j^{[1]}) = \frac{ {\rm tr}(M_j^{[1]}) }{ \|M_j^{[1]}\| }.
\end{align}
Combine with \eqref{ineqnorm}, and 
\begin{align}
&{\rm tr}(M_j^{[1]}) \nonumber \\
=& \sum_{l \neq j} \sigma_l^2 \left( \prod_{q=2}^p \langle \bu_l^{(q)}, \hat{\bu}_j^{(q)[2]} \rangle \right)^2 + \tilde{\sigma}_j^2 + d_1 \nonumber \\
\le & \max_{l \neq j} \left| \sigma_l^2 \langle \bu_l^{(2)}, \hat{\bu}_j^{(2)[2]} \rangle \right| + \sigma_j^2 + d_1  \nonumber \\
\le & A a_n + \sigma_j^2 + d_1 \nonumber
\end{align} 
where the first inequality is by Cauchy inequality. So for large enough $n$,
\begin{align}
\frac{ d_1 }{\sigma_j^2 +1} \le r(M_j^{[1]}) \le \frac{\sigma_j^2 + d_1 +1}{\frac{1}{2} \sigma_j^2 +1}. \label{ineqr}
\end{align}

%\paragraph{Proof for \eqref{eq:b_upper}.}

Combine inequalities \eqref{ineqs1} from Lemma \ref{thm3kolt} with bounds \eqref{ineqnorm}, \eqref{ineqg1} and \eqref{ineqr}, we have that under event $\mathcal{B}_n$, for large enough $n$,
\begin{align}
\left| b_j^{(1)[1]} \right| \le \left\| \bar{\mathbb{E}}(\calS_j^{[1]}) \right\| \le \bar{\mathbb{E}} \left\| \calS_j^{[1]} \right\| \le C  \frac{(\sigma_j^2 +1)^2}{ \frac{1}{4} \sigma_j^4 } \left( \sqrt{ \frac{\sigma_j^2 + d_1 +1}{n(\frac{1}{2} \sigma_j^2 +1)} } \bigvee \frac{\sigma_j^2 + d_1 +1}{n(\frac{1}{2} \sigma_j^2 +1)} \right)^2 ,  \label{ineqb1}
\end{align}
since we treat $\sigma_j$ as fixed, the right hand side is bounded by $C \left(\sqrt{\frac{d}{n}} \bigvee \frac{d}{n} \right)^2\le C \frac{d}{n}$.
Recall that $\lim_{n \to \infty} \mathbb{P} (\mathcal{B}_n) =1$, we have $ \mathbb{P} ( | b_j^{(1)[1]} | \le C \frac{d}{n} ) \to 1 $ as $n \to \infty$.

%\paragraph{Proof for \eqref{part1finished}.}
\subsection{Proof for \eqref{part1finished}.}

In this part, we will use Lemma \ref{thm3kolt} to show that 
\begin{align*}
\sqrt{\frac{n}{2}} \big\langle (\hat{\calP}_j - \bar{\mathbb{E}} \hat{\calP}_j - \calL_j^{[1]}) \bu,\bv \big\rangle  \overset{p}{\to} 0.  
\end{align*}
Let $t= \left( \frac{n}{2 r(M_j^{[1]})} \right)^{\frac{1}{3}}$. First note that on the event $\mathcal{B}_n$, by inequality \eqref{ineqr},
\begin{align*}
t \ge \left( \frac{n(\frac{1}{2} \sigma_j^2 +1)}{2(\sigma_j^2 + d_1 +1)} \right)^{\frac{1}{3}}\ge 1,
\end{align*}
for large enough $n$. By Lemma \ref{thm1kolt},
\begin{align*}
\delta_n(t) &=  \bar{\mathbb{E}} \| \hat{M}_j^{[1]} - M_j^{[1]} \| + C \| M_j^{[1]} \| \sqrt{\frac{t}{n}}  \\
& \le  C \| M_j^{[1]} \| \left( \sqrt{ \frac{r(M_j^{[1]})}{n} } \bigvee \frac{r(M_j^{[1]})}{n} \right) + C \| M_j^{[1]} \| \sqrt{\frac{t}{n}}  \\
& \le C \| M_j^{[1]} \| \left[ \frac{4(\sigma_j^2 + d_1 +1)}{(\frac{1}{2} \sigma_j^2 +1) n} + \sqrt{ \frac{ 2 \left( \frac{n}{2 r(M_j^{[1]})} \right)^{\frac{1}{3}} }{n} } \right] \\
& \le C (\sigma_j^2 +1) \left[ \frac{\sigma_j^2 + d_1 +1}{(\frac{1}{2} \sigma_j^2 +1) n} + n^{-\frac{1}{3}} \right] ,
\end{align*}
on events $\mathcal{B}_n$. The last inequality holds because of \eqref{ineqr} and \eqref{ineqnorm}. Recall that $\bar{g}_1 \ge \frac{1}{2} \sigma_j^2$. We have that for large enough $n$, $\delta_n(t) \le \frac{1}{6} \bar{g}_1$.

Now, by Lemma \ref{thm3kolt} with $t= \left( \frac{n}{2 r(M_j^{[1]})} \right)^{\frac{1}{3}}$ and $\gamma=\frac{1}{2}$, there exists a constant $D:=D_{\frac{1}{2}}$ such that on the event $\mathcal{B}_n$, conditional on $\{ \scrX_{n/2+1},\scrX_{n/2+2},\dots,\scrX_{n} \}$, with probability at least $1-e^{-t} \ge 1 - \exp \left[ - \left( \frac{n(\frac{1}{2} \sigma_j^2 +1)}{2(\sigma_j^2 + d_1 +1)} \right)^{\frac{1}{3}} \right]$ :
\begin{align}
& \sqrt{\frac{n}{2}} \left| \big\langle (\hat{\calP}_j - \bar{\mathbb{E}} \hat{\calP}_j - \calL_j^{[1]}) \bu,\bv \big\rangle \right|  \nonumber \\
\le & D \sqrt{\frac{n}{2}} \frac{\| M_j^{[1]} \|^2}{\bar{g}_1^2} \left( \sqrt{\frac{r(M_j^{[1]})}{n}} \bigvee \sqrt{\frac{t}{n}} \right) \sqrt{\frac{t}{n}}  \nonumber \\
=& D \frac{\| M_j^{[1]} \|^2}{\bar{g}_1^2} \left[ \left( \frac{r(M_j^{[1]})}{n} \right)^{\frac{1}{3}} \bigvee \frac{1}{r(M_j^{[1]})^{\frac{1}{3}} (n/2)^{\frac{1}{6}} } \ \right]  \nonumber \\
\le & 4 D \left( \frac{\sigma_j^2+1}{\frac{1}{2}\sigma_j^2} \right)^2 \left[ \left( \frac{\sigma_j^2 + d_1 +1}{n(\frac{1}{2} \sigma_j^2 +1)} \right)^{\frac{1}{3}} \bigvee \frac{1}{n^{\frac{1}{6}} } \ \right],  \label{5091342}
\end{align}
conditional on $\{ \scrX_k, k=n/2+1,\dots,n \}$ under events $\mathcal{B}_n$, for large enough $n$, with probability at least $1-e^{-t}$. The last inequality holds because of \eqref{ineqnorm}, \eqref{ineqg1} and \eqref{ineqr}.

Since on the event $\mathcal{B}_n$, conditional on $\{ \scrX_{n/2+1},\scrX_{n/2+2},\dots,\scrX_{n} \}$, \eqref{5091342} holds with probability at least $1 - \exp \left[ - \left( \frac{n(\frac{1}{2} \sigma_j^2 +1)}{2(\sigma_j^2 + d_1 +1)} \right)^{\frac{1}{3}} \right]$, we have that on the event $\mathcal{B}_n$, \eqref{5091342} also holds with probability at least $1 - \exp \left[ - \left( \frac{n(\frac{1}{2} \sigma_j^2 +1)}{2(\sigma_j^2 + d_1 +1)} \right)^{\frac{1}{3}} \right]$. Combining with the assumption that ${d}/{n} \to 0$, we have
\begin{align*}
\sqrt{\frac{n}{2}} \left| \big\langle (\hat{\calP}_j - \bar{\mathbb{E}} \hat{\calP}_j - \calL_j^{[1]}) \bu,\bv \big\rangle \right| \cdot \mathbb{I}(\mathcal{B}_n)
\overset{p}{\to} 0.
\end{align*}

%\paragraph{Proof for \eqref{part2finished}.}
\subsection{Proof for \eqref{part2finished}.}

In this part, we will use Lemma \ref{thm4kolt} to show that 
\begin{align*}
\sqrt{\frac{n}{2}} \left\| \bar{\mathbb{E}} \hat{\calP}_j - \calP_j - \calP_j \bar{\mathbb{E}}(\calS_j^{[1]}) \calP_j \right\| \overset{p}{\to} 0.  
\end{align*}
By Lemma \ref{thm1kolt},
\begin{align*}
&  \bar{\mathbb{E}} \| \hat{M}_j^{[1]} - M_j^{[1]} \| + C \| M_j^{[1]} \| \frac{\log n}{n}  \\
\le & C \| M_j^{[1]} \| \left( \sqrt{ \frac{r(M_j^{[1]})}{n} } \bigvee \frac{r(M_j^{[1]})}{n} \right) + C \| M_j^{[1]} \| \frac{\log n}{n}  \\
\le & C (\sigma_j^2 +1) \left[ \frac{\sigma_j^2 + d_1 +1}{(\frac{1}{2} \sigma_j^2 +1) n} + \frac{\log n}{n}  \right] ,
\end{align*}
on events $\mathcal{B}_n$. The last inequality holds because of \eqref{ineqr} and \eqref{ineqnorm}. Remember that $\bar{g}_1 \ge \frac{1}{2} \sigma_j^2$ (inequality \ref{ineqg1}), since $d=o(n)$, we have that for large enough $n$, $\delta_n(t) \le \frac{1}{4} \bar{g}_1$.

Now, by Lemma \ref{thm4kolt}, with $\gamma=\frac{1}{2}$, there exists a constant $D:=D_{\frac{1}{2}}$ such that on the event $\mathcal{B}_n$:
\begin{align}
& \sqrt{\frac{n}{2}} \left\| \bar{\mathbb{E}} \hat{\calP}_j - \calP_j - \calP_j \bar{\mathbb{E}}(\calS_j^{[1]}) \calP_j \right\| \nonumber \\
=& D \sqrt{n/2} \frac{\| M_j^{[1]} \|^2}{\bar{g}_1^2} \frac{1}{\sqrt{n/2}} \left( \sqrt{\frac{r(M_j^{[1]})}{n}} \bigvee \sqrt{\frac{\log n}{n}} \right) \nonumber \\
\le & 4 D \left( \frac{\sigma_j^2+1}{\frac{1}{2}\sigma_j^2} \right)^2 \left[ \sqrt{ \frac{\sigma_j^2 + d_1 +1}{n(\frac{1}{2} \sigma_j^2 +1)} } \bigvee \sqrt{\frac{\log n}{n}} \right], 
\end{align}
for large enough $n$. The last inequality holds because of \eqref{ineqnorm}, \eqref{ineqg1} and \eqref{ineqr}. Equation \eqref{part2finished} then follows.

%\paragraph{Proof for \eqref{part3finished}.}
\subsection{Proof for \eqref{part3finished}.}

Note that for $\bu,\bv \in \mathbb{H}$,
\begin{align}
\langle \calL_j^{[1]} \bu, \bv \rangle = \frac{2}{n} \sum_{k=1}^{n/2} \left[ \langle z_{jk},\calP_j \bv \rangle \langle z_{jk},\calC_j^{[1]} \bu \rangle + \langle z_{jk},\calP_j \bu \rangle \langle z_{jk},\calC_j^{[1]} \bv \rangle \right],
\end{align}
and remember that $\langle L_j \bu, \bv \rangle = \frac{2}{n} \sum_{k=1}^{n/2} \left[ \langle y_{jk},\calP_j \bv \rangle \langle y_{jk},\calC_j \bu \rangle + \langle y_{jk},\calP_j \bu \rangle \langle y_{jk},\calC_j \bv \rangle \right]$, so to prove \eqref{part3finished}, we only need to show: for $\forall \bu, \bv \in \mathbb{H}$, $\|\bu\|, \|\bv\| \le 1$,
\begin{align}
\sqrt{\frac{2}{n}} \sum_{k=1}^{n/2} \left[ \langle z_{jk},\calP_j \bv \rangle \langle z_{jk},\calC_j^{[1]} \bu \rangle - \langle y_{jk},\calP_j \bv \rangle \langle y_{jk},\calC_j \bu \rangle \right] \overset{p}{\to} 0. \label{part3main}
\end{align}
We further break the proof for \eqref{part3main} into three steps.

\paragraph{Step 1.}

\begin{align}
&\langle z_{jk},\calP_j \bv \rangle \langle z_{jk},\calC_j^{[1]} \bu \rangle - \langle y_{jk},\calP_j \bv \rangle \langle z_{jk},\calC_j^{[1]} \bu \rangle  \nonumber \\
=& \langle z_{jk} - y_{jk},\calP_j \bv \rangle \langle z_{jk},\calC_j^{[1]} \bu \rangle
\end{align}
but conditional on $\{ \scrX_{n/2+1},\scrX_{n/2+2},\dots,\scrX_{n} \}$, $\langle z_{jk} - y_{jk},\calP_j \bv \rangle$ and  $\langle z_{jk},\calC_j^{[1]} \bu \rangle$, $k=1,\dots,n$ are mean-$0$ Gaussian random variables and uncorrelated hence independent. So $\bar{\mathbb{E}}( \langle z_{jk} - y_{jk},\calP_j \bv \rangle \langle z_{jk},\calC_j^{[1]} \bu \rangle ) = 0$. Moreover, direct calculation gives us
\begin{align}
&\bar{\mathbb{E}} \langle z_{jk} - y_{jk},\calP_j \bv \rangle ^2 \nonumber \\
=& \langle \bu_j^{(1)}, \bv \rangle^2 \bar{\mathbb{E}} \bigg[ \scrX_k \times_1 \bu_j^{(1)} \times_2 \hat{\bu}_j^{(2)[2]} \dots \times_p \hat{\bu}_j^{(p)[2]}  - \scrX_k \times_1 \bu_j^{(1)} \times_2 \bu_j^{(2)} \dots \times_p \bu_j^{(p)}  \bigg]^2  \nonumber \\
=& \langle \bu_j^{(1)}, \bv \rangle^2 \sigma_j^2 \prod_{q=2}^r \left( 1- \langle \hat{\bu}_j^{(q)[2]}, \bu_j^{(q)} \rangle \right)^2 + \langle \bu_j^{(1)}, \bv \rangle^2 \left\| \hat{\bu}_j^{(2)[2]} \otimes \dots \otimes \hat{\bu}_j^{(p)[2]} - \bu_j^{(2)} \otimes \dots \otimes \bu_j^{(p)} \right\|^2  \nonumber \\
=& \langle \bu_j^{(1)}, \bv \rangle^2 \sigma_j^2 \prod_{q=2}^r \left( 1- \langle \hat{\bu}_j^{(q)[2]}, \bu_j^{(q)} \rangle \right)^2 + \langle \bu_j^{(1)}, \bv \rangle^2 \left( 2- 2 \prod_{q=2}^r \langle \hat{\bu}_j^{(q)[2]}, \bu_j^{(q)} \rangle \right)  \nonumber \\
\le& \sigma_j^2 \left( \frac{1}{4} a_n^4 \right)^{p-1} + 2- 2 \left(1-\frac{1}{2}a_n^2\right)^{p-1}
\end{align}
since $\| \hat{\bu}_j^{(q)[2]} - \bu_j^{(q)} \| \le a_n $ which implies $1- \langle \hat{\bu}_j^{(q)[2]}, \bu_j^{(q)} \rangle \le \frac{1}{2} a_n^2$.
Also we have
\begin{align}
\bar{\mathbb{E}} \left[ \langle z_{jk},\calC_j^{[1]} \bu \rangle \right]^2 = \langle \calC_j^{[1]} M_j^{[1]} \calC_j^{[1]} \bu, \bu \rangle ,
\end{align}
where
\begin{align}
\calC_j^{[1]} M_j^{[1]} \calC_j^{[1]}:=\sum_{l \neq j} \frac{\tilde{\sigma}_l^2}{\left( \tilde{\sigma}_j^2 - \tilde{\sigma}_l^2 \right)^2} \bu_l^{(1)} \otimes \bu_l^{(1)} + \frac{1}{\tilde{\sigma}_j^4}  \calP_{-}.
\end{align}
As already proven, $ (1-a_n^2)^{\frac{p}{2}} \sigma_j^2 \le \tilde{\sigma}_j^2$, and $\max_{l \neq j} \{ \tilde{\sigma}_l^2 \} \le A a_n \to 0$, and for large enough $n$, $ \tilde{\sigma}_j^2 - \tilde{\sigma}_l^2 \ge \frac{1}{2} \sigma_j^2$, so for large enough $n$, we have
\begin{align}
\left\| \calC_j^{[1]} M_j^{[1]} \calC_j^{[1]} \right\| \le \frac{1}{ \left[ (1-a_n^2)^{\frac{p}{2}} \sigma_j^2 \right]^4 }.
\end{align}
Thus,
\begin{align}
&\bar{\mathbb{E}} \left[ \langle z_{jk} - y_{jk},\calP_j \bv \rangle \langle z_{jk},\calC_j^{[1]} \bu \rangle \right]^2  \nonumber\\
\le& \frac{1}{ \left[ (1-a_n^2)^{\frac{p}{2}} \sigma_j^2 \right]^4 }  \left[ \sigma_j^2 \left( \frac{1}{4} a_n^4 \right)^{p-1} + 2- 2 \left(1-\frac{1}{2}a_n^2\right)^{p-1} \right] .  \label{5351611}
\end{align}

\paragraph{Step 2.}

\begin{align}
&\langle y_{jk},\calP_j \bv \rangle \langle z_{jk},\calC_j^{[1]} \bu \rangle - \langle y_{jk},\calP_j \bv \rangle \langle y_{jk},\calC_j^{[1]} \bu \rangle  \nonumber \\
=& \langle y_{jk},\calP_j \bv \rangle \langle z_{jk}-y_{jk},\calC_j^{[1]} \bu \rangle
\end{align}
but conditional on $\{ \scrX_{n/2+1},\scrX_{n/2+2},\dots,\scrX_{n} \}$, $\langle y_{jk},\calP_j \bv \rangle$ and  $\langle z_{jk}-y_{jk},\calC_j^{[1]} \bu \rangle$, $k=1,\dots,n$ are mean-$0$ Gaussian random variables and uncorrelated hence independent. So $\bar{\mathbb{E}}( \langle z_{jk} - y_{jk},\calP_j \bv \rangle \langle z_{jk},\calC_j^{[1]} \bu \rangle ) = 0$.

Moreover,
\begin{align}
&\bar{\mathbb{E}} \langle z_{jk}-y_{jk},\calC_j^{[1]} \bu \rangle^2 \nonumber \\
=& \sum_{l \neq j} \sigma_l^2 \langle \calC_j^{[1]} \bu, \bu_l^{(1)} \rangle^2 \prod_{q=2}^r \left( \langle \hat{\bu}_j^{(q)[2]}, \bu_l^{(q)} \rangle \right)^2  \nonumber\\
&\qquad + \left\| \calC_j^{[1]} \bu \right\|^2 \cdot \left\| \hat{\bu}_j^{(2)[2]} \otimes \dots \otimes \hat{\bu}_j^{(p)[2]} - \bu_j^{(2)} \otimes \dots \otimes \bu_j^{(p)} \right\|^2  \nonumber \\
\le & \left\| \calC_j^{[1]} \right\|^2 \max_{l \neq j} \left| \sigma_l^2 \langle \bu_l^{(2)}, \hat{\bu}_j^{(2)[2]} \rangle \right| \sqrt{ 1- \langle \bu_j^{(2)}, \hat{\bu}_j^{(2)[2]} \rangle^2 } + \left\| \calC_j^{[1]} \right\|^2 \left( 2- 2 \left(1-\frac{1}{2}a_n^2\right)^{p-1} \right)
\end{align}
Since
\begin{align}
\calC_j^{[1]}=\sum_{l \neq j} \frac{1}{\tilde{\sigma}_j^2 - \tilde{\sigma}_l^2} \bu_l^{(1)} \otimes \bu_l^{(1)} + \frac{1}{\tilde{\sigma}_j^2}  \calP_{-},
\end{align}
As already proven, for large enough $n$, $ \tilde{\sigma}_j^2 - \tilde{\sigma}_l^2 \ge \frac{1}{2} \sigma_j^2$, so for large enough $n$, we have
\begin{align}
\left\| \calC_j^{[1]} \right\| \le \frac{2}{ \sigma_j^2 }.
\end{align}
Because $\| \hat{\bu}_j^{(q)[2]} - \bu_j^{(q)} \| \le a_n $, and $ \max_{l \neq j} \left| \sigma_l^2 \langle \bu_l^{(2)}, \hat{\bu}_j^{(2)[2]} \rangle \right| \le A$,
\begin{align}
&\bar{\mathbb{E}} \langle z_{jk}-y_{jk},\calC_j^{[1]} \bu \rangle^2 \nonumber \\
\le & \frac{2}{ \sigma_j^2 } \left[ A a_n + 2- 2 \left(1-\frac{1}{2}a_n^2\right)^{p-1} \right]
\end{align}

and $ E \langle y_{jk},\calP_j \bv \rangle^2 = ( \sigma_j^2 +1 ) \langle \bu_j^{(1)},v \rangle^2$, so
\begin{align}
\bar{\mathbb{E}} \left[ \langle y_{jk},\calP_j \bv \rangle \langle z_{jk}-y_{jk},\calC_j^{[1]} \bu \rangle \right]^2 \le ( \sigma_j^2 +1 ) \frac{2}{ \sigma_j^2 } \left[ A a_n + 2- 2 \left(1-\frac{1}{2}a_n^2\right)^{p-1} \right]  \label{5351612}
\end{align}

\paragraph{Step 3.}

\begin{align}
&\langle y_{jk},\calP_j \bv \rangle \langle y_{jk},\calC_j^{[1]} \bu \rangle - \langle y_{jk},\calP_j \bv \rangle \langle y_{jk},\calC_j \bu \rangle  \nonumber \\
=& \langle y_{jk},\calP_j \bv \rangle \langle y_{jk},(\calC_j^{[1]}-\calC_j) \bu \rangle
\end{align}
but conditional on $\{ \scrX_{n/2+1},\scrX_{n/2+2},\dots,\scrX_{n} \}$, $\langle y_{jk},\calP_j \bv \rangle$ and  $\langle y_{jk},(\calC_j^{[1]}-\calC_j) \bu \rangle$, $k=1,\dots,n$ are mean-$0$ Gaussian random variables and uncorrelated hence independent. So $\bar{\mathbb{E}}( \langle y_{jk},\calP_j \bv \rangle \langle y_{jk},(\calC_j^{[1]}-\calC_j) \bu \rangle ) = 0$.

Moreover,
\begin{align}
\calC_j^{[1]} - \calC_j = \sum_{l \neq j} \left( \frac{1}{\tilde{\sigma}_j^2 - \tilde{\sigma}_l^2} - \frac{1}{\sigma_j^2} \right) \bu_l^{(1)} \otimes \bu_l^{(1)} + \left( \frac{1}{\tilde{\sigma}_j^2}  - \frac{1}{\sigma_j^2} \right)  \calP_{-} - \frac{1}{\sigma_j^2} \left( I_{\mathbb{H}} - I_{d_1} \right)
\end{align}

\begin{align}
&\bar{\mathbb{E}} \langle y_{jk},(\calC_j^{[1]}-\calC_j) \bu \rangle^2 \nonumber \\
=& \left\| \left[ \sum_{l \neq j} \left( \frac{1}{\tilde{\sigma}_j^2 - \tilde{\sigma}_l^2} - \frac{1}{\sigma_j^2} \right) \bu_l^{(1)} \otimes \bu_l^{(1)} + \left( \frac{1}{\tilde{\sigma}_j^2}  - \frac{1}{\sigma_j^2} \right)  \calP_{-} \right] \bu \right\|^2
\end{align}
but as already proven, $ (1-a_n^2)^{\frac{p}{2}} \sigma_j^2 \le \tilde{\sigma}_j^2$, and $\max_{l \neq j} \{ \tilde{\sigma}_l^2 \} \le A a_n \to 0$, and for large enough $n$, $ \tilde{\sigma}_j^2 - \tilde{\sigma}_l^2 \ge \frac{1}{2} \sigma_j^2$, so for large enough $n$, we have
\begin{align}
& \left\| \sum_{l \neq j} \left( \frac{1}{\tilde{\sigma}_j^2 - \tilde{\sigma}_l^2} - \frac{1}{\sigma_j^2} \right) \bu_l^{(1)} \otimes \bu_l^{(1)} + \left( \frac{1}{\tilde{\sigma}_j^2}  - \frac{1}{\sigma_j^2} \right)  \calP_{-} \right\|  \nonumber \\
=& \max_{l \neq j} \left( \frac{1}{\tilde{\sigma}_j^2 - \tilde{\sigma}_l^2} - \frac{1}{\sigma_j^2} \right) \bigvee \left( \frac{1}{\tilde{\sigma}_j^2}  - \frac{1}{\sigma_j^2} \right)  \nonumber \\
=& \max_{l \neq j} \left( \frac{1}{\tilde{\sigma}_j^2 - \tilde{\sigma}_l^2} - \frac{1}{\sigma_j^2} \right) \nonumber \\
\le & \frac{1}{(1-a_n^2)^{\frac{p}{2}} \sigma_j^2 - A a_n} - \frac{1}{\sigma_j^2} 
\end{align}
and $ E \langle y_{jk},\calP_j \bv \rangle^2 = ( \sigma_j^2 +1 ) \langle \bu_j^{(1)},v \rangle^2$,
so
\begin{align}
\bar{\mathbb{E}} \left[ \langle y_{jk},\calP_j \bv \rangle \langle y_{jk},(\calC_j^{[1]}-\calC_j) \bu \rangle \right]^2 \le ( \sigma_j^2 +1 ) \left( \frac{1}{(1-a_n^2)^{\frac{p}{2}} \sigma_j^2 - A a_n} - \frac{1}{\sigma_j^2} \right) .  \label{5351613}
\end{align}

Combining \eqref{5351611}, \eqref{5351612}, \eqref{5351613}, we have that on events $\mathcal{B}_n$, for large enough $n$,
\begin{align*}
& \bar{\mathbb{E}} \left( \langle z_{jk},\calP_j \bv \rangle \langle z_{jk},\calC_j^{[1]} \bu \rangle - \langle y_{jk},\calP_j \bv \rangle \langle y_{jk},\calC_j \bu \rangle  \right)^2   \nonumber \\
\le & \frac{1}{ \left[ (1-a_n^2)^{\frac{p}{2}} \sigma_j^2 \right]^4 }  \left[ \sigma_j^2 \left( \frac{1}{4} a_n^4 \right)^{p-1} + 2- 2 \left(1-\frac{1}{2}a_n^2\right)^{p-1} \right]   \nonumber \\
&+ ( \sigma_j^2 +1 ) \frac{2}{ \sigma_j^2 } \left[ A a_n + 2- 2 \left(1-\frac{1}{2}a_n^2\right)^{p-1} \right]  
+ ( \sigma_j^2 +1 ) \left( \frac{1}{(1-a_n^2)^{\frac{p}{2}} \sigma_j^2 - A a_n} - \frac{1}{\sigma_j^2} \right).
\end{align*}
Note that conditional on $\{ \scrX_{n/2+1},\scrX_{n/2+2},\dots,\scrX_{n} \}$,
\begin{align*}
\langle z_{jk},\calP_j \bv \rangle \langle z_{jk},\calC_j^{[1]} \bu \rangle - \langle y_{jk},\calP_j \bv \rangle \langle y_{jk},\calC_j \bu \rangle
\end{align*}
are independent with each other, so
\begin{align*}
& \bar{\mathbb{E}} \left( \sqrt{\frac{2}{n}} \sum_{k=1}^{n/2} \left[ \langle z_{jk},\calP_j \bv \rangle \langle z_{jk},\calC_j^{[1]} \bu \rangle - \langle y_{jk},\calP_j \bv \rangle \langle y_{jk},\calC_j \bu \rangle \right] \right)^2   \nonumber \\
\le & \frac{1}{ \left[ (1-a_n^2)^{\frac{p}{2}} \sigma_j^2 \right]^4 }  \left[ \sigma_j^2 \left( \frac{1}{4} a_n^4 \right)^{p-1} + 2- 2 \left(1-\frac{1}{2}a_n^2\right)^{p-1} \right]   \nonumber \\
&+ ( \sigma_j^2 +1 ) \frac{2}{ \sigma_j^2 } \left[ A a_n + 2- 2 \left(1-\frac{1}{2}a_n^2\right)^{p-1} \right]  
+ ( \sigma_j^2 +1 ) \left( \frac{1}{(1-a_n^2)^{\frac{p}{2}} \sigma_j^2 - A a_n} - \frac{1}{\sigma_j^2} \right),
\end{align*}
on events $\mathcal{B}_n$, for large enough $n$. Combine with $\lim_n a_n =0$, we have
\begin{align*}
& \mathbb{E} \left\{ \left( \sqrt{\frac{2}{n}} \sum_{k=1}^{n/2} \left[ \langle z_{jk},\calP_j \bv \rangle \langle z_{jk},\calC_j^{[1]} \bu \rangle - \langle y_{jk},\calP_j \bv \rangle \langle y_{jk},\calC_j \bu \rangle \right] \right) \cdot \mathbb{I}(\mathcal{B}_n) \right\}^2  \\
=& \mathbb{E} \left\{ \bar{\mathbb{E}} \left( \sqrt{\frac{2}{n}} \sum_{k=1}^{n/2} \left[ \langle z_{jk},\calP_j \bv \rangle \langle z_{jk},\calC_j^{[1]} \bu \rangle - \langle y_{jk},\calP_j \bv \rangle \langle y_{jk},\calC_j \bu \rangle \right] \right)^2 \cdot \mathbb{I}(\mathcal{B}_n) \right\} \\
\to & \ 0.
\end{align*}
So
\begin{align*}
\left( \sqrt{\frac{2}{n}} \sum_{k=1}^{n/2} \left[ \langle z_{jk},\calP_j \bv \rangle \langle z_{jk},\calC_j^{[1]} \bu \rangle - \langle y_{jk},\calP_j \bv \rangle \langle y_{jk},\calC_j \bu \rangle \right] \right) \cdot \mathbb{I}(\mathcal{B}_n)  
\overset{p}{\to} 0 ,
\end{align*}
thus proving \eqref{part3main}, and \eqref{part3finished} follows.

\end{document}